\documentclass{tacmconf}
\usepackage{graphicx} % [pdftex] for pdf
\usepackage{epsfig}
\usepackage{stmaryrd}
\usepackage{latexsym}
\usepackage{amssymb}
\usepackage{daytime}
\usepackage{eepic}
\usepackage{url}
\def\thisPaperTitle{On the Theory of Structural Subtyping}
\usepackage{defs}
\title{\thisPaperTitle}

\author{Viktor Kuncak and Martin Rinard\\
        Laboratory for Computer Science\\
        Massachusetts Institute of Technology\\
        Cambridge, MA 02139, USA \\
        {\tt $\{$vkuncak, rinard$\}$@lcs.mit.edu} \\ \\
        {\sf MIT-LCS-TR-879, January 2003}
}

\begin{document}

\sloppy

\maketitle

\renewcommand{\thefootnote}{\fnsymbol{footnote}}
\footnotetext[1]{This research was supported in part by
  DARPA Contract F33615-00-C-1692, NSF Grant CCR00-86154,
  NSF Grant CCR00-63513, and the Singapore-MIT Alliance.}
\footnotetext[2]{Draft of \today, \Daytime, \\ see 
\url{http://www.mit.edu/~vkuncak/papers} for later versions.}

\renewcommand{\thefootnote}{\arabic{footnote}}

\begin{abstract}
We show that the first-order theory of structural subtyping
of non-recursive types is decidable.

Let $\Sigma$ be a language consisting of function symbols
(representing type constructors) and $C$ a decidable
structure in the relational language $L$ containing a binary
relation $\leq$.  $C$ represents primitive types; $\leq$
represents a subtype ordering.  We introduce the notion of
\emph{$\Sigma$-term-power} of $C$, which generalizes the
structure arising in structural subtyping.  The domain of
the $\Sigma$-term-power of $C$ is the set of $\Sigma$-terms over the
set of elements of $C$.

We show that the decidability of the first-order theory of
$C$ implies the decidability of the first-order theory of
the $\Sigma$-term-power of $C$.  This result implies the
decidability of the first-order theory of structural
subtyping of non-recursive types.

Our decision procedure is based on quantifier elimination
and makes use of quantifier elimination for term algebras
and Feferman-Vaught construction for products of decidable
structures.

We also explore connections between the theory of structural
subtyping of recursive types and monadic second-order theory
of tree-like structures.  In particular, we give an
embedding of the monadic second-order theory of infinite
binary tree into the first-order theory of structural
subtyping of recursive types.
\end{abstract}

%%% Local Variables: 
%%% mode: latex
%%% TeX-master: "main"
%%% End: 

% LocalWords:  Feferman Vaught


\pagebreak

\tableofcontents

\pagebreak

\section{Introduction}

Subtyping constraints are an important technique for
checking and inferring program properties, used both in type
systems and program analyses
\cite{Mitchell91TypeInferenceSimpleTypes,  
  FreemanPfenning91RefinementML,
  DaviesPfenning00IntersectionTypesComputationalEffects,
  KozenETAL95EfficientRecursiveSubtyping,
  JimPalsberg99TypeInferenceSystemsRecursiveTypesSubtyping,  
  AmadioCardelli93SubtypingRecursiveTypes,
  AikenETAL94SoftTypingConditionalTypes,
  Aiken99SetConstraintsIntro,
  AikenETAL95DecidabilitySetConstraints,  
  HengleinRehof97ComplexitySubtypeEntailmentSimpleTypes,
  Pottier01SimplifyingSubtypingConstraints,
  Frey02SubtypeInequalitiesPolynomialSpace,
  Tiuryn92SubtypeInequalities,
  CharatonikPacholski94SetConstraintsProjections,
  CharatonikPodelski97SetConstraintsIntersection,
  Andersen94ProgramAnalysisSpecializationCProgrammingLanguage,
  Rehof98ComplexitySimpleSubtypingSystems,
  Steensgaard96PointsAnalysisAlmostLinearTime,
  HeintzeTardieu01UltraFastAliasAnalysis}.

This paper presents a decision procedure for the first-order
theory of structural subtyping of non-recursive types.  This
result solves (for the case of non-recursive types) a
problem left open in
\cite{SuETAL02FirstOrderTheorySubtypingConstraints}.
\cite{SuETAL02FirstOrderTheorySubtypingConstraints} provides
the decidability result for structural subtyping of only
unary type constructors, whereas we solve the problem for
any number of constructors of any arity.  Furthermore, we do
not impose any constraints on the subtyping relation $\leq$, it
need not even be a partial order.  The generality of our
construction makes it potentially of independent interest in
logic and model theory.

%%%%

We approach the problem of structural subtyping using
quantifier elimination and, to some extent, using monadic
second-order logic of tree-like structures.  This paper
makes the contributions:
\begin{itemize}
  
\item we give a new presentation of Feferman-Vaught theorem
  for direct products using a multisorted logic
  (Section~\ref{sec:fefermanVaught}); for completeness we
  also include proof of quantifier-elimination for boolean
  algebras of sets (Section~\ref{sec:qeBoolalg});
  
\item we give a new presentation of decidability of the
  first-order theory of term algebras; the proof uses the
  language of both constructor and selector symbols
  (Section~\ref{sec:qeTA});
  
\item as an introduction to main result, we show
  decidability of structural subtyping with one covariant
  binary constructor and two constants
  (Section~\ref{sec:twoconst}), this result does not rely on
  Feferman-Vaught technique;
  
\item we present a new construction, {\em term-power
    algebra} for creating tree-like theories based on
  existing theories (Section~\ref{sec:nconst});
  
\item as a central result, we prove that if the base theory
  is decidable, so is the theory of term-power with
  arbitrary variance of constructors; we give an effective
  decision procedure for quantifier elimination in
  term-power structure; the procedure combines elements of
  quantifier elimination in Feferman-Vaught theorem and
  quantifier elimination in term algebras
  (Sections~\ref{sec:nconst},~\ref{sec:nonrec}).
  
\item we show the decidability of structural subtyping
  non-recursive types as a direct consequence of the main
  result;
      
\item we give a simple embedding of monadic second-order
  theory of infinite binary tree into the theory of
  structural subtyping of recursive types with two primitive
  types (Section~\ref{sec:msolEmbedding});
  
\item we show that structural subtyping of recursive types
  where terms range over constant shapes is decidable
  (Section~\ref{sec:fixedRational});

\end{itemize}

In addition to showing the decidability of structural
subtyping, our hope is to promote the important technique of
quantifier elimination, which forms the basis of our result.

Quantifier elimination \cite[Section
2.7]{Hodges93ModelTheory} is a fruitful technique that was
used to show decidability and classification of boolean
algebras \cite{Skolem19Untersuchungen,
  Tarski49ArithmeticalClassesTypesBooleanAlgebras}
decidability of term algebras \cite[Chapter
23]{Malcev71MetamathematicsAlgebraicSystems},
\cite{Oppen80ReasoningRecursivelyDefinedDataStructures,
  Maher88CompleteAxiomatizationsAlgebrasTrees}, with
membership constraints
\cite{ComonDelor94EquationalFormulaeMembershipConstraints}
and with queues
\cite{RybinaVoronkov01DecisionProcedureTermAlgebrasQueues},
decidability of products
\cite{Mostowski52DirectProductsTheories,
  FefermanVaught59FirstOrderPropertiesProductsAlgebraicSystems},
\cite[Chapter 12]{Malcev71MetamathematicsAlgebraicSystems},
and algebraically closed fields
\cite{Tarski49ArithmeticalClassesTypesAlgebraicallyClosed},

The complexity of the decision problem for the first-order
theory of structural subtyping has a non-elementary lower
bound.  This is a consequence of a general theorem about
pairing functions \cite[Theorem 1.2, Page
163]{FerranteRackoff79ComputationalComplexityLogicalTheories}
and applies to term algebras already, as observed in
\cite{Oppen80ReasoningRecursivelyDefinedDataStructures,
  RybinaVoronkov01DecisionProcedureTermAlgebrasQueues}.

% !! Proper introduction like in LICS missing.

\nocite{Pfenning91UnificationCalculusConstructions}

\section{Preliminaries}

In this section we review some notions used in the this
paper.

If $w$ is a word over some alphabet, we write $|w|$ for the
length of $w$.  We write $w_1 \cdot w_2$ to denote the
concatenation of words $w_1$ and $w_2$.

A node $v$ in a directed graph is a {\em sink} if $v$ has no
outgoing edges.  A node $v$ in a directed graph is a {\em
source} if $v$ has no incoming edges.

We write $E_1 \equiv E_2$ to denote equality of syntactic
entities $E_1$ and $E_2$.

We write $\bar x$ to denote some sequence of variables
$x_1,\ldots,x_n$.

We assume that formulas are built from propositional
connectives $\land$, $\lor$, $\lnot$, the remaining connectives are
defined as shorthands.  Connective $\lnot$ binds the strongest,
followed by $\land$ and $\lor$.

A literal $L$ is an atomic formula $A$ or a negation of an
atomic formula $\lnot A$.  We define complementation of a
literal by $\neglit{A} = \lnot A$ and $\neglit{\lnot A}=A$.

A formula $\psi$ is in prenex form if it is of the form
\[
   Q_1 x_1. \ldots Q_n x_n. \phi
\]
where $Q_i \in \{ \forall,\exists\}$ for $1 \leq i \leq n$
and $\phi$ is a quantifier free formula.  We call $\phi$ a
{\em matrix} of $\psi$.

If $\phi$ is a formula then $\FV(\phi)$ denotes the set of free
variables in $\phi$.

We write $[x_1 \mapsto a_1,\ldots,x_k \mapsto a_k]$ for the substitution $\sigma$
such that $\sigma(x_i)=a_i$ for $1 \leq i \leq k$.

If $\phi$ is a formula and $t_1,\ldots,t_k$ terms, we write
$\phi[x_1:=t_1,\ldots,x_k:=t_k]$ for the result of simultaneously
substituting free occurrences of variables $x_i$ with term
$t_i$, for $1 \leq i \leq k$.

We write $\termHeight{t}$ for the height of term $t$.
$\termHeight{a}=0$ if $a$ is a constant, $\termHeight{x}=0$
if $x$ is a variable.  If $f(t_1,\ldots,t_k)$ is a term then
\[
   \termHeight{f(t_1,\ldots,t_k)} = 
      1 + \max(\termHeight{t_1},\ldots,\termHeight{t_k})
\]
We assume that all function symbols are of finite arity.  If
there are finitely many function symbols then for any
non-negative integer $k$ there is only a finite number of
terms $t$ such that $\termHeight{t} \leq k$.

If $\phi(u)$ is a conjunction of literals, we say that $\phi'$
results from $\exists u. \phi(u)$ by {\em dropping quantified
variable $u$} iff $\phi'$ is the result of eliminating from
$\phi(u)$ all conjunctions containing $u$.  More generally,
if $\psi$ is a formula of form
\[
   Q_1 x_1 \ldots Q u \ldots Q_k x_k.\ \psi_0
\]
then the result of dropping $u$ from $\psi$ is
\[
   Q_1 x_1 \ldots Q_k x_k.\ \psi'_0
\]
where $\psi'_0$ is the result of dropping $u$ from $\exists u. \phi_0$.

An {\em equality} is an atomic formula $t_1 = t_2$ where
$t_1$ and $t_2$ are terms.  A {\em disequality} is negation
of an equality.

We use the usual Tarskian semantics of formulas.  Unless
otherwise stated $\phi \models \psi$ will denote that formula $\phi
\implies \psi$ is true in a fixed relational structure that is
under current consideration.

Occasionally we find it convenient to work with multisorted
logic, where domain is union of disjoint sets called sorts,
and arity specifies the sorts of all operations.  Constants
are operations with zero arguments.  Relations are
operations that return the result in a distinguished sort
$\boolSort$ interpreted over the boolean lattice
$\{\boolFalse,\boolTrue\}$ or over the distributive lattice of
three-valued logic $\{ \boolFalse, \boolTrue, \boolUndef\}$
from Section~\ref{sec:partFun}).

A structure ${\cal C}$ of a given
language $L$ is a pair of domain $C$ and the interpretation
function $\interpretS{\_}{C}$.  Hence, we name operations of
the structure using symbols of the language and the
interpretation function.  If $C$ is clear from the context
we write simply $\interpret{\_}$ for $\interpretS{\_}{C}$.

In Section~\ref{sec:fefermanVaught} and
Section~\ref{sec:nonrec} we
use logic with several kinds of quantifiers.  Our
logic is first-order, but we give higher-order types to
quantifiers.  For example, a quantifier
\begin{equation*}
  Q ::  (A \to B) \to B
\end{equation*}
denotes a quantifier that binds variables of $A$ sort
enclosed within an expression of $B$ sort and returns an
expression of $B$ sort.  If $X$ and $Y$ are sets then $X \to
Y$ denotes the set of all functions from $A$ to $B$.  When
specifying the semantics of the quantifier $Q$ we specify a
function
\begin{equation*}
  \interpret{Q} : (\interpret{A} \to \interpret{B}) \to \interpret{B}
\end{equation*}
The semantics of an expression $M$ of sort $B$ takes an
environment $\sigma$ which is a function from variable names to
elements of $A$ and produces an element of $B$, hence
$\interpret{M}\sigma \in \interpret{B}$.  We define the semantics
of an expression $Q x.\ M$ by:
\begin{equation*}
  \interpret{Q x.\ M}\sigma = \interpret{Q} h
\end{equation*}
where $h : \interpret{A} \to \interpret{B}$ is the function
\begin{equation*}
  h(a) = \interpret{M} (\sigma[x:=a])
\end{equation*}
Here
\begin{equation*}
  \sigma[x:=a](y) = \left\{ \begin{array}{rl}
      \sigma(y), & \mbox{ if } y \not\equiv x \mnl
      a, & \mbox{ if }  y \equiv x
    \end{array} \right.
\end{equation*}
Specifying types for quantifiers allows to express more

%% A higher-order multisorted structure is just simply typed
%% lambda calculus (without polymorphic types).  This allows
%% theories that have several different kinds of quantifiers
%% (like in Feferman-Vaught structure in
%% Section~\ref{sec:fefermanVaught}).  In contrast, we cannot
%% get multiple quantifier in first-order multisorted logic.
  
%% This definition of structure is quite nice, because it
%% allows logic to define it's own quantifiers, but still
%% factors out from definition of structure the environment
%% updates.  We assume the standard interpretation of typed
%% lambda calculus, i.e.\
%% \[
%%    \interpret{\lambda x. M}\sigma \ = \ \lambda d.\ \interpret{M}(\sigma[x:=d])
%% \]
%% We then have
%% \[
%%    \interpret{\forall x. M}\sigma \ = \ \interpret{\forall}\, (\interpret{\lambda x.M}\sigma)
%% \]
%% Note that higher-order functions are immediately consumed by
%% the quantifier, so they are more of a convenience than a
%% necessity.  See Sections~\ref{sec:fefermanVaught}
%% and~\ref{sec:nonrec} for examples of theories with
%% user-defined quantifiers.  

Let $\sigma_A$ be some arbitrary dummy global environment.  If
$F$ is a formula without global variables we write
$\interpret{F}\sigma_A$ to denote the truth value of $F$; clearly
$\interpret{F}\sigma_A$ does not depend on $\sigma_A$ and we denote it
simply $\interpret{F}$ when no ambiguity arises.

We use Hilbert's epsilon as a notational convenience in
metatheory.  If $P(x)$ is a unary predicate, then $\varepsilon x.
P(x)$ denotes an arbitrary element $d$ such that $P(d)$
holds, if such element exists, or an arbitrary object
otherwise.

\subsection{Term Algebra} \label{sec:termAlgebrasDef}

We introduce the notion of term algebra \cite[Page
14]{Hodges93ModelTheory}.

Let $\Nat$ be the set of natural numbers.  Let the signature
$\FreeSig$ be a finite set of function symbols and constants
and let $\ar : \FreeSig \to \Nat$ be a function specifying
arity $\ar(f)$ for every function symbol or constant $f \in
\FreeSig$.  Let $\FT$ denote the set of finite ground terms
over signature $\FreeSig$.  We assume that $\FreeSig$
contains at least one constant $c \in \FreeSig$, $\ar(c)=0$,
and at least one function symbol $f \in \FreeSig$, $\ar(f) >
0$.  Therefore, $\FT$ is countably infinite.

Let $\FreeOps$ be the term algebra interpretation of
signature $\FreeSig$, defined as follows
\cite[Page 14]{Hodges93ModelTheory}.  For every $f \in
\FreeSig$ with $\ar(f)=k$ define $\freeOf{f} \in
\FreeOps$, with $\freeOf{f} : \FT^k \to \FT$ by
\[
   \freeOf{f}(t_1,\ldots,t_k) = f(t_1,\ldots,t_k)
\]
We will write $f$ instead of $\freeOf{f}$ when it causes no
confusion.

\subsection{Terms as Trees}
\label{sec:termsAsTrees}

We define trees representing terms as follows.  

We use sequences of nonegative integers to denote paths in
the tree.  Let $\FreeSig$ be a signature.  A tree over
$\FreeSig$ is a partial function $t$ from the set
$\TreePaths$ of paths to the set $\FreeSig$ of function
symbols such that:
\begin{enumerate}
\item if $w \in \TreePaths$, $x \in \TreeLan$, and $t(w \cdot x)$ is
      defined, then $t(w)$ is defined as well;
\item if $t(w) = f$ with $\ar(f)=k$, then
\[
   \{ i \mid t(w \cdot i) \mbox{ is defined } \} = \{ 1,\ldots,k \}
\]
\end{enumerate}
A {\em finite tree} is a tree with a finite domain.

%%% Local Variables: 
%%% mode: latex
%%% TeX-master: "main"
%%% End: 
% LocalWords:  prenex arity iff Tarskian multisorted Feferman Vaught Hilbert's
% LocalWords:  disequality metatheory nonegative

  \subsection{First Order Structures with Partial Functions}
\label{sec:partFun}

We make use of partial functions in our quantifier
elimination procedures.  In this section we briefly describe
the approach to partial functions we chose to use; other
approaches would work as well, see e.g.\ 
\cite{KerberKohlhase94MechanizationPartialFunctions}.

A language of partial functions $\PartLan$ contains partial
function symbols in addition to total function symbols and
relation symbols.  Consider a structure with the domain $A$
interpreting a language with partial function symbols $\Sigma_1$.
Given some environment $\sigma$, we have $\interpret{t}\sigma \in A \cup \{
\bot \}$ where $\bot \notin A$ is a special value denoting undefined
results.  We require the interpretations of total and
partial function symbols to be strict in $\bot$, i.e.\
$f(a_1,\ldots,a_i,\bot,a_{i+2},\ldots,a_k)=\bot$.

We interpret atomic formulas and their negations over the
three-valued domain $\{\boolFalse,\boolTrue,\boolUndef \}$
using strong Kleene's three-valued logic
\cite{Kleene52IntroductionMetamathematics,
  KerberKohlhase94MechanizationPartialFunctions,
  SagivETAL02Parametric}.  We require that
$\interpret{R}(a_1,\ldots,a_i,\bot,a_{i+2},\ldots,a_k) = \boolUndef$ for
every relational symbol $R$.  Logical connectives in
Kleene's strong three-valued logic are the strongest
``regular'' extension of the corresponding connectives on
the two-valued domain
\cite{Kleene52IntroductionMetamathematics}.  The regularity
requirement means that the three-valued logic is a sound
approximation of two-valued logic in the following sense.
We may obtain the truth tables for three-valued logic by
considering the truth values
$\boolFalse,\boolTrue,\boolUndef$ as shorthands for sets
$\{\boolFalse\}, \{\boolTrue\}, \{\boolFalse,\boolTrue\}$ and
defining each logical operation $*$ by:
\[
    s_1 \ \interpret{*} \ s_2 = \{ b_1 \circ b_2 \mid b_1 \in s_1 \land b_2 \in s_2 \}
\]
where $\circ$ denotes the corresponding operation in the
two-valued logic.  As in a call-by-value semantics of lambda
calculus, variables in the environments ($\sigma$) do not range
over $\bot$.  We interpret quantifiers as ranging over the
domain $A$ or its subset if the logic is multisorted; the
interpretation of quantifiers are similarly the best regular
approximations of the corresponding two-valued
interpretations.

These properties of Kleene's three-valued logic have the
following important consequence.  Suppose that we extend the
definition of all partial functions to make them total
functions on the domain $A$ by assigning arbitrary values
outside the original domain.  Suppose that a formula $\phi$
evaluates to an element of $b \in \{\boolFalse,\boolTrue\}$ in
Kleene's logic.  Then $\phi$ evaluates to the \emph{same}
truth-value $b$ in the new logic of total functions.  This
property of three-valued logic implies that the algorithms
that we use to transform formulas with partial functions
will apply even for the logic that makes all functions total
by completing them with arbitrary elements of $A$.

We say that a formula $\psi$ is {\em well-defined} iff its
truth value is an element of $\{\boolFalse,\boolTrue\}$.
%% Unless otherwise stated we will be assuming that every
%% formula is well-defined.
\begin{example}
  Consider the domain of real numbers.
  The following formulas are not well-defined:
  \[\begin{array}{l}
    3 = 1/0 \mnl
    \forall x.\ 1/x > 0 \ \lor \ 1/x < 0 \ \lor \ 1/x = 0
  \end{array}\]
  The following formulas are well-defined:
  \[\begin{array}{l}
    \exists x.\ 1/x = 3 \mnl
    \forall x.\ 1/x \neq 3 \mnl   
    x = 0 \ \lor\ 1/x > 0 \mnl
  \end{array}\]
\end{example}

We say that a formula $\phi_1$ is {\em equivalent} to a formula
$\phi_2$ and write $\phi_1 \fullyEquiv \phi_2$ iff
\begin{equation*}
  \interpret{\phi_1}\sigma = \interpret{\phi_2}\sigma
\end{equation*}
for all valuations $\sigma$ (including those for which
$\interpret{\phi_1}\sigma = \boolUndef$).

Sections below perform equivalence-preserving
transformations of formulas.  This means that starting from
a well-defined formula we obtain an equivalent well-defined
formula.

When doing equivalence preserving transformations it is
useful to observe that $\land,\lor$ still form a distributive
lattice.  The partial order of this lattice is the chain
$\boolFalse \leq \boolUndef \leq \boolTrue$.  The element
$\boolUndef$ does not have a complement in the lattice;
unary operation $\lnot$ does not denote the lattice complement.
However, the following laws still hold:
\[\begin{array}{l}
    \lnot (x \land y) \fullyEquiv \lnot x \lor \lnot y \mnl
    \lnot (x \lor y) \fullyEquiv \lnot x \land \lnot y \mnl
    \lnot \lnot x \fullyEquiv x
\end{array}\]
The properties of $\land,\lor,\lnot$ are sufficient to transform any
quantifier-free formula into disjunction of conjunctions of
literals using the well-known straightforward technique.
However, this straightforward technique in some cases yields
conjunctions that are not well-defined, even though the
formula as a whole is well-defined.
\begin{example}
  Transforming a negation of well-defined formula:
  \[
  \lnot (x \neq 0 \land (y = 1/x \lor z = x+1))
  \]
  may yield the following disjunction of conjunctions:
  \[
  x = 0 \lor (y \neq 1/x \land z \neq x+1)
  \]
  where $y \neq 1/x \land z \neq x+1$ is not a well-defined
  conjunction for $x=0$.
\end{example}

To enable the transformation of each well-defined formula
into a disjunction of well-defined conjunctions of literals,
we enrich the language of function and relation symbols as
follows.  With each partial function symbol $f \in \PartLan$
of arity $k=\ar(f)$ we associate a \emph{domain description}
$D_f = \tu{\tu{x_1,\ldots,x_k},\phi}$ specifying the domain of $f$.
Here $x_1,\ldots,x_k$ are distinct variables and $\phi$ is an
unnested conjunction of literals such that $\FV(\phi) \subseteq \{
x_1,\ldots,x_k \}$.  We require every interpretation of a
first-order structure with partial function symbols to
satisfy the following property:
\[
    \interpret{f}(a_1,\ldots,a_k) \neq \bot \iff \interpret{\phi}[x_1 \mapsto a_1,\ldots,x_k \mapsto a_k]
\]
for all $a_1,\ldots,a_k \in A$.  We henceforth assume that every
structure with partial functions is equipped with a domain
description $D_f$ for every partial function symbol $f$.  
%% If
%% $t_1,\ldots,t_k$ are terms, we write $D_f(t_1,\ldots,t_k)$ for the
%% formula $\phi[x_1:=t_1,\ldots,x_k:=t_k]$.

The Proposition~\ref{prop:partialDNF} below gives an
algorithm for transforming a given well-defined formula into
a disjunction of well-defined conjunctions.  We first give
some definitions and lemmas.

\begin{definition} \label{def:domainFormula}
  If $\psi$ is a formula with free variables, a \emph{domain
    formula} for $\psi$ is a formula $\phi$ not containing partial
  function symbols such that, for every valuation $\sigma$,
\[
    \interpret{\psi}\sigma \neq \boolUndef  \iff  \interpret{\phi}\sigma = \boolTrue
\]
\end{definition}
From Definition~\ref{def:domainFormula} we obtain the following
Lemma~\ref{lemma:makeDefined}.
\begin{lemma} \label{lemma:makeDefined}
Let $\psi$ be a formula and $\phi$ a domain formula for $\psi$.
Then
\[
    \psi  \fullyEquiv  (\psi \land \phi) \lor (\boolUndef \land \lnot \phi)
\]
\end{lemma}
\begin{proof}
Let $\sigma$ be arbitrary valuation.  Let $v=\interpret{\psi}\sigma$.  If
$v \in \{\boolTrue,\boolFalse\}$ then $\interpret{\phi}\sigma=\boolTrue$ and
\[\begin{array}{l}
   \interpret{(\psi \land \phi) \lor (\boolUndef \land \lnot \phi)}\sigma = \mnl
 \qquad   (v \land \boolTrue) \lor (\boolUndef \land \boolFalse) = v.
\end{array}\]
If $v=\boolUndef$ then $\interpret{\phi}=\boolFalse$, so
\[\begin{array}{l}
   \interpret{(\psi \land \phi) \lor (\boolUndef \land \lnot \phi)}\sigma = \mnl
\qquad (\boolUndef \land \boolFalse) \lor (\boolUndef \land \boolTrue) = \boolUndef.
\end{array}\]
\end{proof}

Observe that $\psi \land \phi$ in Lemma~\ref{lemma:makeDefined} is a
well-defined conjunction.  We use this property to construct
domain formulas using partial function domain descriptions.

Let
\[
    D_f = \tu{\tu{x_1,\ldots,x_k},B^f_1 \land \ldots \land B^f_{l^f}}
\]
for each partial function symbol $f \in \PartLan$ of arity
$k$, where $B^f_1,\ldots,B^f_{l^f}$ are unnested literals.  If
$t_1,\ldots,t_k$ are terms, we write $B^f_i(t_1,\ldots,t_k)$ for
$B^f_i[x_1:=t_1,\ldots,x_k:=t_k]$.  Let $\subterms{t}$ denote the
set of all subterms of term $t$.

For any literal $B(t_1,\ldots,t_n)$ where
$B(t_1,\ldots,t_n) \equiv R(t_1,\ldots,t_n)$ or $B(t_1,\ldots,t_n) \equiv \lnot R(t_1,\ldots,t_n)$,
define
\begin{equation} \label{eqn:domMakeDef}
\begin{array}{l}
    \domFormMake{B(t_1,\ldots,t_n)} = \mnl
\qquad
      \bigwedge\limits_{\begin{array}{c} \scriptstyle
         f(s_1,\ldots,s_k) \in \cup_{1 \leq i \leq n} \subterms{t_i} \mnls \scriptstyle
         1 \leq j \leq l^f
         \end{array}}
         B^f_j(s_1,\ldots,s_k)
\end{array}
\end{equation}

\begin{lemma}
Let $B(t_1,\ldots,t_n)$ be a literal containing partial function
symbols.  Then $\domFormMake{B(t_1,\ldots,t_n)}$ is a domain
formula for $B(t_1,\ldots,t_n)$.
\end{lemma}
\begin{proof}
Let $\sigma$ be a valuation.  By strictness of
interpretations of function and predicate
symbols, $\interpret{B(t_1,\ldots,t_n)}\sigma \neq \boolUndef$ iff
$\interpret{f(s_1,\ldots,s_k)}\sigma \neq \bot$ for every subterm
$f(s_1,\ldots,s_k)$ of every term $t_i$, iff
$\interpret{B^f_j(s_1,\ldots,s_k)}\sigma = \boolTrue$ for every $1 \leq j
\leq l^f$ and every subterm $f(s_1,\ldots,s_k)$.
\end{proof}

\begin{lemma} 
\label{lemma:wellifyLiteral}
Let $B$ be a literal and let
\[
   \domFormMake{B}=F_1 \land \ldots \land F_m.
\]
Then
\[\begin{array}{l}
   B \fullyEquiv 
    \begin{array}[t]{l}
        (B \land F_1 \land \ldots \land F_m)\ \lor \mnl
        \bigvee_{1 \leq i \leq m} (\boolUndef \land \lnot F_i \land \domFormMake{F_i})
       \end{array}
\end{array}\]
\end{lemma}
\begin{proof}
If $\interpret{B}\sigma \neq \boolUndef$, then $\interpret{F_i}\sigma=\boolTrue$
for every $1 \leq i \leq m$, and
\[
    \interpret{\boolUndef \land \lnot F_i \land \domFormMake{F_i}}\sigma = \boolFalse
\]
so the right-hand side evaluates to $\interpret{B}\sigma$ as well.  Now
consider the case when $\interpret{B}\sigma = \boolUndef$.  Then
there exists a term $f(s_1,\ldots,s_k)$ such that
$\interpret{f(s_1,\ldots,s_k)}\sigma = \boolUndef$.  Because $\sigma(x) \neq
\bot$ for every variable $x$, there exists a term
$f(s_1,\ldots,s_k)$ such that $\interpret{f(s_1,\ldots,s_k)}\sigma =
\boolUndef$ and $\interpret{s_i}\sigma \neq \boolUndef$ for $1 \leq i \leq
k$.  Then there exists a formula $F_p$ of form
$B^f_j(s_1,\ldots,s_k)$ such that $\interpret{B^f_j(s_1,\ldots,s_k)}\sigma
= \boolFalse$, and
\[
    \interpret{\boolUndef \land \lnot F_p \land \domFormMake{F_p}}\sigma = \boolUndef.
\]
Because
\[
    \interpret{B \land F_1 \land \ldots \land F_m}\sigma = \boolFalse,
\]
and for every $q$,
\[
    \interpret{\boolUndef \land \lnot F_q \land \domFormMake{F_q}}\sigma \in
        \{ \boolUndef, \boolFalse \},
\]
the right-hand side evaluates to $\boolUndef$.
\end{proof}

\begin{lemma} \label{lemma:matrixUndefElim}
Let $\phi_0(\ybar)$ and $\phi_1(\ybar)$ be well-defined formulas whose
free variables are among $\ybar$ and let
\[
   \psi(\ybar) \ \equiv \ (\boolUndef \land \phi_0(\ybar)) \lor \phi_1(\ybar)
\]
If $\psi(\ybar)$ is well-defined for all values of variables
$\ybar$, then
\[
   \psi(\ybar) \ \fullyEquiv \ \phi_1(\ybar)
\]
\end{lemma}
\begin{proof}
  Consider any valuation $\sigma$.  Let
  \[
     v = \interpret{\phi_1(\ybar)}\sigma
  \]
  and
  \[
     v' = \interpret{\psi(\ybar)}\sigma
  \]
  We need to show $v=v'$.  Because $\phi(\ybar)$ and $\psi(\ybar)$
  are well-defined, $v,v' \in \{ \boolFalse, \boolTrue \}$.  We
  consider two cases.

  \noindent
  {\bf Case 1.} $v = \boolTrue$.  Then also 
  $v' = \boolTrue$.

  \noindent
  {\bf Case 2.} $v = \boolFalse$.  Then
  $v' = \boolUndef \land \phi_0(\ybar)$.
  Because $v' \neq \boolUndef$, we conclude $v' = \boolFalse$.
\end{proof}

\begin{proposition}
\label{prop:partialDNF}
Every well-defined quantifier-free formula $\psi$ can be
transformed into an equivalent disjunction $\psi'$ of
well-defined conjunctions of literals.
\end{proposition}
\begin{proof}
Using the standard procedure, convert $\psi$ to disjunction of conjunctions 
\[
    C_1 \lor \ldots \lor C_n
\]
Let $C_i = B \land C_i'$ where $B$ is a literal and let
$\domFormMake{B}=F_1 \land \ldots \land F_m$.
Replace $B \land C_i'$ by
\[\begin{array}{l}
        (B \land F_1 \land \ldots \land F_m \land C_i')\ \lor \mnl
        \bigvee_{1 \leq i \leq m} (\boolUndef \land \lnot F_i \land \domFormMake{F_i} \land C_i')
\end{array}\]
By Lemma~\ref{lemma:wellifyLiteral} and distributivity, the
result is an equivalent formula.  Repeat this process for
every literal in $C_1 \lor \ldots \lor C_n$.  The result can be written in the form
\begin{equation} \label{eqn:hasUndefs}
  (\boolUndef \land \phi_1) \lor \ldots \lor (\boolUndef \land \phi_p) \lor
  \phi_{p+1} \lor \ldots \lor \phi_{p+q}
\end{equation}
where each $\phi_i$ for $1 \leq i \leq p+q$ is a well-defined conjunction.
Formula~(\ref{eqn:hasUndefs}) is equivalent to
\begin{equation} \label{eqn:hasUndefsSimple}
  (\boolUndef \land (\phi_1 \lor \ldots \lor \phi_p)) \lor
  \phi_{p+1} \lor \ldots \lor \phi_{p+q}
\end{equation}
and is equivalent to the well-defined formula $\psi$, so it is
well-defined.  Formulas $\phi_1 \lor \ldots \lor \phi_p$ and $\phi_{p+1} \lor \ldots \lor
\phi_{p+q}$ are also well-defined.  By
Lemma~\ref{lemma:matrixUndefElim}, we conclude that
formula (\ref{eqn:hasUndefsSimple}) is equivalent to
\begin{equation} \label{eqn:partialDNFres}
  \phi_{p+1} \lor \ldots \lor \phi_{p+q}
\end{equation}
Because~(\ref{eqn:partialDNFres}) is a disjunction of
well-defined formulas, (\ref{eqn:partialDNFres}) is the
desired result $\psi'$.
\end{proof}

The following proposition presents transformation to
unnested form for the structures with equality and partial
function symbols, building on
Proposition~\ref{prop:partialDNF}.  For a similar unnested
form in the first-order logic containing only total function
symbols, see \cite[Page 58]{Hodges93ModelTheory}.
\begin{proposition} \label{prop:unnestedDNFpartial}
  Every well-defined quantifier-free formula $\psi$ in a
  language with equality can be effectively transformed into an
  equivalent formula $\psi'$ where $\psi'$ is a disjunction of
  existentially quantified well-defined conjunctions of the
  following kinds of literals:
  \begin{itemize}
  \item $R(x_1,\ldots,x_k)$ where $R$ is some relational symbol of arity $k$
    and $x_1,\ldots,x_k$ are variables;
  \item $\lnot R(x_1,\ldots,x_k)$ where $R$ is some relational symbol of arity $k$
    and $x_1,\ldots,x_k$ are variables;
  \item $x_1 = x_2$ where $x_1,x_2$ are variables;
  \item $x = f(x_1,\ldots,x_k)$ where $f$ is some partial or
    total function symbol of arity $k$ and $x,x_1,\ldots,x_k$ are
    variables;
  \item $x_1 \neq x_2$ where $x_1$ and $x_2$ are variables.
  \end{itemize}
\end{proposition}
\begin{proof}
  Transform the formula to disjunction of well-formed conjunctions
  of literals as in the proof of Proposition~\ref{prop:partialDNF}.
  
  Then repeatedly perform the following transformation on
  each well-defined conjunction $\phi$.
  Let $A(f(x_1,\ldots,x_k))$ be an atomic formula containing term
  $f(x_1,\ldots,x_k)$.  Replace $\phi \land A(f(x_1,\ldots,x_k))$ with
  \[
      \exists x_0.\ \phi \land x_0 = f(x_1,\ldots,x_k) \land A(x_0)
  \]
  Replace $x \neq f(x_1,\ldots,x_k)$ with
  \[
     x_0 = f(x_1,\ldots,x_k) \land x_0 \neq x
  \]
  Repeat this process until the resulting conjunction $\phi'$
  is in unnested form.  $\phi'$ is clearly equivalent to the
  original conjunction $\phi$ when all partial functions are
  well-defined.  When some partial function is not
  well-defined, then both $\phi$ and $\phi'$ evaluate to
  $\boolFalse$, because by construction of $\phi$ in the proof
  of Proposition~\ref{prop:partialDNF}, each conjunction
  contains conjuncts that evaluate to $\boolFalse$ when some
  application of a function symbol is not well-defined.
\end{proof}

Let a \emph{left-strict} conjunction in Kleene logic be
denoted by  $\lsland$ and defined by
\[
   p \lsland q = (p \land q) \lor (p \land \lnot p)
\]

The correctness of the transformation to unnested form in
Proposition~\ref{prop:unnestedDNFpartial} relies on the
presence of conjuncts that ensure that the entire
conjunction evaluates to $\boolFalse$ whenever some term is
undefined.  The following Lemma~\ref{lemma:generalUnnesting}
enables transformation to unnested form in an arbitrary
context, allowing the transformation to unnested form to be
performed independently from ensuring well-definedness of
conjuncts.
\begin{lemma} \label{lemma:generalUnnesting}
  Let $\phi(x)$ be a formula with free variable $x$ and let $t$
  be a term possibly containing partial function symbols.
  Then
  \begin{enumerate}
  \item
    $
      \begin{array}{l}
        \phi(t) \ \fullyEquiv \
        (\exists x.\ x = t \ \land \ \phi(x)) \ \lor \ (\boolUndef \land \forall x. \lnot \phi(x))
      \end{array}
    $;
  \item 
    $
      \begin{array}{l}
        \phi(t) \ \fullyEquiv \
        \exists x.\ x = t \lsland \phi(x)
      \end{array}
    $;
  \item
    $
      \begin{array}{l}
        \phi(t) \ \fullyEquiv \
        (\exists x.\ x = t \ \land \ \phi(x)) \ \lor \ (t \neq t)
      \end{array}
    $.
  \end{enumerate} 
\end{lemma}
\begin{proof}
Straightforward.
\end{proof}

Proposition~\ref{prop:quantUndefElim} below shows that a
simplification similar to one in
Lemma~\ref{lemma:matrixUndefElim} can be applied even within
the scope of quantifiers.  To show
Proposition~\ref{prop:quantUndefElim} we first show two
lemmas.

\begin{lemma} \label{lemma:existsUndefElim}
  For all formulas 
  $\phi_0(x,\ybar)$ and $\phi_1(x,\ybar)$,
  \[\begin{array}{l}
    \exists x.\ (\boolUndef \land \phi_0(x, \ybar)) \lor \phi_1(x,\ybar) \ \fullyEquiv \mnl
    (\boolUndef \land \exists x. \phi_0(x,\ybar)) \lor \exists x. \phi_1(x,\ybar)
  \end{array}\]
\end{lemma}
\begin{proof}
  By distributivity of quantifiers and propositional
  connectives in Kleene logic we have:
  \[\begin{array}{l}
    \exists x.\ (\boolUndef \land \phi_0(x, \ybar)) \lor \phi_1(x,\ybar) \ \fullyEquiv \mnl
    (\exists x. \boolUndef \land \phi_0(x, \ybar)) \lor \exists x. \phi_1(x,\ybar) \ \fullyEquiv \mnl
    (\boolUndef \land \exists x. \phi_0(x,\ybar)) \lor \exists x. \phi_1(x,\ybar)
  \end{array}\]
\end{proof}

\begin{lemma} \label{lemma:forallUndefElim}
For all formulas $\phi_0(x,\ybar)$ and $\phi_1(x,\ybar)$,
\[\begin{array}{l}
    \forall x.\ (\boolUndef \land \phi_0(x, \ybar)) \lor \phi_1(x,\ybar) \ \fullyEquiv \mnl
    (\boolUndef \land \forall x. \phi_0(x,\ybar) \lor \phi_1(x,\ybar)) \lor
    \forall x. \phi_1(x,\ybar)
\end{array}\]
\end{lemma}
\begin{proof}
  The following sequence of equivalences holds.
\[\begin{array}{l}
    \forall x.\ (\boolUndef \land \phi_0(x, \ybar)) \lor \phi_1(x,\ybar) \ \fullyEquiv \mnl
    \lnot \exists x. \lnot (\boolUndef \land \phi_0(x, \ybar)) \lor \phi_1(x,\ybar) \ \fullyEquiv \mnl
    \lnot \exists x.\ (\boolUndef \lor \lnot \phi_0(x,\ybar)) \land \lnot \phi_1(x,\ybar) \ \fullyEquiv \mnl
    \lnot \exists x.\ (\boolUndef \land \lnot \phi_1(x,\ybar)) \lor 
          (\lnot \phi_0(x,\ybar) \land \lnot \phi_1(x,\ybar)) \ \fullyEquiv \mnl
    \lnot \left( (\boolUndef \land \exists x. \lnot \phi_1(x,\ybar)) \lor 
           (\exists x.\ \lnot \phi_0(x,\ybar) \land \lnot \phi_1(x,\ybar)) \right) \ \fullyEquiv \mnl
    (\boolUndef \lor \forall x. \phi_1(x,\ybar)) \lor 
    (\forall x.\ \phi_0(x,\ybar) \lor \phi_1(x,\ybar)) \ \fullyEquiv \mnl
    (\boolUndef \land \forall x. \phi_0(x,\ybar) \lor \phi_1(x,\ybar)) \lor
    \forall x. \phi_1(x,\ybar)
\end{array}\]
\end{proof}

\begin{proposition} \label{prop:quantUndefElim}
Let $\phi_0(\xbar,\ybar)$ and $\phi_1(\xbar,\ybar)$ be well-defined formulas whose
free variables are among $\ybar$ and let
\[
   \psi(\ybar) \ \equiv \ 
     Q_1 x_1 \ldots Q_n x_n.\
     (\boolUndef \land \phi_0(\xbar,\ybar)) \lor \phi_1(\xbar,\ybar)
\]
where $Q_1,\ldots,Q_n$ are quantifiers.
If $\psi(\ybar)$ is well-defined for all values of variables
$\ybar$, then
\[
   \psi(\ybar) \ \fullyEquiv \ Q_1 x_1 \ldots Q_n x_n.\ \phi_1(\xbar,\ybar)
\]
\end{proposition}
\begin{proof}
Applying successively Lemmas~\ref{lemma:existsUndefElim}
and~\ref{lemma:forallUndefElim} to quantifiers $Q_n,\ldots,Q_1$,
we conclude
\[\begin{array}{l}
   \psi(\ybar) \ \fullyEquiv \ 
    (\boolUndef \land \phi_2(\ybar)) \lor Q_1 x_1 \ldots Q_n x_n.\ \phi_1(\xbar,\ybar)
\end{array}\]
for some formula $\phi_2(\ybar)$.  Then by Lemma~\ref{lemma:matrixUndefElim},
\[
   \psi(\ybar) \ \fullyEquiv \ Q_1 x_1 \ldots Q_n x_n.\ \phi_1(\xbar,\ybar).
\]
\end{proof}

%%% Local Variables: 
%%% mode: latex
%%% TeX-master: "main"
%%% End: 
% LocalWords:  Kleene's SagivETAL multisorted iff arity unnested subterms
% LocalWords:  subterm Kleene definedness

\section{Some Quantifier Elimination Procedures}

As a preparation for the proof of the decidability of term
algebras of decidable theories, we present quantifier
elimination procedures for some theories that are known to
admit quantifier elimination.  We use the results and ideas
from this section to show the new results in
Sections~\ref{sec:twoconst}, \ref{sec:nconst},
\ref{sec:nonrec}.

%%% Local Variables: 
%%% mode: latex
%%% TeX-master: "main"
%%% End: 

  \subsection{Quantifier Elimination} \label{sec:qeIntro}

Our technique for showing decidability of structural
subtyping of recursive types is based on quantifier
elimination.  This section gives some general remarks on
quantifier elimination.

We follow \cite{Hodges93ModelTheory} in describing
quantifier elimination procedures.  According to \cite[Page
70, Lemma~2.7.4]{Hodges93ModelTheory} it suffices to
eliminate $\exists y$ from formulas of the form
\begin{equation} \label{eqn:qelimStep}
    \exists y.\ \bigwedge_{0 \leq i<n} \psi_i(\bar x,y)
\end{equation}
where $\bar x$ is a tuple of variables and $\psi_i(\bar x,y)$
is a literal whose all variables are among $\bar x,y$.  The
reason why eliminating formulas of the
form~(\ref{eqn:qelimStep}) suffices is the following.
Suppose that the formula in prenex form and consider the
innermost quantifier of a formula.  Let $\phi$ be the
subformula containing the quantifier and the subformula that
is the scope of the quantifier.  If $\phi$ is of the form $\forall
x.\ \phi_0$ we may replace $\phi$ with $\lnot \exists x.\lnot \phi_0$.  Hence, we
may assume that $\phi$ is of the form $\exists x.\ \phi_1$.  We then
transform $\phi_1$ into disjunctive normal form and use the
fact
\begin{equation} \label{eqn:existsPropagation}
   \exists x.\ (\phi_2 \lor \phi_3) \iff (\exists x.\ \phi_2) \lor (\exists x.\ \phi_3)
\end{equation}
We conclude that elimination of quantifiers from formulas of
form~(\ref{eqn:qelimStep}) suffices to eliminate the
innermost quantifier.  By repeatedly eliminating innermost
quantifiers we can eliminate all quantifiers from a formula.

We may also assume that $y$ occurs in every literal $\psi_i$,
otherwise we would place the literal outside the existential
quantifier using the fact
\[
   \exists y.\ (A \land B)  \iff (\exists y. A) \land B
\]
for $y$ not occurring in $B$.

To eliminate variables we often use the following identity
of a theory with equality:
\begin{equation} \label{eqn:replacement}
   \exists x. x = t \land \phi(x)  \iff  \phi(t)
\end{equation}
Section~\ref{sec:partFun} presents analogous identities for
partial functions.

Quantifier elimination procedures we give imply the
decidability of the underlying theories.  In this paper the
interpretations of function and relation symbols on some
domain $A$ are effectively computable functions and
relations on $A$.  Therefore, the truth-value of every
formula without variables is computable.  The quantifier
elimination procedures we present are all effective.  To
determine the truth value of a closed formula $\phi$ it
therefore suffices to apply the quantifier elimination
procedure to $\phi$, yielding a quantifier free formula $\psi$, and
then evaluate the truth value of $\psi$.

%%% Local Variables: 
%%% mode: latex
%%% TeX-master: "main"
%%% End: 
% LocalWords:  ModelTheory prenex subformula

  \subsection{Quantifier Elimination for Boolean Algebras}
\label{sec:qeBoolalg}

% !! FIX.  Make this more tutorial, as a 1st example of 
%% quantifier elimination.

This section presents a quantifier elimination procedure for
finite boolean algebras.  This result dates back at least to
\cite{Skolem19Untersuchungen}, see also
\cite{
  Tarski49ArithmeticalClassesTypesBooleanAlgebras,
  Kozen80ComplexityBooleanAlgebras,
  MartinNipkow89BooleanUnification,
  BoergerETAL97ClassicalDecisionProblem,
  Sudan02QuantifierEliminationBooleanAlgebrasTrivial},
\cite[Section 2.7 Exercise 3]{Hodges93ModelTheory}.  Note that the operations union,
intersection and complement are definable in the first-order
language of the subset relation.  Therefore, quantifier
elimination for the first-order theory of the boolean
algebra of sets is no harder than the quantifier
elimination for the first-order theory of the subset
relation.  However, the operations of boolean algebra are
useful in the process of quantifier elimination, so we give
the quantifier elimination procedure for the language
containing boolean algebra operations.

Instead of the first-order theory of the subtype relation we
could consider monadic second-order theory with no relation
or function symbols.  These two languages are equivalent
because the first-order quantifiers can be eliminated from
monadic second-order theory using the subset
relation (see Section~\ref{sec:msolEmbedding}).

Finite boolean algebras are isomorphic to boolean algebras
whose elements are all subsets of some finite set.  We
therefore use the symbols for the set operations as the
language of boolean algebras.  $t_1 \cap t_2$, $t_1 \cup
t_2$, $t_1^c$, $0$, $1$, correspond to set intersection, set
union, set complement, empty set, and full set,
respectively.  We write $t_1 \subseteq t_2$ for $t_1 \cap
t_2 = t_1$, we write $t_1 \subset t_2$ for the conjunction
$t_1 \subseteq t_2 \ \land\ t_1 \neq t_2$.

For every nonnegative integer $k$ we introduce formulas $|t|
\geq k$ expressing that the set denoted by $t$ has at least
$k$ elements, and formulas $|t| = k$ expressing that the set
denoted by $t$ has exactly $k$ elements.  These properties
are first-order definable as follows.
\[\begin{array}{lcl}
|t| \geq 0     &\equiv& \boolTrue \mnl
|t| \geq k{+}1 &\equiv& \exists x.\ x \subset t \ \land\ 
                                  |x| \geq k \mnl
   |t| = k   &\equiv& |t| \geq k \ \land\ \lnot |t| \geq k{+}1
\end{array}\]
We call a language which contains terms $|t|\geq k$ and
$|t|=k$ the language of boolean algebras with finite
cardinality constraints.  Because finite cardinality
constraints are first-order definable, the language with
finite cardinality constraints is equally expressive as the
language of boolean algebras.

Every inequality $t_1 \subseteq t_2$ is equivalent to
the equality $t_1 \cap t_2 = t_1$, and every equality $t_3 =
t_4$ is equivalent to the cardinality constraint
\[
    |(t_3 \cap t_4^c) \cup (t_4 \cap t_3^c)| = 0
\]
It is therefore sufficient to consider the first-order
formulas whose only atomic formulas are of the form $|t|=0$.
For the purpose of quantifier elimination we will
additionally consider formulas that contain atomic formulas
$|t|{=}k$ for all $k \geq 1$, as well as
$|t|{\geq}k$ for $k \geq 0$.

Note that we can eliminate negative literals as follows:
\begin{equation} \label{eqn:atleastExact}
\begin{array}{rcl}
\lnot |t|=k      &\iff&  |t|=0 \ \lor \cdots \lor\ 
                         |t|=k{-}1 \ \lor
                         |t| \geq k{+}1 \mnl
\lnot |t|\geq k  &\iff&  |t|=0 \ \lor \cdots \lor\ |t|=k{-}1
\end{array}
\end{equation}
Every formula in the language of boolean algebras can
therefore be written in prenex normal form where the matrix
of the formulas is a disjunction of conjunctions of atomic
formulas of the form $|t|=k$ and $|t|\geq k$, with no
negative literals.

Note that if a term $t$ contains at least one operation of
arity one or more, we may assume that the constants $0$ and
$1$ do not appear in $t$, because $0$ and $1$ can be
simplified away.  Furthermore, the expression $|0|$ denotes
the integer zero, so all terms of form $|0|=k$ or $|0|\geq
k$ evaluate to $\boolTrue$ or $\boolFalse$.  We can
therefore simplify every nontrivial term $t$ so that it
either $t$ contains no occurrences of constants $0$ and $1$,
or $t \equiv 1$.  

%% Let $\alpha_k$ be a sentence stating that there exist
%% $k$ distinct elements:
%% \[
%%    \alpha_k \ \ \equiv\ \
%%      \exists x_1,\ldots,x_k.\
%%         \distinctF(x_1,\ldots,x_n)
%% \]
%% Then the atomic formula $|1| \geq k$ is equivalent to
%% $\alpha_k$, formula $|1|=0$ is equivalent to $\boolFalse$, and
%% $|1|=k$ for $k \geq 1$ is equivalent to $\alpha_k \land
%% \lnot \alpha_{k-1}$.  For the purpose of quantifier
%% elimination we allow $\alpha_k$ and $\lnot \alpha_k$ to
%% occur as literals.

We next describe a quantifier elimination procedure for
finite boolean algebras.  

We first transform the formula into prenex normal form and
then repeatedly eliminate the innermost quantifier.  As
argued in Section~\ref{sec:qeIntro}, it suffices to show
that we can eliminate an existential quantifier from any
existentially quantified conjunction of literals.  Consider
therefore an arbitrary existentially quantified conjunction
of literals
\[
    \exists y.\ \bigwedge_{1 \leq i \leq n} \psi_i(\bar x,y)    
\]
where $\psi_i$ is of the form $|t| = k$ or of the form
$|t|\geq k$.  We assume that $y$ occurs in every formula
$\psi_i$.  It follows that no $\psi_i$ contains $|0|$ or $|1|$.

Let $x_1,\ldots,x_m,y$ be the set of variables occurring in
formulas $\psi_i$ for $1 \leq i \leq n$.

First consider the more general case $m \geq 1$.  Let for
$i_1,\ldots,i_m \in \{0,1\}$,
\[
   t_{i_1\ldots i_m} = x_1^{i_1} \cap \cdots \cap x_m^{i_m}
\]
where $t^0 = t$ and $t^1 = t^c$.
The terms in the set
\[
   P = \{ t_{i_1\ldots i_m} \mid
           i_1,\ldots,i_m \in \{0,1\} \}
\]
form a partition; moreover every boolean algebra expression
whose variables are among $x_i$ can be written as a disjoint
union of some elements of the partition $P$.  Any boolean
algebra expression containing $y$ can be written, for some
$p,q \geq 0$ as
\[\begin{array}{l}
    (s_1 \cap y) \cup \cdots \cup (s_p \cap y) \cup \mnl
    (t_1 \cap y^c) \cup \cdots \cup (t_q \cap y^c)
\end{array}\]
where $s_1,\ldots,s_p \in P$ are pairwise distinct elements
from the partition and $t_1,\ldots,t_q \in P$ are pairwise
distinct elements from the partition.  Because
\[\begin{array}{l}
    |(s_1 \cap y) \cup \cdots \cup (s_p \cap y) \cup
    (t_1 \cap y^c) \cup \cdots \cup (t_q \cap y^c)| = \mnl
\qquad
    |s_1 \cap y| + \cdots +|s_p \cap y| +
    |t_1 \cap y^c| + \cdots + |t_q \cap y^c|
\end{array}\]
the constraint of form $|t|=k$ can be written as
\[
   \bigvee_{k_1,\ldots,k_p,l_1,\ldots,l_q}
   \begin{array}[t]{l}
   |s_1 \cap y|=k_1 \land \cdots \land |s_p \cap y|=k_p \ \land \mnl
   |t_1 \cap y^c|=l_1 \land \cdots \land |t_q\cap y^c|=l_p
   \end{array}
\]
where the disjunction ranges over nonnegative integers
$k_1,\ldots,k_p,l_1,\ldots,l_q \geq 0$ that satisfy
\[
   k_1+\cdots+k_p +l_1+\cdots+l_q=k
\]
From~(\ref{eqn:atleastExact}) it follows that we can perform
a similar transformation for constraints of form $|t|\geq
k$.  After performing this transformation, we bring the
formula into disjunctive normal form and continue
eliminating the existential quantifier separately for each
disjunct, as argued in Section~\ref{sec:qeIntro}.  We may
therefore assume that all conjuncts $\psi_i$ are of one of
the forms: $|s \cap y|=k$, $|s \cap y^c|=k$, 
$|s \cap y|\geq k$, and $|s \cap y^c| \geq k$ 
where $s \in P$.

If there are two conjuncts both of which contain $|s \cap
y|$ for the same $s$, then either they are contradictory or
one implies the other.  We therefore assume that for any $s
\in P$, there is at most one conjunct $\psi_i$ containing
$|s \cap y|$.  For analogous reasons we assume that for
every $s \in P$ there is at most one conjunct $\psi_i$
containing $|s \cap y^c|$.  The result of eliminating the
variable $y$ is then given in Figure~\ref{fig:BAQERules}.
\begin{figure}
\begin{center}
\[\begin{array}{c|c}
\mbox{original formula} & \mbox{eliminated form} \\ 
\hline
\exists y.\
|s \cap y| \geq k \land |s \cap y^c| \geq l & |s| \geq k+l \mnl
\exists y.\
|s \cap y| = k \land |s \cap y^c| \geq l & |s| \geq k+l \mnl
\exists y.\
|s \cap y| \geq k \land |s \cap y^c| = l & |s| \geq k+l \mnl
\exists y.\
|s \cap y| = k \land |s \cap y^c| = l & |s| = k+l \mnl
\end{array}\]
\end{center}
\caption{Rules for Eliminating Quantifiers\label{fig:BAQERules}}
\end{figure}
The case when a literal containing $|s \cap y|$ does not
occur is covered by the case $|s \cap y|\geq k$ for $k=0$,
similarly for a literal containing $|s \cap y^c|$.

It remains to consider the case $m=0$.  Then $y$ is the only
variable occurring in conjuncts $\psi_i$.  Every cardinality
expression $t$ containing only $y$ reduces to one of $|y|$
or $|y^c|$.  If there are multiple literals containing
$|y|$, they are either contradictory or one implies the
others.  We may therefore assume there is at most one
literal containing $|y|$ and at most one literal containing
$|y^c|$.  We eliminate quantifier by applying rules in
Figure~\ref{fig:BAQERules} putting formally $s=1$ where $1$
is the universal set.

This completes the description of quantifier elimination
from an existentially quantified conjunction.  By repeating
this process for all quantifiers we arrive at a
quantifier-free formula $\psi$.  Hence we have the following
theorem.

\begin{theorem} \label{thm:BAqe}
For every first-order formula $\phi$ in the language of boolean
algebras with finite cardinality constraints there exists a
quantifier-free formula $\psi$ such that $\psi$ is a disjunction
of conjunctions of literals of form $|t|\geq k$ and $|t|=k$
where $t$ are terms of boolean algebra, the free variables
of $\psi$ are a subset of the free variables of $\phi$, and $\psi$ is
equivalent to $\phi$ on all algebras of finite sets.
\end{theorem}

\begin{remark} \label{rem:domainSize}
  Now consider the case when formula $\phi$ has no free
  variables.  By~Theorem~\ref{thm:BAqe}, $\phi$ is
  equivalent to $\psi$ where $\psi$ contains only terms
  without variables.  A term without variables in boolean
  algebra can always be simplified to $0$ or $1$.  Because
  $|0|=0$, the literals with $|0|$ reduce to $\boolTrue$ or
  $\boolFalse$, so we may simplify them away.  The
  expression $|1|$ evaluates to the number of elements in
  the boolean algebra.  We call literals $|1|=k$ and
  $|1|\geq k$ {\em domain cardinality constraints}.  A
  quantifier-free formula $\psi$ can therefore be written as
  a propositional combination of domain cardinality
  constraints.  We can simplify $\psi$ into a disjunction of
  conjunctions of domain cardinality constraints and
  transform each conjunction so that it contains at most one
  literal.  The result $\psi'$ is a single disjunction of
  domain cardinality constraints.  We may further assume
  that the disjunct of form $|1|\geq k$ occurs at most
  once.  Therefore, the truth value of each closed boolean
  algebra formula is characterized by a set $C$ of possible
  cardinalities of the domain.  If $\psi'$ does not contain
  any $|1|\geq k$ literals, the set $C$ is finite.
  Otherwise, $C = C_0 \cup \{k,{k+1},\ldots\}$ for some $k$
  where $C_0$ is a finite subset of $\{1,\ldots,{k-1}\}$.
\end{remark}

% Why wouldn't it work for
% infinite boolean algebras?  I was decomposing cardinalities
% |A+B| = |A| + |B| and perhaps assuming that these are natural numbers.
% Well it probably works for all algebras of sets, but
% not all infinite boolean algebras are algebras of sets!

%%% Local Variables: 
%%% mode: latex
%%% TeX-master: "main"
%%% End: 

% LocalWords:  Kozen ComplexityBooleanAlgebras MartinNipkow BooleanUnification
% LocalWords:  BoergerETAL ClassicalDecisionProblem prenex arity
% LocalWords:  QuantifierEliminationBooleanAlgebrasTrivial

  \subsection{Feferman-Vaught Theorem}
\label{sec:fefermanVaught}

\begin{quote}\em
The Feferman-Vaught technique is a way of discovering
the first-order theories of complex structures by analyzing their
components.  This description is a little vague, and in fact
the Feferman-Vaught technique itself has something of a floating
identity.  It works for direct products, as we shall see.
Clever people can make it work in other situations too.
\leftline{--- \cite{Hodges93ModelTheory}, page 458}
\end{quote}

We next review Feferman-Vaught theorem for direct products
\cite{FefermanVaught59FirstOrderPropertiesProductsAlgebraicSystems}
which implies that the products of structures with decidable
first-order theories have decidable first-order theories.

The result was first obtained for strong and weak powers of
theories in \cite{Mostowski52DirectProductsTheories};
\cite{Mostowski52DirectProductsTheories} also suggests the
generalization to products.  Our sketch here mostly follows
\cite{FefermanVaught59FirstOrderPropertiesProductsAlgebraicSystems}
and \cite{Mostowski52DirectProductsTheories}, see also
\cite[Chapter 12]{Malcev71MetamathematicsAlgebraicSystems}
as well as \cite[Section 9.6]{Hodges93ModelTheory}.
Somewhat specific to our presentation is the fact that we
use a multisorted logic and build into the language the
correspondence between formulas interpreted over $C$ and the
cylindric algebra of sets of positions.

Let $\lanC$ be a relational language.  Let further $I$ be
some nonempty finite or countably infinite index set.  For
each $i \in I$ let $\ThC_i = \tu{C_i,\interpretS{\_}{\ThC_i}}$
be a decidable structure interpreting the language $\lanC$.

We define \emph{direct product} of the family of structures
$\ThC_i$, $i \in I$, as the structure
\[
     \ThP = \Pi_{i \in I} \ThC_i
\]
where $\ThP = \tu{P,\interpretS{\_}{P}}$.  $P$ is the set of
all functions $t$ such that $t(i) \in C_i$ for $i \in I$, and
$\interpretS{\_}{P}$ is defined by
\[
    \interpretS{r}{P}(t_1,\ldots,t_k) \ = \
     \forall i.\ \interpretS{r}{\ThC_i}(t_1(i),\ldots,t_k(i))
\]
for each relation symbol $r \in \lanC$.

\begin{figure}[tbhp]
\newcommand{\tit}[1]{\multicolumn{3}{c}{{\mbox{\bf #1}}}\mnl}
\[\begin{array}{rcl}
%--------------------------------------------------
\tit{inner formula relations for $r \in \lanC$}
   r &::& \tupleSort^k \to \setposSort \mnl
%--------------------------------------------------
\tit{inner logical connectives}
   \landI, \lorI %, \impliesI 
     &::& \setposSort \times \setposSort \to \setposSort \mnl
   \lnotI &::& \setposSort \to \setposSort \mnl
   \trueI, \falseI &::& \setposSort \mnl
%--------------------------------------------------
\tit{inner formula quantifiers}
  \existsI, \forallI &::&
   (\tupleSort \to \setposSort) \to \setposSort \mnl
%--------------------------------------------------
\tit{index set equality}
  \setposEq &::& \setposSort \times \setposSort \to \boolSort \mnl
%--------------------------------------------------
\tit{logical connectives}
   \land_, \lor &::& \boolSort \times \boolSort \to \boolSort \mnl
   \lnot &::& \boolSort \to \boolSort \mnl
   \boolTrue, \boolFalse &::& \boolSort \mnl
%--------------------------------------------------
\tit{index set quantifiers}
  \existsl, \foralll &::&
   (\setposSort \to \boolSort) \to \boolSort \mnl
%--------------------------------------------------
%\tit{tuple equality}
%  \termEq &::& \tupleSort \times \tupleSort \to \boolSort \mnl
%--------------------------------------------------
\tit{tuple quantifiers}
  \exists, \forall &::&
   (\tupleSort \to \boolSort) \to \boolSort \mnl
\end{array}\]
\caption{Operations in product structure
\label{fig:productStructureLogic}}
\end{figure}

For the purpose of quantifier elimination we consider a
richer language of statements about product structure
$\ThP$.  Figure~\ref{fig:productStructureLogic} shows this
richer language.  The corresponding structure $\ThPTwo =
\tu{\PTwo,\interpretS{\_}{\PTwo}}$ contains, in addition to
the function space $P$, a copy of the boolean algebra $2^I$
of subsets of the index set $I$.  We interpret a relation $r
\in \lanC$ by
\[
    \interpretS{r}{\PTwo}(t_1,\ldots,t_k) = 
     \{\, i \mid \interpretS{r}{\ThC_i}(t_1(i),\ldots,t_k(i)) \,\}
\]
We let $\interpretS{\trueI}{\PTwo} = I$ and
write
\[
     r(t_1,\ldots,t_k) \setposEq \trueI
\]
to express $\interpretS{r}{P}(t_1,\ldots,t_k)$.  Hence
$\PTwo$ is at least as expressive as $\ThP$.

Note that Figure~\ref{fig:productStructureLogic} does not
contain an equality relation between tuples.  If we need to
express the equality between tuples, we assume that some
binary relation $r_0 \in \lanC$ in the base structure is
interpreted as equality, and express the equality between
tuples $t_1$ and $t_2$ using the formula:
\begin{equation*}
   r_0(t_1,t_2) \setposEq \trueI.
\end{equation*}

\begin{figure}[tbhp]
\newcommand{\tit}[1]{\multicolumn{3}{c}{{\mbox{\bf #1}}}\mnl}
\newcommand{\sem}[1]{\interpretS{#1}{\PTwo}}
\[\begin{array}{rcl}
%--------------------------------------------------
\tit{inner formula relations for $r \in \lanC$}
\sem{r}(t_1,\ldots,t_k) &=& 
    \{\, i \mid \interpretS{r}{\ThC_i}(t_1(i),\ldots,t_k(i)) \,\} \mnl
%--------------------------------------------------
\tit{inner logical connectives}
\sem{\landI}(A_1,A_2) &=& A_1 \land A_2 \mnl
\sem{\lorI}(A_1,A_2) &=& A_1 \cup A_2 \mnl
%\sem{\impliesI}(A_1,A_2) &=& (I \setminus A_1) \cup A_2 \mnl
\sem{\lnotI}(A)  &=& I \setminus A \mnl
\sem{\trueI} &=& I \mnl
\sem{\falseI} &=& \emptyset \mnl
%--------------------------------------------------
\tit{inner formula quantifiers}
\sem{\existsI} f &=&  \bigcup_{t \in P} f(t) \mnl
\sem{\forallI} f &=&  \bigcap_{t \in P} f(t) \mnl
%--------------------------------------------------
\tit{index set equality}
\sem{\setposEq}(A_1,A_2) &=& (A_1 = A_2) \mnl
%--------------------------------------------------
\tit{logical connectives}
\tit{\rm (interpreted as usual)}
%--------------------------------------------------
\tit{index set quantifiers}
\sem{\existsl} f &=& \bigcup_{A \in 2^I} f(A) \mnl
\sem{\foralll} f &=& \bigcap_{A \in 2^I} f(A) \mnl
%--------------------------------------------------
%% If we need equality on tuples, assume base theory has equality
%\tit{tuple equality}
%\sem{\termEq}(t_1,t_2) &=& (t_1 = t_2) \mnl
%--------------------------------------------------
\tit{tuple quantifiers}
\sem{\exists} f &=& \exists t \in P.\ f(t) \mnl
\sem{\forall} f &=& \forall t \in P.\ f(t) \mnl
\end{array}\]
\caption{Semantics of operations in product structure $\ThPTwo$
\label{fig:productLogicSemantics}}
\end{figure}

Figure~\ref{fig:productLogicSemantics} shows the semantics
of the language in Figure~\ref{fig:productStructureLogic}.
(The logic has no partial functions, so we interpret the sort
$\boolSort$ over the set $\{\boolTrue,\boolFalse\}$.)

We let $A_1 \subseteqI A_2$ stand for $A_1 \landI A_2
\setposEq A_2$.

Note that the interpretations of $\landI$, $\lorI$, $\lnotI$,
$\trueI$, $\falseI$, $\setposEq$, $\existsl$, $\foralll$
form a first-order structure of boolean algebras of subsets
of the set $I$.  We call formulas in this boolean algebra
sublanguage \emph{index-set algebra formulas}.

On the other hand, relations $r$ for $r \in \lanC$, together
with $\landI$, $\lorI$, $\lnotI$, $\existsI$, $\forallI$
form the signature of first-order logic with relation
symbols.  We call formulas built only from these operations
{\em inner formulas}.

Let $\phi$ be a an inner formula with free $\tupleSort$
variables $t_1,\ldots,t_m$ and no free $\setposSort$ variables.
Then $\phi$ specifies a relation $\rho \subseteq D^m$.  Consider the
corresponding first-order formula $\phi'$ interpreted in the
base structure $\ThC$; formula $\phi'$ specifies a relation $\rho'
\subseteq C^m$.  
The following property follows from the
semantics in Figure~\ref{fig:productLogicSemantics}:
\begin{equation}
   \rho(t_1,\ldots,t_m) = \{\, i \in I \mid \rho'(t_1(i),\ldots,t_m(i)) \,\}
\end{equation}
Sort constraints imply that quantifiers $\existsI,\forallI$
are only applied to inner formulas.  Let $\phi$ be a formula
of sort $\boolSort$.  By labelling subformulas of sort
$\setposSort$ with variables $A_1,\ldots,A_n$, we can write
$\phi$ in form $\phi^1$:
\[
\begin{array}{l}
  \existsl A_1,\ldots,A_n. \mnl
\begin{array}[t]{l}
   A_1 \setposEq \phi_1 \ \land \ldots \land \ A_n \setposEq \phi_n \ \land \ \mnl
   \psi(A_1,\ldots,A_n)
\end{array}
\end{array}
\]
where
\[
   \phi = \psi(\phi_1,\ldots,\phi_n)
\]
Furthermore, by defining $B_1,\ldots,B_m$ to be the partition
of $\trueI$ consisting of terms of form
\[
    A_1^{p_1} \ \landI \ldots \landI A_n^{p_n}
\]
for $p_1,\ldots,p_n \in \{0,1\}$,
we can find a formula $\psi'$ and formulas $\phi_1',\ldots,\phi_m'$
such that $\phi^1$ is equivalent to $\phi^2$:
\begin{equation} \label{eqn:unnestedInner}
\begin{array}{l}
  \existsl B_1,\ldots,B_m. \mnl
\begin{array}[t]{l}
   B_1 \setposEq \phi'_1 \ \land \ldots \land \ B_m \setposEq \phi'_m \ \land \ \mnl
   \psi'(B_1,\ldots,B_n)
\end{array}
\end{array}
\end{equation}
and where $\phi_1',\ldots,\phi'_m$ evaluate to sets that form partition
of $\trueI$ for all values of free variables.  (By partition
of $\trueI$ we here mean a family of pairwise disjoint sets
whose union is $\trueI$, but we do not require the sets to
be non-empty.)

Now consider a formula of form $\exists t.\phi$ where $\phi$ is without
$\exists,\forall$ quantifiers (but possibly contains $\existsI,
\forallI$ and $\existsl,\foralll$ quantifiers).
We transform $\phi$ into $\phi^2$ as described, and then replace
\begin{equation} \label{eqn:quantifiedInners}
\begin{array}{l}
  \exists t.\
  \existsl B_1,\ldots,B_m. \mnl
\begin{array}[t]{l}
   B_1 \setposEq \phi'_1 \ \land \ldots \land \ B_m \setposEq \phi'_m \ \land \ \mnl
   \psi'(B_1,\ldots,B_n)
\end{array}
\end{array}
\end{equation}
with
\begin{equation}
\begin{array}{l} \label{eqn:eliminatedInners}
  \existsl D_1,\ldots,D_m.\
  \existsl B_1,\ldots,B_m. \mnl
\begin{array}[t]{l}   
   D_1 \setposEq (\existsI t. \phi'_1) \ \land \ldots \land \ 
   D_m \setposEq (\existsI t. \phi'_m) \ \land \ \mnl
   B_1 \subseteqI D_1 \ \land \ldots \land \ B_m \subseteqI D_m  \ \land \mnl
  \partitionF(B_1,\ldots,B_n) \land \psi'(B_1,\ldots,B_n)
\end{array}
\end{array}
\end{equation}
where $\partitionF(B_1,\ldots,B_n)$ denotes a boolean algebra
expression expressing that sets $B_1,\ldots,B_n$ form the
partition of $\trueI$.

It is easy to see that~\ref{eqn:quantifiedInners}
and~\ref{eqn:eliminatedInners} are equivalent.

By repeating this construction we eliminate all term
quantifiers from a formula.  We then eliminate all set
quantifiers as in Section~\ref{sec:qeBoolalg}.  For that
purpose we extend the language with cardinality constraints.

As the result we obtain cardinality constraints on inner
formulas.  Closed inner formulas evaluate to $\trueI$ or
$\falseI$ depending on their truth value in base structure
$\ThC$.  Hence, if $\ThC$ is decidable, so is
$\ThPTwo$.

\begin{theorem}[Feferman-Vaught]
  Let $\ThC$ be a decidable structure.  Then every formula
  in the language of Figure~\ref{fig:productStructureLogic}
  is equivalent on the structure $\PTwo$ to a propositional
  combination of cardinality constraints of the index-set
  boolean algebra i.e.\ formulas of form $|\phi|\geq k$ and
  $|\phi| = k$ where $\phi$ is an inner formula.
\end{theorem}

\begin{example} \label{exa:feferman}
  Let $r \in \lanC$ be a binary relation on structure $C$.
  Let us eliminate quantifier $\exists t$ from the formula $\phi(t_1,t_2)$:
  \[
  \begin{array}{l}
    \exists t. \existsl A_1,A_2,A_3.\mnl
    \begin{array}{l}
      A_1 \setposEq r(t,t_1) \ \land \
      A_2 \setposEq r(t_1,t) \ \land \
      A_3 \setposEq r(t_2,t) \ \land \mnl
      |\lnotI A_1| = 0 \ \land \ 
      |\lnotI A_2| = 0 \ \land
      |\lnotI A_3| \geq 1
    \end{array}
  \end{array}
  \]
  We first introduce sets $B_0,\ldots,B_7$ that form partition of $\trueI$.
  The formula is then equivalent to $\phi_1$:
  \[
  \begin{array}{l}
    \exists t. \existsl B_0,B_1,B_2,B_3,B_4,B_5,B_6,B_7.\mnl
    \begin{array}{l}
      B_0 \setposEq r(t,t_1) \landI r(t_1,t) \landI r(t_2,t) \ \land \mnl
      B_1 \setposEq \lnotI r(t,t_1) \landI r(t_1,t) \landI r(t_2,t) \ \land \mnl
      B_2 \setposEq r(t,t_1) \landI \lnotI r(t_1,t) \landI r(t_2,t) \ \land \mnl
      B_3 \setposEq \lnotI r(t,t_1) \landI \lnotI r(t_1,t) \landI r(t_2,t) \ \land \mnl
      B_4 \setposEq r(t,t_1) \landI r(t_1,t) \landI \lnotI r(t_2,t) \ \land \mnl
      B_5 \setposEq \lnotI r(t,t_1) \landI r(t_1,t) \landI \lnotI r(t_2,t) \ \land \mnl
      B_6 \setposEq r(t,t_1) \landI \lnotI r(t_1,t) \landI \lnotI r(t_2,t) \ \land \mnl
      B_7 \setposEq \lnotI r(t,t_1) \landI \lnotI r(t_1,t) \landI \lnotI r(t_2,t) \ \land \mnl
      \phi_0
    \end{array}
  \end{array}
  \]
  where
  \[
  \begin{array}{l}
    \phi_0 \ \equiv \ \mnl
    \quad
    \begin{array}{l}
      |B_1| = 0 \ \land \ |B_2| = 0 \ \land \mnl
      |B_3| = 0 \ \land \ |B_5| = 0 \ \land \mnl
      |B_6| = 0 \ \land \ |B_7| = 0 \ \land \mnl
      |B_4| \geq 1
    \end{array}
  \end{array}
  \]
  We now eliminate the quantifier $\exists t$ from the formula
  $\phi_1$, obtaining formula $\phi_2$:
  \[
  \begin{array}{l}
    \begin{array}[t]{l}
      \existsl D_0,D_1,D_2,D_3,D_4,D_5,D_6,D_7. \mnl
      \begin{array}{l}
        D_0 \setposEq \existsI t.\ r(t,t_1) \landI r(t_1,t) \landI r(t_2,t) \ \land \mnl
        D_1 \setposEq \existsI t.\ \lnotI r(t,t_1) \landI r(t_1,t) \landI r(t_2,t) \ \land \mnl
        D_2 \setposEq \existsI t.\ r(t,t_1) \landI \lnotI r(t_1,t) \landI r(t_2,t) \ \land \mnl
        D_3 \setposEq \existsI t.\ \lnotI r(t,t_1) \landI \lnotI r(t_1,t) \landI r(t_2,t) \ \land \mnl
        D_4 \setposEq \existsI t.\ r(t,t_1) \landI r(t_1,t) \landI \lnotI r(t_2,t) \ \land \mnl
        D_5 \setposEq \existsI t.\ \lnotI r(t,t_1) \landI r(t_1,t) \landI \lnotI r(t_2,t) \ \land \mnl
        D_6 \setposEq \existsI t.\ r(t,t_1) \landI \lnotI r(t_1,t) \landI \lnotI r(t_2,t) \ \land \mnl
        D_7 \setposEq \existsI t.\ \lnotI r(t,t_1) \landI \lnotI r(t_1,t) \landI \lnotI r(t_2,t) \ \land \mnl
        \phi_3
      \end{array}
    \end{array}      
  \end{array}
  \]
  where
  \[  
  \begin{array}{l}
    \phi_3 \ \equiv \
    \begin{array}[t]{l}
      \existsl B_0,B_1,B_2,B_3,B_4,B_5,B_6,B_7. \mnl
      B_0 \subseteqI D_0 \ \land \ldots \land \ B_7 \subseteqI D_7  \ \land \mnl
      \phi_0
    \end{array}
  \end{array}
  \]
  We next apply quantifier elimination for boolean algebras to formula
  $\phi_3$ and obtain formula $\phi'_3$:
  \[  
  \begin{array}{l}
    \phi'_3 \ \equiv \ |D_4| \geq 1 \ \land \ |\lnotI D_0 \landI \lnotI D_4| = 0
  \end{array}
  \]
  Hence $\phi(t_1,t_2)$ is equivalent to
  \[
  \begin{array}{l}
    \existsl D_0,D_4.\mnl
    \begin{array}{l}
        D_0 \setposEq \existsI t.\ r(t,t_1) \landI r(t_1,t) \landI r(t_2,t) \ \land \mnl
        D_4 \setposEq \existsI t.\ r(t,t_1) \landI r(t_1,t) \landI \lnotI r(t_2,t) \ \land \mnl
        |D_4| \geq 1 \land |\lnotI D_0 \landI \lnotI D_4| = 0
    \end{array}
  \end{array}
  \]
  After substituting the definitions of $D_0$ and $D_4$,
  formula $\phi(t_1,t_2)$ can be written without quantifiers
  $\exists,\forall,\existsl,\foralll$.
\end{example}

%%% Local Variables: 
%%% mode: latex
%%% TeX-master: "main"
%%% End: 

% LocalWords:  Feferman Vaught multisorted subformulas

  \subsection{Term Algebras}
\label{sec:qeTA}

In this section we present a quantifier elimination
procedure for term algebras (see
Section~\ref{sec:termAlgebrasDef}).  A quantifier
elimination procedure for term algebras implies that the
first-order theory of term algebras is decidable.  In the
sections below we build on the procedure in this section to
define quantifier elimination procedures for structural
subtyping.

The decidability of the first-order theory of term algebras
follows from Mal'cev's work on locally free algebras
\cite[Chapter 23]{Malcev71MetamathematicsAlgebraicSystems}.
\cite{Oppen80ReasoningRecursivelyDefinedDataStructures} also
gives an argument for decidability of term algebra and
presents a unification algorithm based on congruence
closure~\cite{NelsonOppen80DecisionProceduresCongruenceClosure}.
Infinite trees are studied in
\cite{Courcelle83FundamentalPropertiesInfiniteTrees}.
\cite{Maher88CompleteAxiomatizationsAlgebrasTrees} presents
a complete axiomatization for algebra of finite,
infinite and rational trees.  A proof in the style of
\cite{Hodges93ModelTheory} for an extension of free algebra
with queues is presented in
\cite{RybinaVoronkov01DecisionProcedureTermAlgebrasQueues}.
Decidability of an extension of term algebras with
membership tests is presented in
\cite{ComonDelor94EquationalFormulaeMembershipConstraints}
in the form of a terminating term rewriting system.
Unification and disunification problems are special cases of
decision problem for first-order theory of term algebras,
for a survey see e.g.\ \cite{Siekmann89UnificationTheory,
  Comon91Disunification}.

We believe that our proof provides some insight into
different variations of quantifier elimination procedures
for term algebras.  Like~\cite{Hodges93ModelTheory} we use
selector language symbols, but retain the usual constructor
symbols as well.  The advantage of the selector language is
that $\exists y.\ z = f(x,y)$ is equivalent to a quantifier-free
formula $x = f_1(z) \ \land\ \Is{f}(z)$.  On the other hand,
constructor symbols also increase the set of relations on
terms definable via quantifier-free formulas, which can
slightly simplify quantifier-elimination procedure, as will
be seen by comparing Proposition~\ref{prop:baseToConsSel}
and Proposition~\ref{prop:baseToSel}.  Compared to
\cite[Page 70]{Hodges93ModelTheory}, we find that the
termination of our procedure is more evident and the
extension to the term-power algebra in
Section~\ref{sec:nonrec} easier.  Our base formulas somewhat
resemble formulas arising in other quantifier elimination
procedures \cite{Malcev71MetamathematicsAlgebraicSystems,
  ComonLescanne89EquationalProblemsDisunification,
  Maher88CompleteAxiomatizationsAlgebrasTrees}.  Our
terminology also borrows from congruence closure graphs like
those
of~\cite{Oppen80ReasoningRecursivelyDefinedDataStructures,
  NelsonOppen80DecisionProceduresCongruenceClosure},
although we are not primarily concerned with efficiency of
the algorithm described.  Term algebra is an example of a
theory of {\em pairing functions}, and
\cite{FerranteRackoff79ComputationalComplexityLogicalTheories}
shows that non-empty family of theories of pairing functions
as non-elementary lower bound on time complexity.

%%%%%%%%%%%%%%%%%%%%%%%%%%%%%%%%%%%%%%%%%%%%%%%%%%%%%%%%%%%%
\subsubsection{Term Algebra in Selector Language}

To facilitate quantifier elimination we use a {\em selector
  language} $\SelOps$ for term algebra \cite[Page
61]{Hodges93ModelTheory}.  We define term algebra in
selector language as a first-order structure with partial
functions.

The set $\SelOps$ contains, for every function symbol $f \in
\FreeSig$ of arity $\ar(f) = k$, a unary predicate
$\Is{f} \subseteq \FT$ and functions 
$f_1,\ldots,f_k : \FT \to \FT$ such
that
\begin{eqnarray}
  \label{eqn:IsfDef}
  \Is{f}(t) & \iff & \exists t_1,\ldots,t_k.\
                        t = f(t_1,\ldots,t_k) \\
  f_i(f(t_1,\ldots,t_k)) & = & t_i, \qquad 1 \leq i \leq k \\
  f_i(t) & = & \bot, \qquad \lnot \Is{f}(t) \label{eqn:undefDef}
\end{eqnarray}
For every $f \in \FreeSig$ and $1 \leq i \leq \ar(f)$,
expression $f_i(t)$ defined iff $\Is{f}(t)$ holds, so we let
$D_f = \tu{x,\Is{f}(x)}$.

As a special case, if $d$ is a constant, then $\ar(d)=0$ and
$\Is{d}(t) \iff t=d$.

\begin{proposition}
For every formula $\phi_1$ in the language
$\FreeOps$ there exists an equivalent
formula $\phi_2$ in the selector language.
\end{proposition}
\begin{proofsketch}
Because of the presence of equality symbol, every formula in
language $\FreeOps$ can be written in unnested form
such that every atomic formula is of two forms:
$x_1 = x_2$, or $f(x_1,\ldots,x_k) = y$,
where $y$ and $x_i$ are variables.  We keep every formula
$x_1=x_2$ unchanged and transform each formula
\[
    f(x_1,\ldots,x_k) = y
\]
into the well-defined conjunction
\[
    x_1 = f_1(y) \ \land \cdots \land\ x_k = f_k(y)\ \land\
    \Is{f}(y)
\]
\end{proofsketch}

Note that predicates $\Is{f}$ form a partition of the set of
all terms i.e.\ the following formulas are valid:
\begin{equation} \label{eqn:IsfPartit}
\begin{array}{l}
  \forall x.\ \bigvee\limits_{f \in \FreeSig} \Is{f}(x) \mnlb
  \forall x.\ \lnot (\Is{f}(x) \land \Is{g}(x)), \qquad \mbox{ for $f \not\equiv g$}
\end{array}
\end{equation}

A {\em constructor-selector} language contains both
constructor symbols $f \in \FreeOps$ and selector symbols $f_i
\in \SelOps$.

%% Let $t$ be a ground term and $t'$ be tree representing $t$.
%% Let $p \in \TreePaths$, $p = n_1,\ldots,n_k$ such that $t'(p)$ is
%% defined.  Then we can extract a subtree at $p$.  So what.

%%% Local Variables: 
%%% mode: latex
%%% TeX-master: "main"
%%% End: 
 
%%%%%%%%%%%%%%%%%%%%%%%%%%%%%%%%%%%%%%%%%%%%%%%%%%%%%%%%%%%%

\subsubsection{Quantifier Elimination}

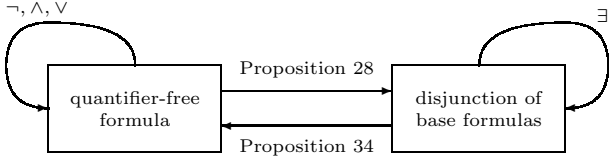
\begin{figure}
\newcommand{\selfLoop}[9]{
\qbezier(#1,#4)(#1,#5)(#3,#5)
\qbezier(#3,#5)(#2,#5)(#2,#4)
\qbezier(#2,#4)(#2,#6)(#3,#6)
\put(#3,#6){\vector(#7,#8){0.0}}
\put(#2,#5){#9}
}

{\scriptsize
\setlength{\unitlength}{0.45in}
\begin{picture}(8,3)
\put(1,1){\framebox(2,1){\begin{minipage}{1in}
                         \centerline{quantifier-free}
                         \centerline{formula}
                         \end{minipage}}}
\put(5,1){\framebox(2,1){\begin{minipage}{1in}
                         \centerline{disjunction of}
                         \centerline{base formulas}
                         \end{minipage}}}
\put(3,1.7){\vector(1,0){2}}
\put(3,1.7){\makebox(2,0.5){Proposition~\ref{prop:qFreeToBaseDisj}}}
\put(5,1.3){\vector(-1,0){2}}
\put(3,0.8){\makebox(2,0.5){Proposition~\ref{prop:baseToConsSel}}}
\selfLoop{2}{0.5}{1}{2}{2.5}{1.5}{1}{0}{\parbox{0.5in}{$\lnot,\land,\lor$\\}}
\selfLoop{6}{7.5}{7}{2}{2.5}{1.5}{-1}{0}{$\!\!\!\exists$}
\end{picture}
}

%%% Local Variables: 
%%% mode: latex
%%% TeX-master: "main"
%%% End: 
\caption{Quantifier Elimination for Term Algebra\label{fig:termAlgScheme}}
\end{figure}

We proceed to quantifier elimination for term algebra.
A schematic view of our proof is in
Figure~\ref{fig:termAlgScheme}.  The basic insight is that
any quantifier-free formula can be written in a particular
unnested form, as a disjunction of {\em base formulas}.
Base formulas trivially permit elimination of an existential
quantifier, yet every base formula can be converted back to
a quantifier-free formula.

A semi-base formula is almost the base formula, except that
it may be cyclic.  We introduce cyclicity after explaining
the graph representation of a semi-base formula.
\begin{definition}[Semi-Base Formula] \label{def:semibase}
A semi-base formula $\beta$ with
\begin{itemize}
\item free variables $x_1,\ldots,x_m$,
\item internal non-parameter variables $u_1,\ldots,u_p$, and 
\item internal parameter variables $u_{p+1},\ldots,u_{p+q}$ 
\end{itemize}
is a formula of form
\[\begin{array}{l}
   \exists u_1,\ldots,u_n\ \mnl
  \qquad \distinctF(u_1,\ldots,u_n) \ \land \mnl
  \qquad \structureF(u_1,\ldots,u_n)\ \land \mnl
  \qquad \labelsF(u_1,\ldots,u_n;x_1,\ldots,x_m)
  \end{array}
\]
$\distinctF(u_1,\ldots,u_n)$ enforces that variables are distinct
\[
    \distinctF(u_1,\ldots,u_n) \equiv 
       \bigwedge_{1 \leq i < j \leq n} u_i \neq u_j \ .
\]
$\structureF(u_1,\ldots,u_n)$ specifies relationships between
terms denoted by variables:
\[\begin{array}{l}
    \structureF(u_1,\ldots,u_n) \equiv {} \mnl
 \qquad \bigwedge\limits_{i=1}^p u_i = t_i(u_1,\ldots,u_n)
\end{array}\]
where each $t_i(u_1,\ldots,u_n)$ is a term of form
$f(u_{l_1},\ldots,u_{l_k})$ for $f \in \FreeSig$,
$k = \ar(f)$.

$\labelsF(u_1,\ldots,u_n; x_1,\ldots,x_m)$ identifies some free
variables with some parameter and non-parameter variables:
\[
    \labelsF(u_1,\ldots,u_n; x_1,\ldots,x_m) \equiv
        \bigwedge_{1 \leq i \leq m}
        x_i = u_{j_i}
\]
for some function $j : \{1,\ldots,m\} \to \{1,\ldots,n\}$.  

We require each semi-base formula to satisfy the following
congruence closure property: there are no two distinct
variables $u_i$ and $u_{i'}$ such that both $u_i =
f(u_{l_1},\ldots,u_{l_k})$ and $u_{i'} = f(u_{l_1},\ldots,u_{l_k})$
occur as conjuncts $\phi_j$ in formula $\structureF$.

We denote by $U$ the set of internal variables of a given
semi-base formula, $U = \{ u_1,\ldots,u_n \}$.
\end{definition}

\begin{definition} \label{def:semibaseSelector}
A semi-base formula in selector language is obtained from the base formula
in constructor language by replacing every conjunct of form
\[
    u_i = f(u_{l_1},\ldots,u_{l_k})
\]
with the well-defined conjunction
\[
    \Is{f}(u_i) \ \land\ u_{l_1} = f_1(u_i)\ \land \cdots \land\
                       u_{l_k} = f_k(u_i)
\]
\end{definition}
A semi-base formula in selector language is clearly a
well-formed conjunction of literals.  All atomic formulas in
a semi-base formula are unnested, in both constructor and
selector language.

We can represent a base formula as a labelled directed graph
with the set of nodes $U$; we call this graph {\em graph
associated with a semi-base formula}.  Nodes of the graph
are in a bijection with internal variables of the semi-base
formula.
% !!! [PICTURE BADLY NEEDED]
We call nodes corresponding to parameter variables
$u_{p+1},\ldots,u_{p+q}$ {\em parameter nodes}; nodes $u_1,\ldots,u_p$
are {\em non-parameter nodes}.  Each non-parameter node is
labelled by a function symbol $f \in \FreeSig$ and has exactly
$\ar(f)$ successors, with edge from $u_k$ to $u_l$ labelled
by the positive integer $i$ iff $f_i(u_k)=u_l$ occurs in the
semi-base formula written in selector language.  A {\em
constant node} is a node labelled by some constant symbol $c
\in \FreeSig$, $\ar(c)=0$.  A constant node is a sink in the
graph; every sink is either a constant or a parameter node.
In addition to the labelling by function symbols, each node $u \in
U$ of the graph is labelled by zero or more free variables
$x$ such that equation $x=u$ occurs in the semi-base
formula.
%We will
%be deliberately confusing the variable and the node
%corresponding to that variable in the associated graph.

\begin{definition}[Base Formula] \label{def:baseFormula}
A semi-base formula $\phi$ is a {\em base formula} iff the
graph associated with $\phi$ is acyclic.
\end{definition}
A semi-base formula whose associated graph is cyclic is
unsatisfiable in the term algebra of finite terms.  Checking
the cyclicity of a base formula corresponds to occur-check
in unification algorithms (see
e.g.~\cite{Lloyd87FoundationsLogicProgramming,
ComonLescanne89EquationalProblemsDisunification}).

\begin{definition}
By {\em height} $\nodeHeight{u}$ of a node $u$ in the
acyclic graph we mean the length of the longest path
starting from $u$.
\end{definition}
A node $u$ is sink iff $\nodeHeight{u}=0$.

\begin{definition}
We say that an internal variable $u_l$ is a {\em source
variable} of a base formula $\beta$ iff $u_l$ is represented by
a node that is source in the directed acyclic graph
corresponding to $\beta$.  Equivalently, if $\beta$ is written in
the selector language, then $u_l$ is a source variable iff
$\beta$ contains no equations of form $u_l = f_i(u_k)$.
\end{definition}

\begin{definition}
If $u_i$ and $u_j$ are internal variables, we write $u_i
\determines u_j$ if there is a path in the underlying graph
from node $u_i$ to node $u_j$.  Equivalently, $u_i
\determines u_j$ iff there exists a term $t(u_i)$ in the
selector language such that $\models \beta \implies u_j = t(u_i)$.
\end{definition}
Relation $\determines$ is a partial order on internal variables
of $\beta$.

% In general, if $x$ and $y$ are free or internal variables we
% write $x \Gdetermines y$ if $\beta \models x=y$ i.e.  if
% $\beta \models u_i = u_j$ for $u_i$ and $u_j$ such that
% $x=u_i$ and $y=u_j$ appear in $\beta$.  $\Gdetermines$ is a
% preorder on the set of all variables occurring in $\beta$.

The following Lemma~\ref{lemma:baseSat} is similar to the
Independence of Disequations Lemma in e.g.\ \cite[Page
178]{ComonDelor94EquationalFormulaeMembershipConstraints}.
\begin{lemma} \label{lemma:baseSat}
Let $\beta$ be a base formula of the form
\[
    \exists u_1,\ldots,u_p, u_{p+1},\ldots,u_{p+q}.\ \beta_0
\]
where $u_{p+1},\ldots,u_{p+q}$ are parameter variables of $\beta$, and $\beta_0$ is
quantifier-free.  Let $S_{p+1},\ldots,S_{p+q}$ be infinite sets
of terms.  Then there exists a valuation $\sigma$ such that
$\interpret{\beta_0}\sigma = \boolTrue$ and $\interpret{u_i}\sigma \in S_i$
for $p+1 \leq i \leq p+q$.
\end{lemma}
\begin{proof}
To construct $\sigma$ assign first the values to parameter
variables, as follows.  Let $h_G$ be the length of the
longest path in the graph associated with $\beta$.  Pick
$\sigma(u_{p+1}) \in S_{p+1}$ so that
$\termHeight{\sigma(u_{p+1})} > h_G$, and for each $i$ where
$p+2 \leq i \leq p+q$ pick $\sigma(u_i) \in S_i$ so that
$\termHeight{\sigma(u_i)} > \termHeight{\sigma(u_{i-1})} + h_G$.  The
set of heights of an infinite set of terms is infinite, so
it is always possible to choose such $\sigma(u_i)$.

Next consider internal nodes $u_1,\ldots,u_{p+q}$ in some
topological order.  For each non-parameter node $u_i$ such
that $u_i = f(u_{l_1},\ldots,u_{l_k})$ occurs in $\beta_0$, let
$\sigma(u_i) = f(\sigma(u_{l_i}),\ldots,\sigma(u_{l_k}))$.

Finally assign the values to free variables by $\sigma(x)
= \sigma(u)$ where $x = u$ occurs in $\beta_0$.

By construction, $\interpret{\structureF}\sigma = \boolTrue$ and
$\interpret{\labelsF}\sigma = \boolTrue$.  It remains to show
$\interpret{\distinctF}\sigma = \boolTrue$ i.e.\ $\sigma(u_i) \neq
\sigma(u_j)$ for $1 \leq \ i,j \ \leq p+q$, $i \neq j$.  We show this
property of $\sigma$ by induction on $m =
\min(\nodeHeight{u_i},\nodeHeight{u_j})$.  Without loss of
generality we assume $\nodeHeight{u_i} \leq \nodeHeight{u_j}$.

Consider first the case $m=0$.  Then $u_i$ is a parameter or
a constant node.  

If $u_i$ is a constant and $u_j$ is a non-parameter variable
then $u_i$ and $u_j$ are labelled by different function
symbols so $\sigma(u_i) \neq \sigma(u_j)$.

If $u_i$ is a constant and $u_j$ is a parameter variable
then $\termHeight{\sigma(u_i)}=0$ whereas
$\termHeight{\sigma(u_j)} > h_G \geq 0$.

Consider the case where $u_i$ is a parameter variable and
$u_j$ is a non-parameter variable.  
Let
\[
    J = \{ j_1 \mid u_{j_1} \mbox{ is a parameter variable s.t. }
              u_j \determines u_{j_1} \}
\]
If $J=\emptyset$, then $\beta_0$ uniquely specifies $\sigma(u_j)$, and
\[
    \termHeight{\sigma(u_j)} = \nodeHeight{u_j} \leq h_G 
  < \termHeight{\sigma(u_i)}
\]
Let $J \neq \emptyset$ and $j_0 = \max J$.
If $i \leq j_0$, then
\[
   \termHeight{\sigma(u_i)} \leq 
   \termHeight{\sigma(u_{j_0})} <
   \termHeight{\sigma(u_j)}
\]
If $j_0 < i$ then
\[
   \termHeight{\sigma(u_j)} \leq
   \termHeight{\sigma(u_{j_0})} + h_G <
   \termHeight{\sigma(u_{j_0+1})} \leq
   \termHeight{\sigma(u_i)}
\]

Now consider the case $m > 0$.  $u_i$ and $u_j$ are
non-parameter nodes, so let $u_i = f(u_{i_1},\ldots,u_{i_k})$ and
$u_j = g(u_{j_1},\ldots,u_{j_l})$.  If $f \neq g$ then clearly
$\sigma(u_i) \neq \sigma(u_j)$.  Otherwise, by congruence
closure property of base formulas, there exists $d$ such
that $u_{i_d} \neq u_{j_d}$.  Then by induction hypothesis
$\sigma(u_{i_d}) \neq \sigma(u_{j_d})$, so $\sigma(u_i) \neq
\sigma(u_j)$.
\end{proof}

\begin{corollary}
Every base formula is satisfiable.
\end{corollary}

\begin{proposition}[Quantification of Base Formula] 
\label{prop:exBaseBase}
If $\beta$ is a base formula and $x$ a free variable in $\beta$,
then there exists a base formula $\beta_1$ equivalent to $\exists x.\beta$.
\end{proposition}
\begin{proof}
Consider a formula $\exists x.\beta$ where $\beta$ is a
base formula.  The only place where $x$ occurs in $\beta$
is $x = u_{s_1}$ in the subformula $\labelsF$.  By dropping the conjunct
$x = u_{s_1}$ from $\beta$ we obtain a base formula
$\beta_1$ where $\beta_1$ is equivalent to $\exists x.\beta$.
\end{proof}

\begin{proposition}[Quantifier-Free to Base]
\label{prop:qFreeToBaseDisj}
Every well-defined quantifier-free formula in
constructor-selector language can be written as $\boolTrue$,
$\boolFalse$, or a disjunction of base formulas.
\end{proposition}
\begin{proofsketch}
Let $\phi$ be a well-defined quantifier-free formula in
constructor-selector language.  By
Proposition~\ref{prop:partialDNF} we can transform $\phi$ into
an equivalent formula in disjunctive normal form
\[
   \psi_1 \lor \cdots \lor \psi_p
\]
where each $\psi_i$ is a well-defined conjunction of literals.
Consider an arbitrary $\psi_i$.  There exists an unnested
quantifier-free formula $\psi'_i$ with additional fresh free
variables $x_1,\ldots,x_q$ such that $\psi_i$ is equivalent to
\[
    \exists x_1,\ldots,x_q.\ \psi'_i
\]
By distributivity
and~(\ref{eqn:existsPropagation}) it suffices to transform
each conjunction of unnested formulas into
disjunction of base formulas.  In the sequel we will assume
transformations based on distributivity
and~(\ref{eqn:existsPropagation}) are applied whenever we
transform conjunction of literals into a formula containing
disjunction.  We also assume that every equation
$f(x_1,\ldots,x_n)=y$ is replaced by the equivalent one
$y=f(x_1,\ldots,x_n)$ and every equation $f_i(x)=y$ is replace by
$y=f_i(x)$.

Because of our assumption that $\FreeSig$ is finite, we can
eliminate every literal of form $\lnot \Is{f}(x)$ using the
equivalence
\begin{equation} \label{eqn:conseqIsfPartit}
    \lnot \Is{f}(x) \iff \bigvee_{g \in \FreeSig \setminus \{f\}}
                          \Is{g}(x)
\end{equation}                        
which follows from~(\ref{eqn:IsfPartit}).
We then transform formula back into disjunctive normal form
and propagate the existential quantifiers to the
conjunctions of literals.  We may therefore assume that
there are no literals of form $\lnot \Is{f}(x)$ in the
conjunction.  Furthermore, $\Is{f}(x) \land \Is{g}(x) \iff
\boolFalse$ for $f \not\equiv g$, so we may assume that for
variable $x$ there is at most one literal $\Is{f}(x)$ for
some $f$.  If $f_i(x)$ occurs in the conjunction, because
the conjunction is well-defined, we may always add the
conjunct $\Is{f}(x)$.  This way we ensure that exactly one
literal of form $\Is{f}(x)$ occurs in the conjunction.

We next ensure that every variable has either
none or all of its components named by variables.  If the
conjunction contains literal $\Is{f}(x)$ but does not
contain $x=f(x_1,\ldots,x_n)$ and does not contain an equation of
form $y=f_i(x)$ for every $i$, $1 \leq i \leq \ar(f)$, we
introduce a fresh existentially quantified variable for each
$i$ such that a term of form $y=f_i(x)$ does not appear in
the conjunction.  At this point we may transform the entire
conjunction into constructor language by replacing
\[
    \Is{f}(u_i) \ \land\ v_{l_1} = f_1(u_i)\ \land \cdots \land\
                    v_{l_k} = f_k(u_i)
\]
with $u_i = f(v_{l_1},\ldots,v_{l_k})$ for $k = \ar(f)$.

We next ensure that for every two variables $x_1$ and $x_2$
occurring in the conjunction exactly one of the conjunct $x_1
= x_2$ or $x_1 \neq x_2$ is present.  Namely if both conjuncts
$x_1 = x_2$ and $x_1 \neq x_2$ are present, the conjunction is
false.  If none of the conjuncts is present, we insert the
disjunction $x_1 = x_2 \lor x_1 \neq x_2$ as one of the conjuncts
and transform the result into disjunction of existentially
quantified conjunctions.

We next perform congruence closure for finite
terms~\cite{NelsonOppen80DecisionProceduresCongruenceClosure}
on the resulting conjunction, using the fact that equality
is reflexive, symmetric, transitive and congruent with
respect to free operations $f \in \FreeOps$ and that $t(x) \neq
x$ for every term $t \not\equiv x$.  Syntactically, the result of
congruence closure can be viewed as adding new equations to
the conjunction.  If the congruence closure procedure
establishes that the formula is unsatisfiable, the result is
$\boolFalse$.  Otherwise, all variables are grouped into
equivalence classes.  If a $u_1=u_2$ occurs in the
conjunction where both $u_1$ and $u_2$ are internal variables,
we replace $u_1$ with $u_2$ in the formula and eliminate the
existential quantifier.  If for some free variable $x$ there
is no internal variable $u$ such that conjunction $x=u$ occurs,
we introduce a new existentially quantified variable and a
conjunct $x=u$.  These transformations ensure that for every
equivalence class there exists exactly one internal variable in
the formula.  It is now easy to pick representative
conjuncts from the conjunction to obtain conjunction of the
syntactic form in Definition~\ref{def:semibase} of semi-base
formula.  The resulting formula is a base formula because
congruence closure algorithm ensures that the associated
graph is acyclic.
\end{proofsketch}

We next turn to the problem of transforming a base formula
into a quantifier-free formula.  We will present two
constructions.  The first construction yields a
quantifier-free formula in {\em constructor-selector
  language} and is sufficient for the purpose of quantifier
elimination.  The second construction yields a
quantifier-free formula in {\em selector language} and is
slightly more involved; we present it to provide additional
insight into the quantifier elimination approach to term
algebras.

We first introduce notions of {\em covered} and {\em
determined} variables of a base formula $\beta$.  The basic
idea behind these notions is that $\beta$ implies a functional
dependence from the free variables of $\beta$ to each of the
determined variables.

In both constructions we use the notion of a a {\em covered
  variable}, which denotes a component of a term denoted by
some free variable.  In the first construction we also use
the notion of {\em determined variable}, which includes
covered variables as well as variables constructed from
covered variables using constructor operations $f \in
\FreeOps$.

\begin{definition} \label{def:coveredVarDef}
Consider an arbitrary base formula $\beta$.  We say that an
internal variable $u$ is {\em covered by a free variable
$x$} iff $x=u'$ occurs in $\beta$ for some $u'$ such that $u
\determines u'$.  An internal variable $u$ is {\em covered} iff
$u$ is covered by $x$ for some free variable $x$ (in
particular, if $x=u$ occurs in $\beta$ then $u$ is covered).
Let $\covered$ denote the set of covered internal variables of
base formula, and let $\uncovered = U \setminus \covered$ where $U$
is the set of all internal variables of $\beta$.
\end{definition}

\begin{lemma}[Covered Base to Selector] \label{lemma:noUncovQFree}
Every base formula without uncovered variables is equivalent
to a quantifier free formula in selector language.
\end{lemma}
\begin{proof}
Consider a base formula $\beta$ where every variable is
covered.  Consider an arbitrary quantified variable $u$.
Because $u$ is covered, there exists variable $x$ free in
$\beta$ such that $u = t(x)$ for some term $t$ in the
selector language.  Replace every occurrence of $u$ in the
matrix of $\beta$ by $t(x)$ and eliminate the quantification
over $u$.  Repeating this process for every variable $u$ we
obtain a quantifier-free formula equivalent to
$\beta$.
\end{proof}

\begin{definition}
Let $\beta$ be a base formula.  The set $\determined$ of
determined variables of $\beta$ is the smallest set $S$ that
contains the set $\covered$ and satisfies the following
condition: if $u$ is a non-parameter node and all successors
$u_1,\ldots,u_k$ ($k \geq 0$) of $u$ in the associated graph are in
$S$, then $u$ is also in $S$.
\end{definition}
In particular, every constant node is determined.  A
parameter node $w$ is determined iff $w$ is covered.

\begin{lemma} \label{lemma:undetermined}
If a node $u$ is not determined, then there exists an
uncovered parameter node $v$ such that $u \determines v$.
\end{lemma}
\begin{proof}
  The proof is by induction on $\nodeHeight(u)$.  If
  $\nodeHeight(u)=0$ then $u$ has no successors, and $u$
  cannot be a constant node because it is not determined.
  Therefore, $u$ is a parameter node, so we may let $v \equiv u$.
  Assume that the statement holds for for every node $u'$
  such that $\nodeHeight(u')=k$ and let
  $\nodeHeight(u)=k+1$.  Because $u$ is not determined,
  there exists a successor $u'$ of $u$ such that $u'$ is not
  determined, so by induction hypothesis there exists an
  uncovered parameter node $v$ such that $u' \determines v$.
  Hence $u \determines u' \determines v$.
\end{proof}

\begin{lemma} \label{lemma:dropUndetermined}
Every base formula $\beta$ is equivalent to a base formula $\beta'$
obtained from $\beta$ by eliminating all nodes that are not determined.
\end{lemma}
\begin{proof}
  Construct $\beta'$ from $\beta$ by eliminating all terms
  containing a variable $u \in U \setminus \determined$ and
  eliminating the corresponding existential quantifiers.
  Then all variables in $\beta'$ are determined.  $\beta'$ has fewer
  conjuncts than $\beta$, so $\models \beta \implies \beta'$. To show $\models \beta'
  \implies \beta $, let $\sigma$ be any assignment of terms to
  determined variables of $\beta$ such that $\beta$ evaluate to true
  under $\sigma$.  As in the proof of Lemma~\ref{lemma:baseSat},
  define the extension $\sigma'$ of $\sigma$ as follows.  Choose
  sufficiently large values $\sigma'(v)$ for every uncovered sink
  variable $v$, so that $\sigma''$ defined as the unique
  extension of $\sigma'$ to the remaining undetermined variables
  assigns different terms to different variables.  This is
  possible because the term model is infinite.  The
  resulting assignment $\sigma''$ satisfies the matrix of the
  base formula $\beta$.  Therefore, $\models \beta' \implies \beta$, so $\beta$
  and $\beta'$ are equivalent base formulas.
\end{proof}

\paragraph{First Construction} 

\begin{proposition}[Base to Constructor-Selector] 
\label{prop:baseToConsSel}
Every base formula $\beta$ is equivalent to a quantifier-free
formula $\phi$ in constructor-selector language.
\end{proposition}
\begin{proof}
By Lemma~\ref{lemma:dropUndetermined} we may assume that all
variables in $\beta$ are determined.  To every
variable $u$ we assign a term $\tau(u)$.  Term
$\tau(u)$ is in constructor-selector language and the variables
of $\tau(u)$ are among the free variables of $\beta$.  If $u \in
\covered$, we assign $\tau(u)$ as in the proof of
Lemma~\ref{lemma:noUncovQFree}.  If $u_1,\ldots,u_k$ are the
successors of a determined node $u$, we put
\[
   \tau(u) = f(\tau(u_1),\ldots,\tau(u_k))
\]
where $f$ is the label of node $u$.  This definition
uniquely determines $\tau(u)$ for all $u \in \determined$.  We
obtain the quantifier-free formula $\phi$ by replacing every
variable $u$ with $\tau(u)$ and eliminating all quantifiers.

For every $u$ we have $\models \beta \implies u=\tau(u)$, so $\models \beta
\implies \phi$.  Conversely, if $\phi$ is satisfied then $\tau$
defines an assignment for $u$ variables which makes the
matrix of $\beta$ true.  Therefore $\beta$ and $\phi$ are
equivalent.
\end{proof}

\paragraph{Second Construction}

The reason for using constructor symbols $f \in \FreeOps$ in
the first construction is to preserve the constraints of
form $u \neq v$ when eliminating node $u$ with successors
$u_1,\ldots,u_k$.  Using constructor symbols we would obtain the
constraint $f(u_1,\ldots,u_k) \neq v$.  Our second construction
avoids introducing constructor operations by decomposing
$f(u_1,\ldots,u_k) \neq v$ into disjunction of inequalities of form
$u_i \neq f_i(v)$.  When $v$ is a parameter node, the presence
of term $f_i(v)$ potentially requires introducing a new node
in the associated graph, we call this process {\em parameter
expansion}.  Parameter expansion may increase the total
number of nodes in the graph, but it decreases the number of
uncovered nodes, so the process of converting a base formula
to a quantifier-free formula in the selector language
terminates.

\begin{lemma} \label{lemma:uncoveredSource}
Let $\beta$ be an arbitrary base formula.
\begin{enumerate}
\item If $u$ is covered and $u \determines u'$ then $u'$ is
covered as well.
\item If $u'$ is uncovered and $u'$ is not a source, then
there exists $u \not\equiv u'$ such that $u \determines u'$ and
$u$ is also uncovered.
\item If $\beta$ contains an uncovered variable then $\beta$
contains an uncovered variable that is a source.
\end{enumerate}
\end{lemma}
\begin{proof}
By definition.
\end{proof}

\paragraph{Parameter Expansion}  We define the operation of 
expanding a parameter node in a base formula as follows.
Let $\beta$ be an arbitrary base formula and $w$ a parameter
variable in $\beta$.  The result of expansion of $w$ is a
disjunction of base formulas $\beta'$ generated by
applying~(\ref{eqn:IsfDef}) to $w$.  In each of the
resulting formulas $\beta'$ variable $w$ is not a parameter any
more.  Each $\beta'$ contains $\Is{f}(w)$ for some $f \in
\FreeSig$ and node $w$ has successors $u_1,\ldots,u_k$ for $k =
\ar(f)$.  Each successor $u_i$ is either an existing
internal variable or a fresh variable.  For a given $\beta$,
sink expansion generates disjunction of formulas $\beta'$ for
every choice of $f \in \FreeSig$ and every choice of
successors $u_i$, subject to congruence closure so that $\beta'$
is a base formula: we discard the choices of successors of
$w$ that yield formulas $\beta'$ violating congruence of
equality.  (This process is similar to converting
quantifier-free formulas into disjunction of base formulas
in the proof of Proposition~\ref{prop:qFreeToBaseDisj}.)
The following lemma shows the correctness of parameter
expansion.

\begin{lemma}[Parameter expansion soundness] \label{lemma:paramExpSound}
Let $\Delta = \beta_1' \lor \cdots \beta_k'$ be the disjunction generated by
parameter expansion of a base formula $\beta$.  Then $\Delta$ is
equivalent to $\beta$.
\end{lemma}
Lemma~\ref{lemma:paramExpSound} justifies the use of
parameter expansion in the following
Lemma~\ref{lemma:uncoveredElim}.
\begin{lemma} \label{lemma:uncoveredElim}
Every base formula $\beta$ can be written as a disjunction of
base formulas without uncovered variables.
\end{lemma}
\begin{proofsketch}
By Lemma~\ref{lemma:dropUndetermined} we may assume that all
variables of $\beta$ are determined.  Suppose $\beta$ contains an
uncovered variable.  Then by
Lemma~\ref{lemma:uncoveredSource}, $\beta$ contains an uncovered
variable $u_0$ such that $u_0$ is a source.  Because $u_0$
is uncovered and determined, it is not a parameter node.  We
show how to eliminate $u_0$ without introducing new
uncovered variables.

% WRONG:
%% $u$ is not a sink, otherwise it would be undetermined.
%% Therefore, $u$ has descendants in the associated graph.

Our goal is to eliminate $u_0$ from the associated graph.
We need to preserve information that $u_0$ is distinct
from variables $u \in U \setminus \{u_0\}$ in the graph.  
We consider two cases.

If $u$ is not a parameter node, then by congruence closure
either $u_0$ and $u$ are labelled by different function
symbols, or they are labelled by the same function symbol $f
\in \FreeSig$ with $\ar(f)=k$ and there exists $i$, $1 \leq i \leq
k$ and variables $u_i = f_i(u_0)$ and $u_i'=f_i(u')$ such
that $u_i \not\equiv u_i'$.  Hence the constraint $u_0 \neq u$ is
deducible from the inequalities of other variables in $\beta$
and we can eliminate $u_0$ without changing the truth value
of $\beta$.

Next consider the case when $u$ is a parameter node.  By
assumption $u$ is determined, and because it is parameter,
it is covered.  We then perform parameter node expansion as
described above.  The result of elimination of $u_0$ in $\beta$
is a disjunction of base formulas $\beta'$, in each $\beta'$ every
parameter node is expanded.  If $u$ is a parameter node in
$\beta$ then the constraint $u_0 \neq u$ is preserved in each $\beta'$
because $u$ is not a parameter node in $\beta'$ so the previous
argument applies.

Because the parameter nodes being expanded are covered, so
are their successor nodes introduced by parameter expansion.
Therefore, by repeatedly applying elimination of uncovered
variables for every uncovered variable $u_0$, we obtain a
disjunction $\Delta$ of formulas $\beta'$ where each $\beta'$ has no
uncovered variables, and $\Delta$ is equivalent to $\beta$.
\end{proofsketch}

\begin{proposition}[Base to Selector]
\label{prop:baseToSel}
For every base formula $\beta$ there exists an equivalent quantifier-free
formula $\psi$ in selector language.
\end{proposition}
\begin{proof}
By Lemma~\ref{lemma:uncoveredElim}, $\beta$ is equivalent to a
disjunction $\beta_1 \lor \cdots \lor \beta_n$ where each $\beta_i$ has no
uncovered variables.  By Lemma~\ref{lemma:noUncovQFree},
each $\beta_i$ is equivalent to some quantifier free formula
$\psi_i$, so $\beta$ is equivalent to the quantifier-free formula
$\psi_1 \lor \cdots \lor \psi_n$.
\end{proof}

\noindent
The final theorem in this section summarizes quantifier
elimination for term algebra.

\begin{theorem}[Term Algebra Quantifier Elimination]
\label{thm:termAlgQElim}
There exist algorithms $A$, $B$, $C$ such that for a given formula 
$\phi$ in constructor-selector language of term algebras:
\begin{enumerate}
\item[a)] $A$ produces a quantifier-free formula $\phi'$ 
          in constructor-selector language
\item[b)] $B$ produces a quantifier-free formula $\phi'$
          in selector language
\item[c)] $C$ produces a disjunction $\phi'$ of base formulas
\end{enumerate}
\end{theorem}
\begin{proof}
$a)$: Transform formula $\phi$ into prenex form
\[
    Q_1 x_1 \ldots Q_{n-1} x_{n-1} Q_n x_n. \phi^{*}
\]
where $\phi^{*}$ is quantifier free, as in
Section~\ref{sec:qeIntro}.  We eliminate the innermost
quantifier $Q_n$ as follows.

Suppose first that $Q_n$ is $\exists$.  Transform the matrix
$\phi^{*}$ into disjunctive normal form $C_1 \lor \cdots \lor C_n$.  By
Proposition~\ref{prop:qFreeToBaseDisj}, transform $C_1 \lor \cdots \lor
C_n$ into disjunction $\beta_1 \lor \cdots \lor \beta_m$ of base formulas.
Then propagate $\exists$ into individual disjuncts, using
\[
    \exists x_n.\ \beta_1 \lor \cdots \lor \beta_m  \ \iff \ (\exists x_n. \beta_1) \lor \cdots \lor (\exists x_n. \beta_m)
\]
By Proposition~\ref{prop:exBaseBase}, an existentially
quantified base formula is again a base formula, so $\exists
x_n. \beta_i \iff \beta'_i$ for some $\beta'_i$.  We thus obtain the
\begin{equation} \label{eqn:baseDisj}
    Q_1 x_1 \ldots Q_{n-1} x_{n-1}.\ \beta'_1 \lor \cdots \lor \beta'_m
\end{equation}
By Proposition~\ref{prop:baseToConsSel}, every
base formula is equivalent to a quantifier-free formula in
selector language, so~\ref{eqn:baseDisj} is equivalent to
\[
    Q_1 x_1 \ldots Q_{n-1} x_{n-1}. \psi
\]
where $\psi$ is a quantifier free formula.  Hence, we have
eliminated the innermost existential quantifier.

Next consider the case when $Q_n$ is $\forall$.  Then $\phi$ is equivalent to
\[
    Q_1 x_1 \ldots Q_{n-1} x_{n-1} \lnot \exists x_n. \lnot \phi^{*}
\]
Apply the procedure for eliminating $x_n$ to $\lnot \phi^{*}$.  The
result is formula of form
\begin{equation} \label{eqn:negElim}
    Q_1 x_1 \ldots Q_{n-1} x_{n-1}.\ \lnot \psi
\end{equation}
where $\psi$ is quantifier free.  But $\lnot \psi$ is also quantifier
free, so we have eliminated the innermost universal
quantifier.  By repeating this process we eliminate all
quantifiers, yielding the desired formula $\phi'$.

The direct construction for showing $b)$ is analogous to
$a)$, but uses Proposition~\ref{prop:baseToSel} in place of
Proposition~\ref{prop:baseToConsSel}.  To show $c)$, apply
e.g.\ construction $a)$ to obtain a quantifier-free formula
$\psi$ and then transform $\psi$ into disjunction of base formulas
using~Proposition~\ref{prop:qFreeToBaseDisj}.
\end{proof}

This completes our description of quantifier elimination for
term algebras.

We remark that there are alternative ways to define base
formula.  In particular the requirement on disequality of
all variables is not necessary.  This requirement may lead
to unnecessary case analysis when converting a
quantifier-free formula to disjunction of base formulas, but
we believe that it simplifies the correctness argument.

%%% Local Variables: 
%%% mode: latex
%%% TeX-master: "main"
%%% End: 

% LocalWords:  disequality Mal'cev's disunification Maher unnested cyclicity
% LocalWords:  CompleteAxiomatizationsAlgebrasTrees iff ComonLescanne preorder
% LocalWords:  Disequations subformula prenex

\section{The Pair Constructor and Two Constants}
\label{sec:twoconst}

In this section we give a quantifier elimination procedure
for structural subtyping of non-recursive types with two
constant symbols and one covariant binary constructor.  Two
constants corresponds to two primitive types; one binary
covariant constructor corresponds to the pair constructor
for building products of types.

The construction in this section is an introduction to the
more general construction in Section~\ref{sec:nconst}, where
we give a quantifier elimination procedure for any number of
constant symbols and relations between them.  The
construction in this section demonstrates the interaction
between the term and boolean algebra components of the
structural subtyping.  We therefore believe the construction
captures the essence of the general result of
Section~\ref{sec:nconst}.

The basic observation behind the quantifier elimination
procedure for two constant symbols is that the structure of
terms in this language is isomorphic to a disjoint union of
boolean algebras with some additional term structure
connecting elements from different boolean algebras.  As we
argue below, the structural subtyping structure contains one
copy of boolean algebra for every equivalence class of terms
that have the same ``shape'' i.e.\ are same up to the
constants in the leaves.

Consider a signature $\FreeSig = \{ a, b, g \}$ where $a$ and
$b$ are constant symbols and $g$ is a function symbol of
arity 2.  We define a partial order $\leq$ on the set $\FT$ of
ground terms over $\FreeSig$ as the least reflexive partial
order relation $\mrho$ satisfying
\begin{enumerate}
\item $a \mrho b$;
\item $(s_1 \mrho t_1) \land (s_2 \mrho t_2) \implies
       g(s_1,s_2) \mrho g(t_1,t_2)$.
\end{enumerate}
The structure with equality in the language $\{a,b,g,\leq\}$,
where $\leq$ is interpreted as above and $a,b,g$ are
interpreted as free operations on term algebra corresponds
to the structural subtyping with two base types $a$ and $b$
and one binary type constructor $g$, with $g$ covariant in
both arguments.  We denote this structure by
$\binarySubtyping$.  We proceed to show that
$\binarySubtyping$ admits quantifier elimination and is
therefore decidable.

\subsection{Boolean Algebras on Equivalent Terms}
\label{sec:baEquivTerms}

In preparation for the quantifier elimination procedure we
define certain operations and relations on terms.  We also
establish some fundamental properties of the structure
$\binarySubtyping$.

Define a new signature $\FreeSigConst = \{ \cs, \gs \}$ as an
abstraction of signature $\FreeSig = \{ a, b, g \}$.  Define
function $\shapifiedSym : \FreeSig \to \FreeSigConst$ by
\[\begin{array}{l}
    \shapified{a} = \cs \mnl
    \shapified{b} = \cs \mnl
    \shapified{g} = \gs
\end{array}\]
Let $\ar(\shapified{f}) = \ar(f)$ for each $f \in
\FreeSig$; in this case $\cs$ is a constant and $\gs$ is a
binary function symbol.  Let $\FTConst$ be the set of ground
terms over the signature $\FreeSigConst$.
Define {\em shape} of a term $t$, as the function
$\termShapeF : \FT \to \FTConst$, by letting
\[\begin{array}{l}
   \termShape{f(t_1,\ldots,t_k)}= \mnl
 \qquad \shapified{f}(\termShape{t_1},\ldots,\termShape{t_k})
\end{array}\]
for $k = \ar(f)$.  In this case we have
\[\begin{array}{rcl}
  \termShape{a} &=& \cs \mnl 
  \termShape{b} &=& \cs \mnl
  \termShape{g(t_1,t_2)} &=&
  \gs(\termShape{t_1},\termShape{t_2})
\end{array}\]
Define $t_1 \sim t_2$ iff $\termShape{t_1}=\termShape{t_2}$.
Then $\sim$ is the smallest equivalence relation $\mrho$ such
that
\begin{enumerate}
\item $a \mrho b$;
\item $(s_1 \mrho t_1) \land (s_2 \mrho t_2) \implies
       g(s_1,s_2) \mrho g(t_1,t_2)$.
\end{enumerate}
For every term $t$ define the word $\termContent{t} \in
\{0,1\}^{*}$ by letting
\[\begin{array}{rcl}
  \termContent{a} &=& 0 \mnl
  \termContent{b} &=& 1 \mnl
  \termContent{f(t_1,t_2)} &=& \termContent{t_1} \cdot \termContent{t_2} \mnl
\end{array}\]
The set of all words $w \in \{0,1\}^n$ is isomorphic the boolean
algebra of $\BoolAlg{n}$ of all subsets of some finite sets
of cardinality $n$, so we write $w_1\cap w_2$, $w_1\cup w_2$,
$w^c$ for operations corresponding to intersection, union,
and set complement in the set of words $w \in \{0,1\}^n$.  We
write $w_1 \subseteq w_2$ for $w_1 \cap w_2 = w_1$.

Define function $\delta$ by
\[
  \delta(t) = \tu{\termShape{t},\termContent{t}}
\]
For term $t$ in any language containing constant symbols,
let $\termLength{t}$ denote the number of occurrences of
constant symbols in $t$.  If $w$ is a sequence of elements
of some set, let $\seqLength{w}$ denote the length of the
sequence.  Observe that
$\seqLength{\termContent{t}}=\termLength{t}$ and
$\termLength{\termShape{t}} = \termLength{t}$.  Moreover,
$t_1 \sim t_2$ implies
$\seqLength{\termContent{t_1}}=\seqLength{\termContent{t_2}}$.
Define the set $B$ by
\[
  \Bset = \{ \tu{s,w} \mid s \in \FTConst, w \in \{0,1\}^{*}, 
                  \termLength{s}=\seqLength{w} \}
\]
Function $\delta$ is a bijection from the set $\FT$ to the set
$\Bset$.  For $b_1,b_2 \in \Bset$ define $b_1 \leq b_2$ iff
$\delta^{-1}(b_1) \leq \delta^{-1}(b_2)$.  From the definitions it
follows
\[
    \tu{s_1,w_1} \leq \tu{s_2,w_2} \ \iff \ s_1=s_2 \land w_1 \subseteq w_2
\]
If $g$ is defined on $\Bset$ via isomorphism $\delta$ we
also have
\[
   g(\tu{s_1,w_1},\tu{s_2,w_2}) = \tu{\gs(s_1,s_2),w_1 \cdot w_2}
\]
For any fixed $s \in \FTConst$, the set
\begin{equation} \label{eqn:bSubSet}
  \BSubSet{s_0} = \{ \tu{s,w} \in B \mid s = s_0 \}
\end{equation}
is isomorphic to the boolean algebra $\BoolAlg{n}$, where $n
= \termLength{s}$.  Accordingly, we introduce on each
$\BSubSet{s}$ the set operations $t_1\cap_s t_2$, $t_1\cup_s t_2$,
${t_1^c}_s$.  Expressions $t_1\cap_s t_2$ and $t_1\cup_s t_2$ are
defined iff $\termShape{t_1}=s$ and $\termShape{t_2}=s$, whereas
expression ${t_1^c}_s$ is defined iff $\termShape{t_1} = s$.

We also introduce cardinality expressions as in
Section~\ref{sec:qeBoolalg}.  If $t$ denotes a term, then
the expression $|t|_s$ denotes the number of elements of the
set corresponding to $t$.  Here we require $s =
\termShape{t}$.  We use expressions $|t|_s =k$ and $|t|_s \geq
k$ as atomic formulas for constant integer $k \geq 0$.  Note
that
\begin{equation} \label{eqn:termRelElim}
\begin{array}{l}
   t_1 \leq t_2 \iff \termShape{t_1}= \termShape{t_2} \ \land\
              |t_1\cap t_2^c|_{\termShape{t_1}}=0
\end{array}
\end{equation}
\begin{equation} \label{eqn:termDiseqElim}
\begin{array}{l}
   t_1 = t_2  \iff 
\begin{array}[t]{l}
 \termShape{t_1}=\termShape{t_2} \ \land \mnl
 |(t_1\cap t_2^c) \cup (t_1^c\cap t_2)|_{\termShape{t_1}}=0
\end{array}
\end{array}
\end{equation}
%%    t_1 \leq t_2 \iff \mnl \qquad \termShape{t_1}= \termShape{t_2} \land
%%                           |t_1\cap t_2^c|_{\termShape{t_1}}=0 \mnl
%%    t_1 = t_2 \iff \mnl \qquad \exists s.\ \termShape{t_1}=s \land \termShape{t_2}=s\ \land \mnl
%%                    \qquad \ \ \ \ |t_1\cap t_2^c|_s=0 \land |t_1^c\cap t_2|_s=0

Let $\termShape{t_1}=s_1$, $\termShape{t_2}=s_2$, and
$s=\gs(s_1,s_2)$.
Then
\begin{equation} \label{eqn:decompose}
 |g(t_1,t_2)|_s = |t_1|_{s_1} + |t_2|_{s_2}
\end{equation}
Equation~\ref{eqn:decompose} allows decomposing formulas of
form $|g(t_1,t_2)|_s\geq k$ into propositional combinations of
formulas of form $|t_1|_{s_1}\geq k$ and $|t_2|_{s_2}\geq k$.

Note further that the following equations hold:
\[\begin{array}{rcl} \label{eqn:tupleProduct}
 g(t_1,t_2) \cap g(t'_1,t'_2) &=& g(t_1 \cap t'_1, t_2 \cap t'_2) \mnl
 g(t_1,t_2) \cup g(t'_1,t'_2) &=& g(t_1 \cup t'_1, t_2 \cup t'_2) \mnl
 g(t_1,t_2)^c &=& g(t_1^c,t_2^c)
\end{array}\]
If $E(x_1,\ldots,x_n)$ denotes an expression consisting only of
operations of boolean algebra, then
from~(\ref{eqn:tupleProduct}) by induction follows that
\begin{equation} \label{eqn:termHom}
   E(g(t_1^1,t_1^2),\ldots,g(t_n^1,t_n^2)) =
    g(E(t_1^1,\ldots,t_n^1),E(t_1^2,\ldots,t_n^2))
\end{equation}
Equations~(\ref{eqn:termHom}) and (\ref{eqn:decompose}) imply
\begin{equation} \label{eqn:anyDecompose}
|E(g(t_1^1,t_1^2),\ldots,g(t_n^1,t_n^2))| = 
 |E(t_1^1,\ldots,t_n^1)| + |E(t_1^2,\ldots,t_n^2)|
\end{equation}
Boolean algebra $\BSubSet{\gs(s_1,s_2)}$ is isomorphic to
the product of boolean algebras $\BSubSet{s_1}$ and
$\BSubSet{s_2}$; the constructor $g$ acts as union of
disjoint sets.

\subsection{A Multisorted Logic}
\label{sec:twoConstLogic}

To show the decidability of structure $\binarySubtyping$, we
give a quantifier elimination procedure for an extended
structure, denoted $\twoSorted$.  We use a first-order
two-sorted logic with sorts $\termSort$ and $\shapeSort$
interpreted over $\twoSorted$.

The domain of structure $\twoSorted$ is $\FT \cup \FTConst$
with elements $\FT$ having sort $\termSort$ and elements
$\FTConst$ having sort $\shapeSort$.  Variables in $\TVar$
have $\termSort$ sort, variables in $\SVar$ have
$\shapeSort$ sort.  In general, if $t$ denotes an element of
$\twoSorted$, we write $t^S$ to indicate that the element
has sort $\shapeSort$.

\begin{figure}
\[\begin{array}{rcl}
   a, b &::& \termSort \mnl
   g &::& \termSort \times \termSort \to \termSort \mnl
   \Is{g} &::& \termSort \to \boolSort \mnl
   g_1, g_2 &::& \termSort \to \termSort \mnl
   \termEq &::& \termSort \times \termSort \to \boolSort \mnl
   \leq &::& \termSort \times \termSort \to \boolSort \mnl
   \cs &::& \shapeSort \mnl
   \gs &::& \shapeSort \times \shapeSort \to \shapeSort \mnl
   \Isgs &::& \shapeSort \to \boolSort \mnl
   \gsOne, \gsTwo &::& \shapeSort \to \shapeSort \mnl
   \termShapeF &::& \termSort \to \shapeSort \mnl
   \shapeEq &::& \shapeSort \times \shapeSort \to \boolSort \mnl
   \cap_{\_}, \cup_{\_} &::& \shapeSort \times \termSort \times \termSort \to \termSort \mnl
   \_^c_{\_} &::& \shapeSort \times \termSort \to \termSort \mnl
   \oneTerm_{\_}, \zeroTerm_{\_} &::& \shapeSort \to \termSort \mnl
   |\_|_{\_} \geq k, \ |\_|_{\_} = k &::& \shapeSort \times \termSort \to \boolSort
\end{array}\]
\caption{Operations and relations in structure $\twoSorted$
\label{fig:twoSortedSig}}
\end{figure}
% !! More BLAH, explain all symbols
Figure~\ref{fig:twoSortedSig} shows operations and relations
in $\twoSorted$ with their sort declarations.  The signature
is infinite because operations $|t|_s\geq k$ and $|t|_s = k$
are parameterized by a non-negative integer $k$.

We require all terms to be well-sorted.  Functions $g_1$ and
$g_2$ are interpreted as partial selector functions in the
term constructor-selector language, 
so $D_{g_1} = D_{g_2} =
\tu{\tu{x},\Is{g}(x)}$.  Similarly, $\gsOne$ and $\gsTwo$
are partial selector functions in the shape
constructor-selector language, so $D_{\gsOne} = D_{\gsTwo} =
\tu{\tu{x},\Isgs(x)}$.  
The expressions $t_1\cap_st_2$ and
$t_1\cup_st_2$ are defined iff
$\termShape{t_1}=\termShape{t_2}=s$, and $t^c_s$ is defined
iff $\termShape{t}=s$.  We therefore let
\[\begin{array}{l}
    D_{\cap_s} = D_{\cup_s} = \mnl
 \qquad \tu{\tu{\ys,x_1,x_2},\termShape{x_1}=\ys \land \termShape{x_2}=\ys}
\end{array}\]
and
\[
   D_{\_^c_{\_}} = \tu{\tu{\ys,x},\termShape{x}=\ys}
\]
For atomic formulas $|t|_s\geq k$ and $|t|_s=k$ we require
atomic formula $\termShape{t}=s$ to ensure well-definedness:
\[
   D_{|\_|\_=k} = D_{|\_|\_\geq k} = \tu{\tu{\ys,x},\termShape{x}=\ys}
\]

Note that the language of Figure~\ref{fig:twoSortedSig}
subsumes the language $\{a,b,g,\leq\}$ for the structural
subtyping structure.  The quantifier-elimination procedure
we present in Section~\ref{sec:twoConstQE} is therefore
sufficient for quantifier elimination in the first-order
logic interpreted over the structural subtyping structure
$\twoSorted$.

\subsection{Quantifier Elimination for Two Constants}
\label{sec:twoConstQE}

We are now ready to present a quantifier elimination
procedure for the structure $\twoSorted$.  The quantifier
elimination procedure is based on the quantifier elimination
for term algebras of Section~\ref{sec:qeTA} as well as the
quantifier elimination for boolean algebras of
Section~\ref{sec:qeBoolalg}.

We first define an auxiliary notion of a $\us$-term as a
term formed starting from shape $\us$ term variables and
shape $\us$ constants, using operations $\cap_{\us}$,
$\cup_{\us}$, and $\_^c_{\us}$.
\begin{definition}[$\us$-terms] \label{def:usterms}
Let $\us \in \SVar$ be a shape variable.  The set of $\us$-terms
$\usterm{\us}$ is the least set such that:
\begin{enumerate}
\item $\TVar \subseteq \usterm{\us}$
\item $0_{\us}, 1_{\us} \in \usterm{\us}$
\item if $t,t' \in \usterm{\us}$, then also
\[\begin{array}{l}
t\cap_{\us} t' \in \usterm{\us}, \mnl
t\cup_{\us}t' \in \usterm{\us}, \mbox{ and} \mnl
t^c_{\us} \in \usterm{\us}
\end{array}\]
\end{enumerate}
\end{definition}

Similarly to base formulas of Section~\ref{sec:qeTA}, we
define {\em structural base formulas} for
$\twoSorted$ structure.  A structural base formula contains
a copy of a base formula for the shape sort ($\shapeBaseF$),
a base formula for the term sort without term disequalities
($\termBaseF$), a formula expressing mapping of term
variables to shape variables ($\homF$), and cardinality
constraints on term parameter nodes of the term base formula
($\cardinF$).

\begin{definition}[Structural Base Formula] \ \\
\label{def:structuralBase}
A {\em structural base formula} with:
\begin{itemize}
\item free term variables $x_1,\ldots,x_m$;
\item internal non-parameter term variables $u_1,\ldots,u_p$;
\item internal parameter term variables $u_{p+1},\ldots,u_{p+q}$;
\item free shape variables $\xs_1,\ldots,\xs_{\ms}$;
\item internal non-parameter shape variables $\us_1,\ldots,\us_{\ps}$;
\item internal parameter shape variables $\us_{\ps},\ldots,\us_{\ps+\qs}$
\end{itemize}
is a formula of form:
\[\begin{array}{l}
  \exists  u_1,\ldots,u_n, \us_1,\ldots,\us_{\ns}. \mnl
    \qquad  \shapeBaseF(\us_1,\ldots,\us_{\ns},\xs_1,\ldots,\xs_{\ms}) \land {} \mnl
    \qquad  \termBaseF(u_1,\ldots,u_n,x_1,\ldots,x_m) \land {} \mnl
    \qquad  \homF(u_1,\ldots,u_n,\us_1,\ldots,\us_{\ns}) \land {} \mnl
    \qquad  \cardinF(u_{p+1},\ldots,u_n,\us_{\ps+1},\ldots,\us_{\ns})
\end{array}\]
where $n=p+q$, $\ns = \ps + \qs$, and formulas
$\shapeBaseF$, $\termBaseF$, $\homF$, and $\cardinF$ are
defined as follows.
\[\begin{array}{l}
\shapeBaseF(\us_1,\ldots,\us_{\ns},\xs_1,\ldots,\xs_{\ms}) = \mnl
\qquad \bigwedge\limits_{i=1}^{\ps} \us_i = t_i(\us_1,\ldots,\us_n) \ \land\
       \bigwedge\limits_{i=1}^{\ms} \xs_i = \us_{j_i} \mnl
\qquad \land \ \distinctF(\us_1,\ldots,\us_n)
\end{array}\]
where each $t_i$ is a shape term of form $f(\us_{i_1},\ldots,\us_{i_k})$ for
some $f \in \FreeSigConst$, $k=\ar(f)$, and 
$j : \{ 1,\ldots,\ms \} \to \{ 1,\ldots,\ns\}$ is a function mapping
indices of free shape variables to indices of internal shape variables.
\[\begin{array}{l}
\termBaseF(u_1,\ldots,u_n,x_1,\ldots,x_m) = \mnl
 \qquad \bigwedge\limits_{i=1}^p u_i = t_i(u_1,\ldots,u_n) \ \land\ \bigwedge\limits_{i=1}^m x_i = u_{j_i}
\end{array}\]
where each $t_i$ is a term of form $f(u_{i_1},\ldots,u_{i_k})$ for
some $f \in \FreeSig$, $k = \ar(f)$, and 
$j : \{ 1,\ldots,m \} \to \{ 1,\ldots,n\}$ is a function mapping
indices of free term variables to indices of internal term variables.
\[\begin{array}{ll}
\homF(u_1,\ldots,u_n,\us_1,\ldots,\us_{\ns}) = &
  \bigwedge\limits_{i=1}^n \termShape{u_i} = \us_{j_i}
\end{array}\]
where $j : \{ 1,\ldots,n \} \to \{ 1,\ldots,\ns\}$ is some function such
that $\{ j_1,\ldots,j_p \} \subseteq \{ 1,\ldots,\ps \}$ and $\{ j_{p+1},\ldots,j_{p+q}
\} \subseteq \{ \ps+1,\ldots,\ps+\qs \}$ 
(a term variable is a parameter
variable iff its shape is a parameter shape variable).
\[\begin{array}{ll}
\cardinF(u_{p+1},\ldots,u_{p+q},\us_{\ps+1},\ldots,\us_{\ps+\qs}) = \psi_1 \land \cdots \land \psi_r
\end{array}\]
where each $\psi_i$ is of form
\[
   |t(u_{p+1},\ldots,u_{p+q})|_{\us}=k
\]
or
\[
   |t(u_{p+1},\ldots,u_{p+q})|_{\us}\geq k
\]
for some $\us$-term
$t(u_{p+1},\ldots,u_{p+q})$ that contains no variables other
than some of the variables $u_{p+1},\ldots,u_{p+q}$,
and the following
condition holds:
\begin{equation} \label{eqn:wellDefinednessCondition}
\begin{minipage}{2.5in}
If a variable $u_{p+j}$ occurs in term
$t(u_{p+1},\ldots,u_{p+q})$, then $\termShape{u_{p+j}}=\us$
occurs in formula $\homF$.
\end{minipage}
\end{equation}

We require each structural base formula to satisfy the following
conditions:
\begin{enumerate}

\item[P0)] the graph associated with shape base formula 
      \[
         \exists  \us_1,\ldots,\us_{\ns}.\
            \shapeBaseF(\us_1,\ldots,\us_{\ns},\xs_1,\ldots,\xs_{\ms})
      \]
      is acyclic (compare to Definition~\ref{def:baseFormula});

\item[P1)] congruence closure property for $\shapeBaseF$ subformula:
      there are no two distinct variables $\us_i$ and
      $\us_j$ such that both $\us_i =
      f(\us_{l_1},\ldots,\us_{l_k})$ and $\us_j =
      f(\us_{l_1},\ldots,\us_{l_k})$ occur as conjuncts in
      formula $\shapeBaseF$;

\item[P2)] congruence closure property for $\termBaseF$ subformula:
      there are no two distinct variables $u_i$ and $u_j$
      such that both $u_i = f(u_{l_1},\ldots,u_{l_k})$ and $u_j =
      f(u_{l_1},\ldots,u_{l_k})$ occur as conjuncts in formula
      $\termBaseF$;

\item[P3)] homomorphism property of $\termShapeF$:
      for every non-parameter term variable $u$ such that $u =
      f(u_{i_1},\ldots,u_{i_k})$ occurs in $\termBaseF$, if
      the conjunct $\termShape{u}=\us$ occurs in $\homF$,
      then for some shape variables
      $\us_{j_1},\ldots,\us_{j_k}$ the term $\us =
      \fs(\us_{j_1},\ldots,\us_{j_k})$ occurs in $\shapeBaseF$
      where $\fs=\shapified{f}$ and for every $r$ where $1 \leq
      r \leq k$, conjunct $\termShape{u_{i_r}}=\us_{j_r}$
      occurs in $\homF$.

\end{enumerate}
\end{definition}
According to Definition~\ref{def:structuralBase} a
structural base formula contains no selector function symbols.
Formulation using selector symbols is also possible, as in
Definition~\ref{def:semibaseSelector}.  The only partial
function symbols occurring in a structural base formula of
Definition~\ref{def:structuralBase} are in $\cardinF$
subformula.  Condition~(\ref{eqn:wellDefinednessCondition})
therefore ensures that functions in $\cardinF$ and thus the
entire base formula are well-defined.

Note that acyclicity of shape base formula $\shapeBaseF$
(condition P0) implies acyclicity of term base formula as
well.  Namely, condition P3 ensures that any cycle in
$\termBaseF$ implies a cycle in $\shapeBaseF$.

As in Section~\ref{sec:qeTA} we proceed to show that each
quantifier-free formula can be written as a disjunction of
base formulas and each base formula can be written as a
quantifier-free formula.

We strongly encourage the reader to study the following
example because it illustrates the idea behind our
quantifier-elimination decision procedure.
\begin{example} \label{exa:twoConstMain}
The following sentence is true in structure $\twoSorted$.
\begin{equation}
\label{eqn:exampleSentence}
\begin{array}{l}
\forall x, y.\ x \leq y \implies \mnl
\begin{array}[t]{l}
\exists z.\ z \leq x \land z \leq y \ \land \mnl
\begin{array}[t]{l}
\forall w.\ w \leq x \land w \leq y \implies \mnl
\quad \forall v.\ g(v,z) \leq g(z,v) \land \Is{g}(v) \land \Is{g}(w) \implies g_1(w) \leq g_1(v)
\end{array}
\end{array}
\end{array}
\end{equation}
An informal proof of sentence~(\ref{eqn:exampleSentence}) is
as follows.  Suppose that $x \leq y$.  Then
$\termShape{x}=\termShape{y}=\xs$.  Let $z = x \cap_{\xs} y$.
Now consider some $w$ such that $w \leq x$ and $w \leq y$.  Then
$\termShape{w}=\xs$, so $w \leq z$.  Suppose that $v$ is such
that $g(v,z) \leq g(z,v)$.  Then by covariance of $g$ we have
$z \leq v$, so $w \leq v$.  If we assume $\Is{g}(w)$ and
$\Is{g}(v)$, then $g_1(w)$ and $g_1(v)$ are well defined and
by covariance of $g$ we conclude $g_1(w) \leq g_1(v)$, as
desired.

We now give an alternative argument that shows that
sentence~(\ref{eqn:exampleSentence}) is true.  This
alternative argument illustrates the idea behind
our quantifier-elimination decision procedure.  For the sake
of brevity we perform some additional simplifications along
the way that are not part of the procedure we present
(although they could be incorporated to improve efficiency),
and we skip consideration of some uninteresting cases during
the case analyses.

\begin{figure}
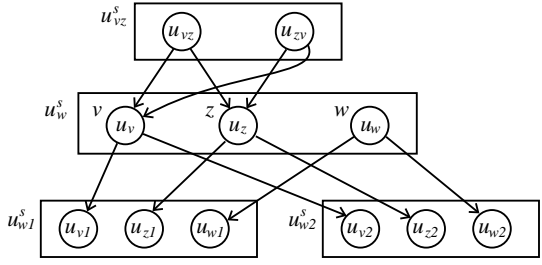

\begin{center}
\mypic{exaStrucBase}
\end{center}
\caption{One of the Base Formulas Resulting from (\ref{eqn:exaInnerMost})
\label{fig:exaStrucBase}}
\end{figure}

Let us first eliminate the quantifier from formula
\begin{equation} \label{eqn:exaInnerMost}
\forall v.\ g(v,z) \leq g(z,v) \land \Is{g}(v) \land \Is{g}(w) \implies g_1(w) \leq g_1(v)
\end{equation}
Formula~(\ref{eqn:exaInnerMost}) is equivalent to
$\lnot \exists v. \phi_1$ where
\begin{equation}
\phi_1 \equiv 
  g(v,z) \leq g(z,v) \land \Is{g}(v) \land \Is{g}(w) \land
  \lnot (g_1(w) \leq g_1(v))
\end{equation}
We next use~(\ref{eqn:termRelElim}) to
eliminate atomic formulas $t_1 \leq t_2$ and replace them
with cardinality constraints, resulting in formula $\phi_2$ equivalent
to $\phi_1$:
\[
   \phi_2 \equiv \phi_{2,1} \land \phi_{2,2}
\]
where
\begin{equation}
\begin{array}{l}
\phi_{2,1} \equiv
\begin{array}[t]{l}
|g(v,z) \cap g(z,v)^c|_{\termShape{g(v,z)}} = 0 \ \land \mnl
\termShape{g(v,z)}=\termShape{g(z,v)} \ \land \mnl
\Is{g}(v) \land \Is{g}(w)
\end{array}
\end{array}
\end{equation}
and
\begin{equation}
\begin{array}{l}
\phi_{2,2} \equiv \mnl
\begin{array}{l}
\lnot \left(|g_1(w) \cap g_1(v)^c|_{\termShape{g_1(w)}} = 0 \ \land \
   \termShape{g_1(w)}=\termShape{g_1(v)}\right)
\end{array}
\end{array}
\end{equation}
Here we have written e.g.\
\[
     |g(v,z) \cap g(z,v)^c|_{\termShape{g(v,z)}} = 0
\]
as a shorthand for
\[
    |g(v,z) \cap_{\termShape{g(v,z)}} g(z,v)^c_{\termShape{g(v,z)}}|_{\termShape{g(v,z)}} = 0
\]
(In general, we omit term shape arguments for boolean algebra
operations if the arguments are identical to the enclosing
term shape argument of the cardinality constraint.)

We next transform $\phi_2$ into disjunction of well-defined
conjunctions.  Following the ideas in Proposition~\ref{prop:partialDNF},
we transform $\phi_{2,2}$ into $\phi_{3,1} \lor \phi_{3,2}$
where
\begin{equation}
\begin{array}{l}
\phi_{3,1} \equiv \mnl
\begin{array}{l}
|g_1(w) \cap g_1(v)^c|_{\termShape{g_1(w)}} \geq 1 \ \land \
   \termShape{g_1(w)}=\termShape{g_1(v)}
\end{array}
\end{array}
\end{equation}
and
\[
   \phi_{3,2} \ \equiv\ \termShape{g_1(w)} \neq \termShape{g_1(v)}
\]
and then transform $\phi_{2,1} \land \phi_{2,2}$ into
\[
   (\phi_{2,1} \land \phi_{3,1}) \lor (\phi_{2,1} \land \phi_{3,2})
\]
For the sake of brevity we ignore the case $\phi_{2,1} \land
\phi_{3,2}$; it is possible to show that $\phi_{2,1} \land \phi_{3,2}$ is
equivalent to $\boolFalse$ in the context of the entire
formula.

We transform $\phi_{2,1} \land \phi_{3,1}$ into unnested form,
introducing fresh existentially quantified variables
$u_{vz}$,$u_{zv}$,$u_{w1}$,$u_{v1}$, $\us_{vz}$, $\us_{w1}$
that denote terms occurring in $\phi_{2,1} \land \phi_{3,1}$.  The
result is formula $\phi_4$ where
\begin{equation}
\begin{array}{l}
\phi_4 \ \equiv \
\begin{array}[t]{l}
\exists u_{vz},u_{zv},u_{w1},u_{v1}, \us_{vz}, \us_{w1}. \mnl
\begin{array}{l}
u_{vz} = g(v,z) \land u_{zv} = g(z,v) \ \land \mnl
u_{w1} = g_1(w) \land u_{v1} = g_1(v) \ \land \mnl
\us_{vz} = \termShape{u_{vz}} \land \us_{w1} = \termShape{u_{w1}} \ \land \mnl
\termShape{u_{zv}} = \us_{vz} \land \termShape{u_{v1}} = \us_{w1} \ \land \mnl
\Is{g}(v) \land \Is{g}(w) \ \land \mnl
|u_{vz} \cap u^c_{zv}|_{\us_{vz}} = 0 \ \land \
|u_{w1} \cap u^c_{v1}|_{\us_{w1}} \geq 1
\end{array}
\end{array}
\end{array}
\end{equation}
To transform $\phi_4$ into disjunction of structural base
formulas we keep introducing new existentially quantified
variables and adding derived conjuncts to satisfy the
invariants of Definition~\ref{def:structuralBase}.  

Because $\Is{g}(v)$ and $\Is{g}(w)$ appear in the conjunct,
we give names to the remaining successors of $v$, $w$, by
introducing $u_{w2} = g_2(w)$, $u_{v2} = g_2(v)$.  We may now
write the constraints in constructor language,
using e.g.\ conjunct $v=g(u_{v1},u_{v2})$ instead of
\[
     \Is{g}(v) \ \land \ u_{v1} = g_1(v) \ \land \ u_{v2} = g_2(v)
\]
To ensure that every term variable has an associated shape
variable, we introduce fresh variables $\us_v$, $\us_w$,
$\us_z$, $\us_{w2}$, $\us_{v2}$ with conjuncts $\us_v =\termShape{v}$,
$\us_w = \termShape{w}$, $\us_z = \termShape{z}$, $\us_{w2}
= \termShape{u_{w2}}$, $\us_{v2}=\termShape{u_{v2}}$.

Note that base formula contains $\distinctF(\us_1,\ldots,\us_n)$
subformula.  In the case when the current conjunction is not
strong enough to entail the disequality between shape
variables $\us_i$ and $\us_j$, we perform case analysis,
considering the case $\us_i=\us_j$ (then $\us_i$ can be
replaced by $\us_j$), and the case $\us_i \neq \us_j$.  This
case analysis will lead to a disjunction of structural base
formulas (unless some of the formulas is shown contradictory
in the transformation process).  In contrast to {\em shape}
variables, we do not not perform case analysis for
disequality of {\em term} variables, because $\termBaseF$ in
Definition~\ref{def:structuralBase} does not contain a
$\distinctF$ subformula.

In this example we perform case analysis on whether $\us_w =
\us_z$ and $\us_w = \us_v$ should hold.  For the sake of the
example let us consider the case when $\us_w = \us_z =
\us_v$, $\us_{v2}=\us_{w2}$ and
$\us_{vz},\us_{w},\us_{w1},\us_{w2}$ are all distinct.  In
that case shape variables $\us_w,\us_z,\us_v$ denote the
same shape, so let us replace e.g.\ $\us_z$ and $\us_v$ with
$\us_w$.  Similarly, we replace $\us_{v2}$ with $\us_{w2}$.
We obtain conjuncts $\termShape{v} = \us_w$,
$\termShape{z} = \us_w$, $\termShape{u_{v2}}=\us_{w2}.$

We next ensure homomorphism property P3 in
Definition~\ref{def:structuralBase}.  From conjuncts $u_{vz}
= g(v,z)$, $\termShape{u_{vz}}=\us_{vz}$, and
$\termShape{v}=\us_w$, we conclude
\[\begin{array}{rl}
   \us_{vz} = 
    \begin{array}[t]{l}
    \termShape{u_{vz}} = \mnl
    \termShape{g(v,z)} = \mnl
    \gs(\termShape{v},\termShape{z}) = \mnl
    \gs(\us_w,\us_w)
  \end{array}
\end{array}
\]
so we add the conjunct $\us_{vz}=\gs(\us_w,\us_w)$ to the
formula.  Similarly, from $w=g(u_{w1},u_{w2})$,
$\termShape{w}=\us_w$, $\termShape{u_{w1}}=\us_{w1}$,
$\termShape{u_{w2}}=\us_{w2}$ we conclude $\us_w =
g(u_{w1},u_{w2})$ and add this conjunct to the formula.
Adding these two conjuncts makes property P3 hold.  (Note
that, had we decided to consider the case where
$\termShape{v} \neq \termShape{z}$ we would have arrived at a
contradiction due to
$\termShape{\us_{vz}}=\termShape{\us_{zv}}$.)

We next apply rule~(\ref{eqn:anyDecompose}) to reduce all
cardinality constraints into cardinality constraints on
parameter nodes (nodes $u$ for which there there is
no conjunct of form $u = f(u_{i_1},\ldots,u_{i_k})$).
We replace $|u_{vz} \cap u^c_{zv}|_{\us_{vz}} = 0$ with
\begin{equation} \label{eqn:exaDecomposed}
    |u_v \cap u^c_z|_{\us_w} = 0 \land |u_z \cap u^c_v|_{\us_w} = 0
\end{equation}
Variable $v$ is a parameter variable, but $z$ is not, which
prevents application of~(\ref{eqn:anyDecompose}).  We
therefore introduce $u_{z1}$ and $u_{z2}$ such that $z =
g(u_{z1},u_{z2})$.  Because $\termShape{z}=\us_w$, we have
$\termShape{u_{z1}}=\us_{w1}$ and
$\termShape{u_{z2}}=\us_{w2}$ by homomorphism property.  We
can now continue applying rule~(\ref{eqn:anyDecompose}) to
(\ref{eqn:exaDecomposed}).  The result is:
\[
\begin{array}{l}
  |u_{v1} \cap u^c_{z1}|_{\us_{w1}} = 0 \ \land\ |u_{z1} \cap u^c_{v1}|_{\us_{w1}} = 0 \ \land \mnl
  |u_{v2} \cap u^c_{z2}|_{\us_{w2}} = 0 \ \land\ |u_{z2} \cap u^c_{v2}|_{\us_{w2}} = 0
\end{array}
\]
To make the formula conform to
Definition~\ref{def:structuralBase} we introduce internal
variables $u_v,u_z,u_w$ corresponding to free variables
$v,z,w$, respectively.  The resulting structural base
formula is
\begin{equation} \label{eqn:exaBase}
\begin{array}{l}
  \exists  u_{vz},u_{zv},u_v,u_z,u_w,u_{v1},u_{v2},u_{z1},u_{z2},u_{w1},u_{w2}, \mnl
\quad  \us_{vz},\us_w,\us_{w1},\us_{w2}. \mnl
    \qquad  \shapeBaseF_1 \land \termBaseF_1 \ \land \mnl
    \qquad  \homF_1 \land \cardinF_1
\end{array}
\end{equation}
where
\[
\begin{array}{l}
\shapeBaseF_1 =
\begin{array}[t]{l}
\us_{vz} = \gs(\us_w,\us_w) \land \us_w = \gs(\us_{w1},\us_{w2}) \ \land \mnl
\distinctF(\us_{vz},\us_w,\us_{w1},\us_{w2})
\end{array}
\end{array}
\]
\[
\begin{array}{l}
\termBaseF_1 =
\begin{array}[t]{l}
u_{vz} = g(u_v,u_z) \land u_{zv} = g(u_z,u_v) \ \land \mnl
u_v = g(u_{v1},u_{v2}) \land u_z = g(u_{z1},u_{z2}) \ \land \mnl
u_w = g(u_{w1},u_{w2}) \ \land \mnl
v = u_v \land z = u_z \land w = u_w
\end{array}
\end{array}
\]
\[
\begin{array}{l}
\homF_1 = \mnl
\begin{array}[t]{l}
\termShape{u_{vz}}=\us_{vz} \land
\termShape{u_{zv}}=\us_{vz} \ \land \mnl
\termShape{u_v}=\us_w \land 
\termShape{u_z}=\us_w \land 
\termShape{u_w}=\us_w \ \land \mnl
\termShape{u_{v1}}=\us_{w1} \land 
\termShape{u_{z1}}=\us_{w1} \land 
\termShape{u_{w1}}=\us_{w1} \ \land \mnl
\termShape{u_{v2}}=\us_{w2} \land 
\termShape{u_{z2}}=\us_{w2} \land 
\termShape{u_{w2}}=\us_{w2}
\end{array}
\end{array}
\]
\[
\begin{array}{l}
\cardinF_1 = 
\begin{array}[t]{l}
  |u_{v1} \cap u^c_{z1}|_{\us_{w1}} = 0 \ \land\ |u_{z1} \cap u^c_{v1}|_{\us_{w1}} = 0 \ \land \mnl
  |u_{v2} \cap u^c_{z2}|_{\us_{w2}} = 0 \ \land\ |u_{z2} \cap u^c_{v2}|_{\us_{w2}} = 0 \ \land \mnl
  |u_{w1} \cap u^c_{v1}|_{\us_{w1}} \geq 1
\end{array}
\end{array}
\]
Figure~\ref{fig:exaStrucBase} shows a graph representation
of the subformulas $\shapeBaseF_1$, $\termBaseF_1$, and
$\homF_1$ of the resulting structural base formula.

Recall that we are eliminating the quantification over $v$
from $\lnot \exists v.\phi_1$.  We can now existentially quantify over
$v$.  As in Proposition~\ref{prop:exBaseBase}, we simply
remove the conjunct $v=u_v$ from $\termBaseF$ and the
quantifier $\exists v$.

As in Figure~\ref{fig:termAlgScheme} of
Section~\ref{sec:qeTA} the structural base formula form
allows us to eliminate an existential quantifier, whereas
the quantifier-free form allows us to eliminate a negation.
We transform the structural base formula~(\ref{eqn:exaBase})
into a quantifier-free formula as follows.  

We first use rule~(\ref{eqn:replacement}) to eliminate
variable $u_{vz}$, replacing it with $g(v,z)$.  In the
resulting formula $g(v,z)$ occurs only in $\homF_1$ in the
form
\begin{equation} \label{eqn:afterReplace}
\termShape{g(u_v,u_z)} = \us_{vz}
\end{equation}
But~(\ref{eqn:afterReplace}) is a consequence of conjuncts
$\us_{vz}=\gs(\us_w,\us_w)$, $\termShape{u_v}=\us_w$ and
$\termShape{u_w}=\us_w$, so we omit~(\ref{eqn:afterReplace})
from the formula.  In analogous way we eliminate variable
$u_{zv}$ and the conjuncts that contain it.  
We also eliminate $u_v$,
analogously to $u_{vz}$ and $u_{zv}$.
%% !! FIX
%% Why am I eliminating shape variable here in the middle?
%% The general technique is to do it at the very end, after
% eliminating all parameter term variables.
In the resulting formula $\us_{vz}$ occurs only in
$\distinctF$ subformula of $\shapeBaseF$.  Conjuncts
$\us_{vz} \neq \us_w$, $\us_{vz} \neq \us_{w1}$, and $\us_{vz} \neq
\us_{w2}$ follow from the remaining conjuncts in
$\shapeBaseF$ by acyclicity.  Hence we may replace
$\distinctF(\us_{vz},\us_w,\us_{w1},\us_{w2})$ by
$\distinctF(\us_w,\us_{w1},\us_{w2})$.  Now $\us_{vz}$ does
not occur in the matrix of the formula, so we may eliminate
$\exists \us_{vz}$ altogether.

The resulting formula
is:
\begin{equation}
\begin{array}{l}
\phi_5 \ \equiv \   \exists  u_z,u_w,u_{v1},u_{v2},u_{z1},u_{z2},u_{w1},u_{w2},
              \us_w,\us_{w1},\us_{w2}. \mnl
\begin{array}{l}
\us_w = \gs(\us_{w1},\us_{w2}) \ \land \ \distinctF(\us_w,\us_{w1},\us_{w2}) \ \land \mnl
u_z = g(u_{z1},u_{z2}) \ \land \ u_w = g(u_{w1},u_{w2}) \ \land \mnl
z = u_z \land w = u_w \ \land \mnl
\termShape{u_z}=\us_w \land 
\termShape{u_w}=\us_w \ \land \mnl
\termShape{u_{v1}}=\us_{w1} \land 
\termShape{u_{z1}}=\us_{w1} \land 
\termShape{u_{w1}}=\us_{w1} \ \land \mnl
\termShape{u_{v2}}=\us_{w2} \land 
\termShape{u_{z2}}=\us_{w2} \land 
\termShape{u_{w2}}=\us_{w2} \ \land \mnl
  |u_{v1} \cap u^c_{z1}|_{\us_{w1}} = 0 \ \land\ |u_{z1} \cap u^c_{v1}|_{\us_{w1}} = 0 \ \land \mnl
  |u_{v2} \cap u^c_{z2}|_{\us_{w2}} = 0 \ \land\ |u_{z2} \cap u^c_{v2}|_{\us_{w2}} = 0 \ \land \mnl
  |u_{w1} \cap u^c_{v1}|_{\us_{w1}} \geq 1
\end{array}
\end{array}
\end{equation}
We next eliminate $u_{v1}$.  It suffices to eliminate it
from conjuncts where it occurs, so we consider formula
$\phi_{5,1}$:
\begin{equation}
\begin{array}{l}
\phi_{5,1} \ \equiv \   \exists  u_{v1}. \mnl
\begin{array}{l}
  \termShape{u_{v1}}=\us_{w1} \land 
  \termShape{u_{z1}}=\us_{w1} \land 
  \termShape{u_{w1}}=\us_{w1} \ \land \mnl
  |u_{v1} \cap u^c_{z1}|_{\us_{w1}} = 0 \ \land\ |u_{z1} \cap u^c_{v1}|_{\us_{w1}} = 0 \ \land \mnl
  |u_{w1} \cap u^c_{v1}|_{\us_{w1}} \geq 1
\end{array}
\end{array}
\end{equation}
Note that all variables from $\phi_{5,1}$ belong to
$\BSubSet{s}$ where $s$ is the value of shape variable
$\us_{w1}$ (see (\ref{eqn:bSubSet})).  This means that we
can apply quantifier elimination for boolean algebra
(Section~\ref{sec:qeBoolalg}) to eliminate $u_{v1}$.  The
result is
\begin{equation}
\phi_{5,2} \ \equiv \ 
\begin{array}[t]{l}
  \termShape{u_{z1}}=\us_{w1} \land \termShape{u_{w1}}=\us_{w1} \ \land \mnl
  |u_{w1} \cap u^c_{z1}|_{\us_{w1}} \geq 1
\end{array}
\end{equation}
Similarly, to eliminate $u_{v2}$ we consider formula $\phi_{5,3}$:
\begin{equation}
\begin{array}{l}
\phi_{5,3} \ \equiv \   \exists  u_{v2}. \mnl
\begin{array}{l}
  \termShape{u_{v2}}=\us_{w2} \land 
  \termShape{u_{z2}}=\us_{w2} \land 
  \termShape{u_{w2}}=\us_{w2} \ \land \mnl
  |u_{v2} \cap u^c_{z2}|_{\us_{w2}} = 0 \ \land\ |u_{z2} \cap u^c_{v2}|_{\us_{w2}} = 0
\end{array}
\end{array}
\end{equation}
The result of boolean algebra quantifier elimination on
$\phi_{5,3}$ is $\boolTrue$ (indeed, one may let
$u_{v2}=u_{z2}$).  The resulting base formula with $u_{v1}$
and $u_{v2}$ eliminated is $\phi_6:$
\begin{equation} \label{eqn:intermBase}
\begin{array}{l}
\phi_6 \ \equiv \   \exists  u_z,u_w,u_{z1},u_{z2},u_{w1},u_{w2},
              \us_w,\us_{w1},\us_{w2}. \mnl
\begin{array}{l}
\us_w = \gs(\us_{w1},\us_{w2}) \ \land \ \distinctF(\us_w,\us_{w1},\us_{w2}) \ \land \mnl
u_z = g(u_{z1},u_{z2}) \ \land \ u_w = g(u_{w1},u_{w2}) \ \land \mnl
z = u_z \land w = u_w \ \land \mnl
\termShape{u_z}=\us_w \land 
\termShape{u_w}=\us_w \ \land \mnl
\termShape{u_{z1}}=\us_{w1} \land 
\termShape{u_{w1}}=\us_{w1} \ \land \mnl
\termShape{u_{z2}}=\us_{w2} \land 
\termShape{u_{w2}}=\us_{w2} \ \land \mnl
  |u_{w1} \cap u^c_{z1}|_{\us_{w1}} \geq 1
\end{array}
\end{array}
\end{equation}
Observe that the equalities in $\phi_6$ are
sufficient to express all variables bound in $\phi_6$ in terms
of free variables (all internal variables are
``covered''):
\begin{equation}
\begin{array}{r@{\>}c@{\>}l@{\ \ }r@{\>}c@{\>}l}
u_z &=& z &
u_w &=& w \mnl
u_{z1} &=& g_1(z) &
u_{z2} &=& g_2(z) \mnl
u_{w1} &=& g_1(w) &
u_{w2} &=& g_2(w) \mnl
\us_w &=& \termShape{w} \mnl
\us_{w1} &=& \gs_1(\termShape{w}) &
\us_{w2} &=& \gs_2(\termShape{w})
\end{array}
\end{equation}
Structural base formula $\phi_6$
is therefore equivalent to the quantifier-free formula $\phi_{7,1}$:
\begin{equation}
\begin{array}{l}
\phi_{7,1} \ \equiv \
\begin{array}[t]{l}
\Isgs(\termShape{w}) \land \Is{g}(w) \land \Is{g}(z) \ \land \mnl
\distinctF(\gs_1(\termShape{w}),\gs_2(\termShape{w})) \mnl
\termShape{z} = \termShape{w} \ \land \
|g_1(w) \cap g_1(z)^c|_{\gs_1(\termShape{w})} \geq 1
\end{array}
\end{array}
\end{equation}
When transforming formula $\phi_4$ we chose the case $\us_{w1} \neq \us_{w1}$.
If we choose the case $\us_{w1} = \us_{w2}$, we obtain
quantifier-free formula $\phi_{7,2}$:
\begin{equation}
\begin{array}{l}
\phi_{7,2} \ \equiv \
\begin{array}[t]{l}
\Isgs(\termShape{w}) \land \Is{g}(w) \land \Is{g}(z) \ \land \mnl
\termShape{z} = \termShape{w} \ \land \ 
\gs_1(\termShape{w}) = \gs_2(\termShape{w}) \ \land  \mnl
|g_1(w) \cap g_1(z)^c|_{\gs_1(\termShape{w})} \geq 1
\end{array}
\end{array}
\end{equation}
Our quantifier elimination would also consider the case
$\termShape{g_2(w)} \neq \termShape{g_2(z)}$.  The procedure
finds the case contradictory in a larger context, when
eliminating $\exists z$, because
$\termShape{z}=\termShape{x}=\termShape{w}$ follows from $z
\leq x$ and $w \leq x$.  Ignoring this case, we observe that $\phi_{7,1} \lor \phi_{7,2}$ is equivalent to the quantifier-free formula $\phi_8$, where
\begin{equation}
\begin{array}{l}
\phi_8 \ \equiv \
\begin{array}[t]{l}
\Isgs(\termShape{w}) \land \Is{g}(w) \land \Is{g}(z) \ \land \mnl
\termShape{z} = \termShape{w} \ \land \
|g_1(w) \cap g_1(z)^c|_{\gs_1(\termShape{w})} \geq 1
\end{array}
\end{array}
\end{equation}
Let us therefore assume that the result of quantifier elimination
in~(\ref{eqn:exaInnerMost}) is $\lnot \phi_8$.

We proceed to eliminate the next quantifier, $\forall w$, from
\begin{equation} \label{eqn:nextQuant}
\forall w.\ w \leq x \land w \leq y \implies \lnot \phi_8
\end{equation}
(\ref{eqn:nextQuant}) is equivalent to
\[
   \lnot \exists w.\ w \leq x \land w \leq y \land \phi_8
\]
After eliminating $\leq$ we obtain
\begin{equation}
\begin{array}{l}
\lnot \exists  w.\
\begin{array}[t]{l}
|w \cap x^c|_{\termShape{w}}=0 \ \land \ \termShape{x}=\termShape{w} \ \land \mnl
|w \cap y^c|_{\termShape{w}}=0 \ \land \ \termShape{y}=\termShape{w} \ \land \mnl
\Isgs(\termShape{w}) \land \Is{g}(w) \land \Is{g}(z) \ \land \mnl
\termShape{z} = \termShape{w} \ \land \
|g_1(w) \cap g_1(z)^c|_{\gs_1(\termShape{w})} \geq 1
\end{array}
\end{array}
\end{equation}
We now proceed similarly as in eliminating variable $v$.
The result is $\lnot \phi_9$ where
\begin{equation}
\begin{array}{l}
\phi_9 \ \equiv \ 
\begin{array}[t]{l}
\termShape{x} = \termShape{z} \ \land \
\termShape{y} = \termShape{z} \ \land \mnl
\Is{g}(x) \land \Is{g}(y) \land \Is{g}(z) \land \Isgs(\termShape{z}) \ \land \mnl
|g_1(x) \cap g_1(y) \cap g_1(z)^c|_{\gs_1(\termShape{z})} \geq 1
\end{array}
\end{array}
\end{equation}
The remaining quantifiers that bind $z$, $y$, and $x$ are
eliminated similarly.

To eliminate the quantifier $\exists z$, we need to transform $\lnot
\phi_9$ into disjunction of base formulas.  This transformation
requires negation of $\phi_9$ and creates several disjuncts.
We consider only the two cases, $\phi_{10}$ and $\phi_{11}$, that
are not contradictory in the enclosing context of conjuncts
$z \leq x$ and $z \leq y$:
\begin{equation}
\begin{array}{l}
\phi_{10} \ \equiv \ 
\begin{array}[t]{l}
\termShape{x} = \termShape{z} \ \land \
\termShape{y} = \termShape{z} \ \land \
\lnot \Isgs(\termShape{z})
\end{array}
\end{array}
\end{equation}
\begin{equation}
\begin{array}{l}
\phi_{11} \ \equiv \ 
\begin{array}[t]{l}
\termShape{x} = \termShape{z} \ \land \
\termShape{y} = \termShape{z} \ \land \mnl
\Is{g}(x) \land \Is{g}(y) \land \Is{g}(z) \land \Isgs(\termShape{z}) \ \land \mnl
|g_1(x) \cap g_1(y) \cap g_1(z)^c|_{\gs_1(\termShape{z})} = 0
\end{array}
\end{array}
\end{equation}
$\phi_{10}$ is equivalent to
\begin{equation}
\termShape{x} = \cs \land \termShape{y} = \cs \land \termShape{z} = \cs
\end{equation}
The result of eliminating $\exists z$ from
\[
    \exists z.\ |z \cap x^c|_{\termShape{z}} = 0 \ \land\ |z \cap y^c|_{\termShape{z}} = 0 \ \land \ \phi_{10}
\]
is therefore
\begin{equation}
\begin{array}{l}
\phi_{10,2} \ \equiv \ 
\begin{array}[t]{l}
\termShape{x} = \termShape{y} \ \land \ \lnot \Isgs(\termShape{x})
\end{array}
\end{array}
\end{equation}
The result of eliminating $\exists z$
from
\[
    \exists z.\ |z \cap x^c|_{\termShape{z}} = 0 \ \land\ |z \cap y^c|_{\termShape{z}} = 0 \ \land \ \phi_{11}
\]
is
\[
   \phi_{11,2} \ \equiv \  \termShape{x}=\termShape{y} \ \land \ \Isgs(\termShape{x})
\]
$\phi_{10,2} \lor \phi_{11,2}$ is equivalent to $\termShape{x} = \termShape{y}$.
Converting
\[
    |x \cap y^c|_{\termShape{x}} = 0 \land \termShape{x}=\termShape{y} \implies
     \termShape{x}=\termShape{y}
\]
to structural base formula yields $\boolTrue$.  We conclude
that~(\ref{eqn:exampleSentence}) is a true sentence in
the structure $\twoSorted$, which 
completes our quantifier elimination procedure example.
\end{example}
Formulas in the Example~\ref{exa:twoConstMain} do not
contain disequalities between terms variables, only
disequalities between shape variables.  If a conjunction
contains disequalities between term variables, we eliminate
the disequalities using rule~(\ref{eqn:termDiseqElim}) in the
process of converting formula to disjunction of structural
base formulas.  The following Example~\ref{exa:disEq}
illustrates this process.
\begin{example} \label{exa:disEq}
Consider the formula 
\[
   \phi'_6 \ \ \equiv \ \ \phi_6 \ \land\ u_z \neq u_w
\]
Where $\phi_6$ is given by~(\ref{eqn:intermBase}).
By~(\ref{eqn:termDiseqElim}), literal
$u_z \neq u_w$ is equivalent to $\psi_1 \lor \psi_2$ where
\begin{equation}
\psi_1 \ \equiv \ \termShape{u_z} \neq \termShape{u_w}
\end{equation}
and
\begin{equation}
\begin{array}{l}
\psi_2 \ \equiv 
\begin{array}[t]{l}
 \termShape{u_z} = \termShape{u_w} \ \land \mnl
  |(u_z\cap u_w^c) \cup (u_z^c\cap u_w)|_{\termShape{u_z}} \geq 1
\end{array}
\end{array}
\end{equation}
In this case, formula $\phi_6 \land \psi_1$ is contradictory.  Formula
$\phi_6 \land \psi_2$ is equivalent to $\phi''_6$ where
\begin{equation}
\begin{array}{l}
\phi''_6 \ \equiv \   \exists  u_z,u_w,u_{z1},u_{z2},u_{w1},u_{w2},
              \us_w,\us_{w1},\us_{w2}. \mnl
\begin{array}{l}
\us_w = \gs(\us_{w1},\us_{w2}) \ \land \ \distinctF(\us_w,\us_{w1},\us_{w2}) \ \land \mnl
u_z = g(u_{z1},u_{z2}) \ \land \ u_w = g(u_{w1},u_{w2}) \ \land \mnl
z = u_z \land w = u_w \ \land \mnl
\termShape{u_z}=\us_w \land 
\termShape{u_w}=\us_w \ \land \mnl
\termShape{u_{z1}}=\us_{w1} \land 
\termShape{u_{w1}}=\us_{w1} \ \land \mnl
\termShape{u_{z2}}=\us_{w2} \land 
\termShape{u_{w2}}=\us_{w2} \ \land \mnl
  |u_{w1} \cap u^c_{z1}|_{\us_{w1}} \geq 1 \ \land \mnl
  |(u_z\cap u_w^c) \cup (u_z^c\cap u_w)|_{\us_w} \geq 1
\end{array}
\end{array}
\end{equation}
As in
Example~\ref{exa:twoConstMain}, 
we now apply rule~(\ref{eqn:anyDecompose}) to
\[
  |(u_z\cap u_w^c) \cup (u_z^c\cap u_w)|_{\us_w} \geq 1
\]
and transform $\phi''_6$ into a disjunction of base formulas.
\end{example}
We proceed to sketch the general case of quantifier
elimination.  The following
Proposition~\ref{prop:StructexBaseBase} is analogous to
Proposition~\ref{prop:exBaseBase}; the proof is again
straightforward.
\begin{proposition}[Quantification of Structural Base]
\label{prop:StructexBaseBase}
If $\beta$ is a structural base formula and $x$ a free term
variable in $\beta$, then there exists a base structural formula
$\beta_1$ equivalent to $\exists x.\beta$.
\end{proposition}
The following Proposition~\ref{prop:StructqFreeToBaseDisj} corresponds
to Proposition~\ref{prop:qFreeToBaseDisj}.
\begin{proposition}[Quantifier-Free to Structural Base]
\label{prop:StructqFreeToBaseDisj}
Every well-defined quantifier-free formula $\phi$ in the language
of Figure~\ref{fig:twoSortedSig} can be written as
$\boolTrue$, $\boolFalse$, or a disjunction of structural
base formulas.
\end{proposition}
\begin{proofsketch}
Let $\phi$ be a well-defined quantifier-free formula in the
language of Figure~\ref{fig:twoSortedSig}.

We first use rule~(\ref{eqn:termRelElim}) to eliminate
occurrences of $\leq$ in the formula replacing them with
cardinality constraints.

We then convert formula into disjunction $\phi_1 \lor \cdots \lor \phi_n$ of
well-formed conjunctions of literals.  We next describe how
to transform each conjunction $\phi_i$ into a disjunction of
base formulas.

\begin{figure}
\[
\begin{array}{l|l}
\mbox{unnested form} & \mbox{cardinality constraint} \\ \hline
x = x_1 \cap_s x_2 &
|x + (x_1 \cap x_2)|_s = 0 \mnl
x = x_1 \cup_s x_2 &
|x + (x_1 \cup x_2)|_s = 0 \mnl
x = {x_1^c}_s &
|x + x_1^c|_s = 0
\end{array}
\]
\caption{Elimination of Boolean Algebra Unnested Formulas
\label{fig:baUnnestElim}.  Expression $x+y$ is a shorthand
for $(x \cap y^c)\cup(y \cap x^c)$.}
\end{figure}

Let $\phi_i$ be a conjunction of literals.  Using the technique
of Proposition~\ref{prop:unnestedDNFpartial}, we convert the
formula to unnested form, adding existential quantifiers.
We then eliminate unnested conjuncts that contain boolean
algebra operations, according to
Figure~\ref{fig:baUnnestElim}.  The only atomic formulas in
the resulting existentially quantified conjunction are of
form $x = a$, $x=b$, $x=g(x_1,x_2)$, $\Is{g}(x)$, $x_1 =
g_1(x)$, $x_2 = g_2(x)$, $x_1=x_2$, $\xs = \cs$, $\xs =
\gs(\xs_1,\xs_2)$, $\Isgs(\xs)$, $\xs_1 = \gs_1(\xs)$,
$\xs_2 = \gs_2(\xs)$, $\xs_1 = \xs_2$, $\xs =
\termShape{x}$, as well as $|t_1|_{\xs} \geq k$ and
$|t_2|_{\xs} = k$ for some $\xs$-terms $t_1$ and $t_2$.  The
only negated atomic formulas are of form $x_1 \neq x_2$, $\xs_1
\neq \xs_2$, $\lnot \Is{g}(x)$ and $\lnot \Isgs(\xs)$.  As in the proof
of Proposition~\ref{prop:qFreeToBaseDisj}, we use
(\ref{eqn:conseqIsfPartit}) to eliminate $\lnot \Is{g}(x)$ and
$\lnot \Isgs(\xs)$.  This process leaves formulas of form $x_1 \neq
x_2$ and $\xs_1 \neq \xs_2$ as the only negated atomic
formulas.

In the sequel, whenever we perform case analysis and generate
a disjunction of conjunctions, existential quantifiers
propagate to the conjunctions, so we keep working with a
existentially quantified conjunction.  The existentially
quantified variables will become internal variables of a
structural base formula.

We next convert conjuncts that contain only term variables
to a base formula, and convert shape part to base formula,
as in the proof of Proposition~\ref{prop:qFreeToBaseDisj}.
We simultaneously make sure every term variable has an
associated shape variable, introducing new shape variables
if needed.  (This process is interleaved with conversion to
base formula, to ensure that there is always a conjunct
stating that newly introduced shape variables are distinct.)
We also ensure homomorphism requirement by replacing internal
variables when we entail their equality.  Another condition
we ensure is that parameter term variables map to parameter
shape variables, and non-parameter term variables to
non-parameter shape variables; we do this by performing
expansion of term and shape variables.  We perform expansion
of shape variables as in Section~\ref{sec:qeBoolalg}.
Expansion of term variables is even simpler because there is
no need to do case analysis on equality of term variable
with other variables.

The resulting existentially quantified conjunction might
contain disequalities $u \neq u'$ between term variables.  We
eliminate these disequalities as explained in
Example~\ref{exa:disEq}, by converting each disequality into
a cardinality constraint using~(\ref{eqn:termDiseqElim}).
In general, we need to consider the case when $\termShape{u}
\neq \termShape{u'}$ and generate another disjunct.  

%%% There is no need for this special case at all.
%% In the special case where one of the variables is labelled
%% by a constant, in the case $\termShape{u}=\termShape{u'}$ we
%% cannot apply~(\ref{eqn:termDiseqElim}).  However, in that
%% case the constraint reduces to either $u \neq a$ or $u \neq b$.
%% We then represent $u \neq a$ with $|u|_{\us}\geq 1$ and represent
%% $u \neq b$ with $|u|_{\us} = 0$, where $\us =
%% \termShape{u}=\cs$.

Elimination of disequalities might violate previously
established homomorphism invariants, so we may need to
reestablish these invariants by repeating the previously
described steps.  The overall process terminates because we
never introduce new inequalities between term variables.

As a final step, we convert all cardinality constraints into
constraints on parameter term variables,
using~(\ref{eqn:anyDecompose}).  
% This works only for finite base theories:
In the case when the shape of cardinality constraint is
$\cs$, we cannot apply~(\ref{eqn:anyDecompose}).  However,
in that case the homomorphism condition ensures that each of
the participating variables is equal to $a$ or equal to $b$.
This means that we can simply evaluate the cardinality
constraint in the boolean algebra $\{a,b\}$.  If the result
is $\boolTrue$ we simply drop the constraint, otherwise the
entire base formula becomes $\boolFalse$.

This completes our sketch of transforming a quantifier-free
formula into disjunction of structural base formulas.
\end{proofsketch}

We introduce the notion of covered variables in structural
base formula by generalizing
Definition~\ref{def:coveredVarDef}.
\begin{definition} \label{def:structCovering}
The set $\covering$ of variable coverings of a structural
base formula $\beta$ is the least set $S$ of pairs $\tu{u,t}$
where $u$ is an internal (shape or term) variable and $t$ is a term over
the free variables of $\beta$, such such that:
\begin{enumerate}

\item if $x=u$ occurs in $\termBaseF$ then $\tu{u,x} \in S$;

\item if $\xs=\us$ occurs in $\shapeBaseF$ then
      $\tu{\us,\xs} \in S$;

\item if $\tu{u,t} \in S$ and $u = f(u_1,\ldots,u_k)$ occurs in
      $\termBaseF$ for some $f \in \FreeSig$ then
      $\{\tu{u_1,f_1(t)},\ldots,\tu{u_k,f_k(t)}\} \subseteq S$;

\item if $\tu{\us,\ts} \in S$ and 
      $\us = \fs(\us_1,\ldots,\us_k)$ occurs in
      $\shapeBaseF$ then
      $\{\tu{\us_1,\fs_1(\ts)},\ldots,\tu{\us_k,\fs_k(\ts)}\} \subseteq S$;

\item if $\tu{u,t} \in S$ and $\termShape{u}=\us$ occurs in $\homF$
      then $\tu{\us,\termShape{t}} \in S$.

      % !! unfortunate confusion of object and meta level possible
      
\end{enumerate}
\end{definition}

\begin{definition}
An internal term variable $u$ is covered iff there exists a
term $t$ such that $\tu{u,t} \in S$.  An internal shape
variable $\us$ is covered iff there exists a term $\ts$ such
that $\tu{\us,\ts} \in S$.
\end{definition}

\begin{lemma} \label{lemma:coveringWorks}
Let $\beta$ be a structural base formula with matrix $\beta_0$ and
let $\covering$ be the covering of $\beta$.
\begin{enumerate}
\item If $\tu{u,t} \in S$ then $\models \beta_0 \implies u=t$.
\item If $\tu{\us,\ts} \in S$ then $\models \beta_0 \implies \us=\ts$.
\end{enumerate}
\end{lemma}
\begin{proof}
By induction, using Definition~\ref{def:structCovering}.
\end{proof}

\begin{corollary} \label{cor:coveredToQfree}
Let $\beta$ be a structural base formula such that every
internal variable is covered.  Then $\beta$ is equivalent to
a well-defined quantifier-free formula.
\end{corollary}
\begin{proof}
By Lemma~\ref{lemma:coveringWorks} using~(\ref{eqn:replacement}).
\end{proof}

\begin{lemma} \label{lemma:strucElimTermNonparam}
Let $u$ be an uncovered non-parameter term variable in a
structural base formula $\beta$ such that $u$ is a source
i.e.\ no conjunct of form
\[
   u' = f(u_1,\ldots,u,\ldots,u_k)
\]
occurs in $\termBaseF$.  Let $\beta'$ be the result of dropping
$u$ from $\beta$.  Then $\beta$ is equivalent to $\beta'$.
\end{lemma}
\begin{proof}
Let $u$ occur in $\termBaseF$ in form
\[
    u = f(u_1,\ldots,u_k)
\]
The only other occurrence of $u$ in $\beta$ is in $\homF$ and has
the form $\termShape{u}=\us$.  Because non-parameter term
variables are mapped to non-parameter shape variables, $\shapeBaseF$
contains formula
\begin{equation} \label{eqn:usShape}
   \us = \shapified{f}(\us_1,\ldots,\us_k)
\end{equation}
where $\us_1,\ldots,\us_k$ are such that, by homomorphism
property, $\termShape{u_i}=\us_i$ occurs in $\homF$.  This
means that the conjunct $\termShape{u}=\us$ is a consequence
of the remaining conjuncts, so it may be omitted.  After
that, applying~(\ref{eqn:replacement}) yields a structural
base formula $\beta'$ not containing $u$, where $\beta'$ is equivalent
to $\beta$.
\end{proof}

\begin{corollary} \label{cor:strucElimTermNonparam}
Every base formula is equivalent to a base formula without
uncovered non-parameter term variables.
\end{corollary}
\begin{proof}
If a structural base formula has an uncovered non-parameter
term variable, then it has an uncovered non-parameter
term variable that is a source.  By repeated application of
Lemma~(\ref{lemma:strucElimTermNonparam}) we eliminate all
uncovered non-parameter term variables.
\end{proof}

The next example illustrates how we deal with cardinality
constraints $|1_s|_s \geq k$ and $|1_s| = k$, which contain no
term variables.  These constraints restrict the size of
shape $s$.  Luckily, we can be translate them into shape
base formula constraints.
\begin{example}({\bf Shape Term Size Constraints}) \\
Let $x < y$ denote conjunction $x \leq y \land x \neq y$.  Let us
eliminate quantifiers from formula $\exists x.\phi(x)$ where
\begin{equation}
\phi(x) \ \equiv 
\begin{array}[t]{l}
\lnot (\exists y. \exists z.\ x < y \land y < z) \ \land \mnl
\lnot (\exists u.\ u < x)
\end{array}
\end{equation}
Eliminating variables $y,z$ from the first conjunct and
variable $u$ from the second conjunct yields
\[
    \lnot |x^c|_{\termShape{x}} \geq 2 \land \lnot |x|_{\termShape{x}} \geq 1
\]
which is equivalent to
\[
       (|x^c|_{\termShape{x}} = 0 \lor |x^c|_{\termShape{x}} = 1) \land |x|_{\termShape{x}} = 0
\]
and further to disjunction
\[
\begin{array}{l}
(|x^c|_{\termShape{x}} = 0 \land |x|_{\termShape{x}}=0) \ \lor \
(|x^c|_{\termShape{x}} = 1 \land |x|_{\termShape{x}}=0)
\end{array}
\]
The first disjunct can be shown contradictory.  Let us
transform the second disjunct into a structural base formula.
After introducing $u = x$ and $\us = \termShape{u}$, we
obtain
\[
   \exists u, \us.\ x = u \ \land \ \termShape{u}=\us \ \land \
   |u|_{\us}=0 \ \land \ |u^c|_{\us} = 1
\]
Then $\exists x.\phi(x)$ is equivalent to
\[
   \exists u, \us.\ \termShape{u}=\us \ \land \
   |u|_{\us}=0 \ \land \ |u^c|_{\us} = 1
\]
Eliminating parameter term variable $u$ yields
\[
    \exists \us.\ |1|_{\us} = 1
\]
Constraint $|1|_{\us}=1$ means that the largest set in the
boolean algebra $B(s)$ where $s$ is the value of $\us$ has
size one.  There exists exactly one boolean algebra of size
one in the structure $\twoSorted$, namely $\{ a, b \}$.
Therefore, $|1|_{\us}=1$ is equivalent to $\us=\cs$.
We may now eliminate $\us$ by letting $\us=\cs$.
We conclude that the sentence $\exists x. \phi(x)$ is true.

Notice that we have also established that formula $\phi(x)$
is equivalent to $\termShape{x}=\cs$, as a consequence of
\[
    |1_{\termShape{x}}|_{\termShape{x}} = 1
\]
\end{example}

The following Proposition~\ref{prop:StructbaseToSel}
corresponds to % Proposition~\ref{prop:baseToConsSel} or
Proposition~\ref{prop:baseToSel}.
\begin{proposition}[Struct.\ Base to Quantifier-Free]
\label{prop:StructbaseToSel}
Every structural base formula $\beta$ is equivalent to a
quantifier-free formula $\phi$ in the language of
Figure~\ref{fig:twoSortedSig}.
\end{proposition}
\begin{proofsketch}
By Corollary~\ref{cor:strucElimTermNonparam} we may assume
that $\beta$ has no uncovered non-parameter term variables.  By
Corollary~\ref{cor:coveredToQfree} we are done if there are
no uncovered variables, so it suffices to eliminate
uncovered parameter term variables and uncovered shape
variables.

Let $u$ be an uncovered parameter term variable.  Then
$u$ does not occur in $\termBaseF$.  Indeed, suppose for
the sake of contradiction that $u$
occurs in $\termBaseF$ in some formula
\[
    u' = f(u_1,\ldots,u,\ldots,u_k)
\]
Then $u'$ is an uncovered non-parameter variable in $\beta$,
which is a contradiction because we have assumed $\beta$ has no
uncovered non-parameter variables.  Therefore, $u$ does no
occur in $\termBaseF$, it occurs only in $\homF$ and
$\cardinF$.  Let $\termShape{u}=\us$ occur in $\homF$.  Let
$\psi_1,\ldots,\psi_p$ be all conjuncts of $\cardinF$ that contain $u$.
Each $\psi_i$ is of form $|t_i|_{\us} \geq k_i$ or $|t_i|_{\us} =
k_i$ for some $\us$-term $t_i$.  Let $u_{j_1},\ldots,u_{j_q}$ be
all term variables appearing in $t_i$ terms other than $u$.
Conjunct $\termShape{u_{j_r}}=\us$ occurs in $\homF$ for
each $r$ where $1 \leq r \leq q$.  The base formula can therefore
be written in form
\[
    \beta_1 \ \equiv \ \exists x_1,\ldots,x_e,\xs_1,\ldots,\xs_f.\ \phi \land \phi_1
\]
where
\begin{equation}
\begin{array}{l}
\phi_1 \ \equiv \ \exists u.\ 
\begin{array}[t]{l}
\termShape{u}=\us \ \land \mnl
\termShape{u_{j_1}}=\us \ \land \ldots \land \ \termShape{u_{j_q}}=\us \ \land \mnl
\psi_1 \ \land \ldots \land \ \psi_p
\end{array}
\end{array}
\end{equation}
All term variables in $\psi_1,\ldots,\psi_k$ range over terms of shape
$\us$. Therefore, $\phi_1$ defines a relation in the boolean
algebra $\BSubSet{\interpret{\us}}$.  This allows us to apply
construction in Section~\ref{sec:qeBoolalg}.  We eliminate
$u$ from $\psi_1 \ \land \ldots \land \ \psi_p$ and obtain a propositional
combination $\psi_0$ of cardinality constraints with
$\us$-terms.  $\phi_0$ does not contain variable $u$.  We may
assume that $\psi_0$ is in disjunctive normal form
\[
    \psi_0 \ \equiv \ \alpha_1 \lor \ \ldots \ \lor \alpha_w
\]
Let
\[
   \phi_{1,i} \ \equiv \ 
    \termShape{u_{j_1}}=\us \ \land \ldots \land \ \termShape{u_{j_q}}=\us \ \land \ \alpha_i
\]
for $1 \leq i \leq w$.  Base formula $\beta_1$ is equivalent to
disjunction of base formulas $\beta_{1,i}$ where
\[
   \beta_{1,i} \ \equiv \ \exists x_1,\ldots,x_e,\xs_1,\ldots,\xs_f.\ \phi \land \phi_{1,i}
\]
We have thus eliminated an uncovered parameter term variable
$u$ from $\beta_1$.  By repeating this process we eliminate all
uncovered parameter term variables from a base formula.  The
resulting formula contains no uncovered term variables.

It remains to eliminate uncovered shape variables.  This
process is similar to term algebra quantifier elimination in
Section~\ref{sec:qeTA}.  An essential part of construction
in Section~\ref{sec:qeTA} is Lemma~\ref{lemma:baseSat},
which relies on the fact that uncovered parameter variables
may take on infinitely many values.  We therefore ensure
that uncovered parameter shape variables are not constrained
by term variables through conjuncts outside $\shapeBaseF$.

Suppose that $\us$ is an uncovered parameter shape variable
in a base formula $\beta$.  $\us$ does not occur in
$\termBaseF$.  $\us$ does not occur in $\homF$
either, because all term variables are covered, and a
conjunct $\termShape{u}=\us$ would imply that $\us$ is
covered.  The only possible occurrence of $\us$ is in
cardinality constraint $\psi$ of subformula $\cardinF$, where
$\psi$ is of form $|t|_{\us} = k$ or of form $|t|_{\us} \geq k$.
Suppose there is some term variable $u$ occurring in $t$.
Then $\termShape{u}=\us$ so $\us$ is covered, which is
a contradiction.  Therefore, $t$ has no variables.  $t$ can
thus be simplified to either $0_{\us}$ or $1_{\us}$.  In
general, a constraint of form $|1|_{\us}=k$ or $|1|_{\us}\geq
k$ is a domain cardinality constraint for boolean algebra
$\BSubSet{\interpret{\us}}$ (see Remark~\ref{rem:domainSize}
as well as (\ref{eqn:bSubSet})).  A constraint containing
$|0_{\us}|$ is equivalent to $\boolTrue$ or $\boolFalse$.
A constraint $|1|_{\us}=0$ is equivalent to $\boolFalse$.  A
constraint $|1|_{\us}=k$ for $k \geq 1$ is equivalent to
\[
   \us = \ts_1 \lor \ \cdots \ \lor \us = \ts_p
\]
where $\ts_1,\ldots,\ts_p$ is the list of all ground terms in
signature $\FreeSigConst$ that have exactly $k$ occurrences
of constant $\cs$.  We therefore generate a disjunction of
base formulas $\beta_1,\ldots,\beta_p$ where $\beta_i$ results from $\beta$ by
replacing $|1|_{\us}=k$ with $\us=\ts_i$.  We
convert each $\beta_i$ to a disjunction of base formulas by
labelling subterms of $t_i$ by internal shape variables and
doing case analysis on the equality between new internal
shape variables to ensure the invariants of a base formula.
The result is a disjunction of base formulas where variable
$\us$ occurs only in $\shapeBaseF$ subformula.

Similarly, $|1|_{\us} \geq k+1$ is equivalent to $\lnot (|1|_{\us} = k)$ and
thus to
\begin{equation} \label{eqn:geqCardinRepl}
   \us \neq \ts_1 \land \ \cdots \ \land \us \neq \ts_q
\end{equation}
where $\ts_1,\ldots,\ts_q$ is the list of all ground terms in
signature $\FreeSigConst$ that have at most $k$ occurrences
of constant $\cs$.  We replace $|1|_{\us}\geq k+1$
by~(\ref{eqn:geqCardinRepl}) and again convert the result to
a disjunction of base formulas where $\us$ occurs only in
$\shapeBaseF$ subformula.

Each of the resulting base formulas $\beta^1$ are such that
every uncovered variable in $\beta^1$ is a shape variable that
occurs only in $\shapeBaseF$.  Let
\[
\begin{array}{l}
\beta^1 \ \equiv \ 
\begin{array}[t]{l}
  \exists  u_1,\ldots,u_n, \us_1,\ldots,\us_{\ps},\us_{\ps+1},\ldots,\us_{\ps+\qs}. \mnl
    \qquad  \shapeBaseF(\us_1,\ldots,\us_{\ns},\xs_1,\ldots,\xs_{\ms}) \land {} \mnl
    \qquad  \termBaseF(u_1,\ldots,u_n,x_1,\ldots,x_m) \land {} \mnl
    \qquad  \homF(u_1,\ldots,u_n,\us_1,\ldots,\us_n) \land {} \mnl
    \qquad  \cardinF(u_{p+1},\ldots,u_{p+q},\us_{\ps+1},\ldots,\us_{\ps+\qs})
\end{array}
\end{array}\]
where $\us_1,\ldots,\us_{\ps}$ are uncovered shape variables.  Then
$\beta^1$ is equivalent to $\beta^2$:
\[
\begin{array}{l}
\beta^2 \ \equiv \ 
\begin{array}[t]{l}
  \exists  u_1,\ldots,u_n, \us_{\ps+1},\ldots,\us_{\ps+\qs}. \mnl
    \qquad  \phi^2(\us_{\ps+1},\ldots,\us_{\ps+\qs},\xs_1,\ldots,\xs_{\ms}) \mnl
    \qquad  \termBaseF(u_1,\ldots,u_n,x_1,\ldots,x_m) \land {} \mnl
    \qquad  \homF(u_1,\ldots,u_n,\us_1,\ldots,\us_n) \land {} \mnl
    \qquad  \cardinF(u_{p+1},\ldots,u_{p+q},\us_{\ps+1},\ldots,\us_{\ps+\qs})
\end{array}
\end{array}\]
Here $\phi^2$ is a base formula (Definitions \ref{def:semibase}
and \ref{def:baseFormula}) whose free variables are
variables free in $\beta^2$ as well as all covered shape variables:
\[
\begin{array}{l}
\phi^2(\us_{\ps+1},\ldots,\us_{\ps+\qs},\xs_1,\ldots,\xs_{\ms}) \ \equiv \ 
  \exists \us_1,\ldots,\us_{\ps}. \mnl
\qquad  \shapeBaseF(\us_1,\ldots,\us_{\ps},\us_{\ps+1},\us_{\ps+\qs},
                    \xs_1,\ldots,\xs_{\ms})
\end{array}
\]
Applying Lemma~\ref{lemma:uncoveredElim} we conclude that
$\phi^2$ is equivalent to some disjunction 
\[
   \bigvee_{i=1}^k \phi^{3,i}
\]
of base formulas without uncovered variables.
Let $\beta^{3,i}$ be the result of
replacing $\phi^2$ with $\phi^{3,i}$ in $\beta^2$.  Then $\beta^2$ is equivalent
to
\[
   \bigvee_{i=1}^k \beta^{3,i}
\]
and each $\beta^{3,i}$ has no uncovered variables either,
because every free variable of $\phi^{3,i}$ is either free or
covered in $\beta^{3,i}$.  By Corollary~\ref{cor:coveredToQfree}
each $\beta^{3,i}$ can be written as a quantifier free formula.
\end{proofsketch}

The following Theorem~\ref{thm:twoConstQElim} corresponds
to~\ref{thm:termAlgQElim} of Section~\ref{sec:qeTA}.

\begin{theorem}[Two Constants Quant.\ Elimination]
\label{thm:twoConstQElim}
There exist algorithms $A$, $B$ such that for a given formula 
$\phi$ in the language of Figure~\ref{fig:twoSortedSig}:
\begin{enumerate}
\item[a)] $A$ produces a quantifier-free formula $\phi'$
          in selector language
\item[b)] $B$ produces a disjunction $\phi'$ of structural base formulas
\end{enumerate}
\end{theorem}
\begin{proof}
Analogous to proof of Theorem~\ref{thm:termAlgQElim},
using Proposition~\ref{prop:StructqFreeToBaseDisj}
in place of Proposition~\ref{prop:qFreeToBaseDisj}
and Proposition~\ref{prop:StructbaseToSel}
in place of Proposition~\ref{prop:baseToSel}.
\end{proof}

%% Now remark why this proof fails for the recursive case,
%% assuming, as Maher and Mal'cev show, that we know how to
%% solve shape base formulas, there is only equality there.  It
%% must fail, otherwise Gurevich et.al. would have easier proof
%% for MSOL decidability.]
%% -- See \refday{21}{December}{2002}.

\begin{corollary}
The first-order theory of the structure $\twoSorted$ is decidable.
\end{corollary}

This completes description of our quantifier elimination for
the first-order theory of structure $\twoSorted$, which
models structural subtyping with two base types and one
binary constructor.  It is straightforward to extend the
construction of this section to any number of covariant
constructors if the base formula has only two constants.  In
Section~\ref{sec:nconst} we extend the result to any number
of constants as well.  Finally, in Section~\ref{sec:nonrec}
we extend the result to allow arbitrary decidable structures
for primitive types, even if the number of primitive types
is infinite.

%%% Local Variables: 
%%% mode: latex
%%% TeX-master: "main"
%%% End: 
% LocalWords:  arity iff Multisorted BLAH definedness disequalities subformula
% LocalWords:  acyclicity exaStrucBase unnested disequality subformulas Struct
% LocalWords:  subterms Quant Maher Mal'cev Gurevich al MSOL

\section{A Finite Number of Constants}
\label{sec:nconst}

In this section we prove the decidability of structural
subtyping of any finite number of constant symbols
(primitive types) and any number of function symbols
(constructors).  We first show the result when all
constructors are covariant, we then show the result when
some of the constructors are contravariant.

We introduce the notion of $\Sigma$-term-power of some structure
$\ThC$ as a generalization of the structure of structural
subtyping.

We represent primitive types in structural subtyping as a
structure $\ThC$ with a finite carrier $C$.  We call $\ThC$
the \emph{base structure}.  Without loss of generality, we
assume that $\ThC$ has only relations; functions and
constants are definable using relations.  Let $\lanC$ be a
set of relation symbols and let ${\order} \in \lanC$ be a
distinguished binary relation symbol.  $\order$ represents
the subtype ordering between types.  $C$ is finite, so
$\ThC$ is decidable (see Section~\ref{sec:nonrec} for the
case when $C$ is infinite but decidable).

We represent type constructors as free operations in the term
algebra with signature $\FreeSig$.  To represent the
variance of constructors we define for each constructor $f \in
\FreeSig$ of arity $\ar(f)=k$ and each argument $1 \leq i \leq k$ the
value $\variance(f,i) \in \{-1,1\}$.  The constructor $f$ is
covariant in argument $i$ iff $\variance(f,i)=1$.  For
convenience we assume $\ar(f) \geq 1$ for each $f \in \FreeSig$.

The $\Sigma$-term-power of $\ThC$ is a structure $\ThP$ defined
as follows.  Let $\FreeSigExt = \FreeSig \cup C$.  The domain
of $\ThP$ is the set $\DomP$ of finite ground
$\FreeSigExt$-terms.  Elements of $C$ are viewed as
constants of arity $1$.  The structure $\ThP$ has signature
$\FreeSig \cup \lanC$.  The constructors $f \in \FreeSig$ are
interpreted in $\ThP$ as in a free term algebra:
\[
    \interpretS{f}{\ThP}(t_1,\ldots,t_k) = f(t_1,\ldots,t_k)
\]
A relation $r \in \lanC \setminus \{ \order \}$ is interpreted pointwise
on the terms of same ``shape'' as follows. $\interpretS{r}{\ThP}$
is the least relation $\rho$ such that:
\begin{enumerate} \compr
\item if $\interpretS{r}{\ThC}(c_1,\ldots,c_n)$ then
  $\rho(c_1,\ldots,c_n)$
\item if $\rho(t_{i1},\ldots,t_{in})$ for all $i$ where $1 \leq i \leq k$,
then
\[
   \rho(f(t_{11},\ldots,t_{1k}),\ldots,f(t_{n1},\ldots,t_{nk}))
\]
\end{enumerate}
The relation ${\order} \in \lanC$ is interpreted similarly,
but taking into account the variance.
$\interpretS{\order}{\ThP}$ is the least relation $\rho$
such that
\begin{enumerate} \compr
\item if $\interpretS{\order}{\ThC}(c_1,c_2)$ then
  $\rho(c_1,c_2)$
\item if
\[
   \rho^{\variance(f,i)}(t_{i1},\ldots,t_{in})
\]
for all $i$ where $1 \leq i \leq k$, then
\[
    \rho(f(t_{11},\ldots,t_{1k}),\ldots,f(t_{n1},\ldots,t_{nk}))
\]
\end{enumerate}
Here we use the notation $\rho^v$ for $v \in \{-1,1\}$ with the meaning:
$\rho^1 = \rho$ and $\rho^{-1} = \{ \tu{y,x} \mid \tu{x,y} \in \rho\}$.

%%% Local Variables: 
%%% mode: latex
%%% TeX-master: "main"
%%% End: 
% LocalWords:  contravariant arity iff pointwise

  %\subsection{The Decidability Result}
\label{zzz:sec:decidabilityResult}

We next sketch the decidability of structural subtyping for
any finite number of primitive types $C$.  For now we assume
that all constructors $f \in \FreeSig$ are covariant, the
relation $\order$ thus does not play a special role.

\subsection{Extended Term-Power Structure}

For the purpose of quantifier elimination we define the
structure $\ThPe$ by extending the domain and the set of
operations of the term-power structure $\ThP$.

The domain of $\ThPe$ is $\DomPe = \DomP \cup \DomShapes$ where
$\DomShapes$ is the set of \emph{shapes} defined as follows.
Let $\FreeSigShape = \{ \cs \} \cup \{ \fs \mid f \in \FreeSig \}$ be a
set of function symbols such that $\cs$ is a fresh constant
symbol with $\ar(\cs)=0$ and $\fs$ are fresh distinct
constant symbols with $\ar(\fs)=\ar(f)$ for each $f \in
\FreeSig$.  The set of shapes $\DomShapes$ is the set of
ground $\FreeSigShape$-terms.  When referring to elements of
$\ThPe$ by \emph{term} we mean an element of $\DomP$; by
\emph{shape} we mean an element of $\DomShapes$.  We
write $X^{\shp}$ to denote an entity pertaining to shapes as
opposed to terms, so $\xs,\us$ denote variables ranging over
shapes, and $\ts$ to denotes terms that evaluate to shapes.

The extended structure $\ThPe$ contains term algebra
operations on terms and shapes (including selector
operations and tests, \cite[Page 61]{Hodges93ModelTheory}),
the homomorphism $\termShapeF$, and cardinality constraint
relations $|\phi|_{\ts} = k$ and $|\phi|_{\ts} \geq k$:

\begin{raggedright}
\begin{enumerate} \compr
\item constructors in the term algebra of terms, $f \in \FreeSigExt$
  $\interpretS{f}{\ThPe}(t_1,\ldots,t_k) = f(t_1,\ldots,t_k)$;
\item selectors in term the algebra of terms,
  $\interpretS{f_i}{\ThPe}(f(t_1,\ldots,t_k)) = t_i$;
\item constructor tests in the term algebra of terms,
  $\interpretS{\Is{f}}{\ThPe}(t) =
    \exists t_1,\ldots,t_k.\ t = f(t_1,\ldots,t_k)$;
\item constructors in the term algebra of shapes, $\fs \in \FreeSigShape$
  $\interpretS{\fs}{\ThPe}(\ts_1,\ldots,\ts_k) = \fs(\ts_1,\ldots,\ts_k)$;
\item selectors in the term algebra of shapes,
  $\interpretS{\fs_i}{\ThPe}(\fs(\ts_1,\ldots,\ts_k)) = \ts_i$;
\item constructor tests in the term algebra of shapes,
  $\interpretS{\Isfs}{\ThPe}(\ts) =
    \exists \ts_1,\ldots,\ts_k.\ \ts = \fs(\ts_1,\ldots,\ts_k)$;
\item the homomorphism mapping terms to shapes such that:
\begin{equation}
\begin{array}{l}
\interpretS{\termShapeF}{\ThPe}(f(t_1,\ldots,t_n)) = \mnl
\quad \shapified{f}(\interpretS{\termShapeF}{\ThPe}(t_1),\ldots,
                    \interpretS{\termShapeF}{\ThPe}(t_n))
\end{array}
\end{equation}
where
\begin{equation}
\begin{array}{rcll}
\shapified{x} &=& \cs, & \mbox{ if } x \in C \mnl
\shapified{f} &=& \fs, & \mbox{ if } f \in \FreeSig
\end{array}
\end{equation}
\item cardinality constraint relations
  \begin{equation}
    \begin{array}{l}
      \interpretS{|\phi(x_1,\ldots,x_k)|_{\ts}=k}{\ThPe}(t_1,\ldots,t_k) = \mnl
      \qquad |\interpretS{\phi(x_1,\ldots,x_k)}{\ThPe}(t_1,\ldots,t_k)| = k
    \end{array}
  \end{equation}
  and
  \begin{equation}
    \begin{array}{l}
      \interpretS{|\phi(x_1,\ldots,x_k)|_{\ts}\geq k}{\ThPe}(t_1,\ldots,t_k) = \mnl
      \qquad |\interpretS{\phi(x_1,\ldots,x_k)}{\ThPe}(t_1,\ldots,t_k)| \geq k
    \end{array}
  \end{equation}
  where $\phi(x_1,\ldots,x_k)$ is is a first-order formula over the
  base-structure language $\lanC$ with free variables $x_1,\ldots,x_k$,
  term $\ts$ denotes a shape,
  and $k$ is a nonnegative integer constant.
\end{enumerate}
\end{raggedright}
It remains to complete the semantics of cardinality
constraint relations, by defining the set
$\interpretS{\phi(x_1,\ldots,x_k)}{\ThPe}(t_1,\ldots,t_k)$.  If $s$ is a
shape, we call the set of positions of constant $\cs$ in $s$
\emph{leaves} of $s$, and denote it by $\leaves{s}$.  We represent
a leaf as a sequence of pairs $\tu{f,i}$ where $f$ is a constructor
of arity $k$ and $1 \leq i \leq k$.
If  $l \in \leaves{s}$ and $\termShape{t}=s$, then $t[l]$ denotes
the element $c \in C$ at position $l$ in term $t$ i.e.\ if
$l = \tu{f^1,i^1} \ldots \tu{f^n,i^n}$ then
\begin{equation}
  t[l] = f^n_{i^n}(\ldots f^2_{i^2}(f^1_{i^1}(t))\ldots)
\end{equation}
We define:
\begin{equation} \label{zzz:eqn:leafsetCardinDef}
  \begin{array}{l}
    \interpretS{\phi(x_1,\ldots,x_k)}{\ThPe}(t_1,\ldots,t_k) = \mnl
    \qquad \{ l \mid  \interpretS{\phi(x_1,\ldots,x_k)}{\ThC}(t_1[l],\ldots,t_k[l]) \}
  \end{array}
\end{equation}
The following equations follow from~(\ref{zzz:eqn:leafsetCardinDef})
and can be used as an equivalent alternative definition for
cardinality relations:
\begin{equation}
  \begin{array}{l}
    |\interpretS{\phi(x_1,\ldots,x_k)}{\ThPe}(c_1,\ldots,c_k)| = \mnl
    \qquad
    \left\{\begin{array}{rl}
        1, & \interpretS{\phi(x_1,\ldots,x_k)}{\ThC}(c_1,\ldots,c_k) \mnl
        0, & \lnot \interpretS{\phi(x_1,\ldots,x_k)}{\ThC}(c_1,\ldots,c_k)
      \end{array}\right.
  \end{array}
\end{equation}
\begin{equation} \label{zzz:eqn:decompose}
  \begin{array}{l}
    |\interpretS{\phi(x_1,\ldots,x_k)}{\ThPe}(f(t_{11},\ldots,t_{1l}),\ldots,f(t_{k1},\ldots,t_{kl}))|
    \mnl
    \qquad = \ |\interpretS{\phi(x_1,\ldots,x_k)}{\ThPe}(t_{11},\ldots,t_{k1})| + \ldots \mnl
    \qquad + \ |\interpretS{\phi(x_1,\ldots,x_k)}{\ThPe}(t_{1l},\ldots,t_{kl})|
  \end{array}
\end{equation}
Definition~(\ref{zzz:eqn:leafsetCardinDef}) generalizes
\cite[Definition 2.1, Page
63]{FefermanVaught59FirstOrderPropertiesProductsAlgebraicSystems}.
We write $|\phi(t_1,\ldots,t_k)|_{\ts}=k$ as a shorthand for the
atomic formula $(|\phi(x_1,\ldots,x_k)|_{\ts}=k)(t_1,\ldots,t_k)$,
similarly for $|\phi(t_1,\ldots,t_k)|_{\ts}\geq k$.  This is more than
a notational convenience, see Section~\ref{sec:nonrec} for
an approach which introduces sets of leaves as elements of
the domain of $\ThPe$ and defines a cylindric algebra
interpreted over sets of leaves.  The approach in this
section follows \cite{Mostowski52DirectProductsTheories} in
merging the quantifier elimination for products and
quantifier elimination for boolean algebras.

Some of the operations in $\ThPe$ are partial.  We use the
definitions and results of Section~\ref{sec:partFun} to deal
with partial functions.  $f_i(t)$ is defined iff $\Is{f}(t)$
holds, $\fs_i(\ts)$ is defined iff $\Isfs(\ts)$ holds.
Cardinality constraints $|\phi(t_1,\ldots,t_k)|_{\ts}=k$ and
$|\phi(t_1,\ldots,t_k)|_{\ts} \geq k$ are defined iff
$\termShape{t_1}=\ldots=\termShape{t_k}=\ts$ holds.

The structure $\ThPe$ is at least as expressive as $\ThP$ because
the only operations or relations present in $\ThP$ but not
in $\ThPe$ are $\interpretS{r}{\ThP}$ for $r \in \lanC$,
and we can express $\interpretS{r}{\ThP}(t_1,\ldots,t_k)$ as
$
  |\lnot \, r(t_1,\ldots,t_k)|_{\termShape{t_1}}=0
$.

Our goal is to give a quantifier elimination for first-order
formulas of structure $\ThPe$.  By a quantifier-free formula
we mean a formula without quantifiers outside cardinality
constraints, e.g.\ the formula $|\forall x. x \leq t|_{\xs} = k$ is
quantifier-free.

\subsection{Structural Base Formulas}

In this section we define the notion of structural base
formulas for any base structure $\ThC$ with a finite
carrier.

Definition~\ref{zzz:def:nonrecStructuralBase} of structural
base formula for quantifier elimination in $\ThPe$ differs
from Definition~\ref{def:structuralBase} in the conjuncts of
$\cardinF$ subformula.  Instead of cardinality constraints
on boolean algebra terms,
Definition~\ref{zzz:def:nonrecStructuralBase} contains
cardinality constraints on first-order formulas.

The notion of base formula and Lemma~\ref{lemma:baseSat}
apply to terms $\DomP$ as well as shapes $\DomShapes$ in the
structure $\DomPe$ because shapes are also terms over the
alphabet $\FreeSigShape$.  For brevity we write $\ustar$ for
an internal shape or term variable, and similarly $\xstar$
for a free shape or term variable, $\tstar$ for terms,
$\fstar$ for term or shape term algebra constructor and
$\fstar_i$ for a term or shape term algebra selector.

\begin{raggedright}
\begin{definition}[Structural Base Formula] \ \\
\label{zzz:def:nonrecStructuralBase}
A {\em structural base formula} with:
\begin{itemize} \compr
\item free term variables $x_1,\ldots,x_m$;
\item internal non-parameter term variables $u_1,\ldots,u_p$;
\item internal parameter term variables $u_{p+1},\ldots,u_{p+q}$;
\item free shape variables $\xs_1,\ldots,\xs_{\ms}$;
\item internal non-parameter shape variables $\us_1,\ldots,\us_{\ps}$;
\item internal parameter shape variables $\us_{\ps},\ldots,\us_{\ps+\qs}$
\end{itemize}
is a formula of the form:
\[\begin{array}{l}
  \exists  u_1,\ldots,u_n, \us_1,\ldots,\us_{\ns}. \mnl
\begin{array}[t]{l}
    \shapeBaseF(\us_1,\ldots,\us_{\ns},\xs_1,\ldots,\xs_{\ms}) \ \land \mnl
    \termBaseF(u_1,\ldots,u_n,x_1,\ldots,x_m) \ \land \mnl
    \termHomF(u_1,\ldots,u_n,\us_1,\ldots,\us_{\ns}) \ \land \mnl
    \cardinF(u_{p+1},\ldots,u_n,\us_{\ps+1},\ldots,\us_{\ns})
\end{array}
\end{array}\]
where $n=p+q$, $\ns = \ps + \qs$, and formulas
$\shapeBaseF$, $\termBaseF$, $\termHomF$, $\cardinF$ are
defined as follows.
\[\begin{array}{l}
\shapeBaseF(\us_1,\ldots,\us_{\ns},\xs_1,\ldots,\xs_{\ms}) = \mnl
\qquad \bigwedge\limits_{i=1}^{\ps} \us_i = t_i(\us_1,\ldots,\us_{\ns}) \ \land\
       \bigwedge\limits_{i=1}^{\ms} \xs_i = \us_{j_i} \mnl
\qquad \land \ \distinctF(\us_1,\ldots,\us_n)
\end{array}\]
where each $t_i$ is a shape term of the form $\fs(\us_{i_1},\ldots,\us_{i_k})$ for
some $f \in \FreeSigConst$, $k=\ar(f)$, and 
$j : \{ 1,\ldots,\ms \} \to \{ 1,\ldots,\ns\}$ is a function mapping
indices of free shape variables to indices of internal shape variables.

\[\begin{array}{l}
\termBaseF(u_1,\ldots,u_n,x_1,\ldots,x_m) = \mnl
 \qquad \bigwedge\limits_{i=1}^p u_i = t_i(u_1,\ldots,u_n) \ \land\ \bigwedge\limits_{i=1}^m x_i = u_{j_i}
\end{array}\]
where each $t_i$ is a term of the form $f(u_{i_1},\ldots,u_{i_k})$ for
some $f \in \FreeSig$, $k = \ar(f)$, and 
$j : \{ 1,\ldots,m \} \to \{ 1,\ldots,n\}$ is a function mapping
indices of free term variables to indices of internal term variables.

\[\begin{array}{ll}
\termHomF(u_1,\ldots,u_n,\us_1,\ldots,\us_{\ns}) = &
  \bigwedge\limits_{i=1}^n \termShape{u_i} = \us_{j_i}
\end{array}\]
where $j : \{ 1,\ldots,n \} \to \{ 1,\ldots,\ns\}$ is some function such
that $\{ j_1,\ldots,j_p \} \subseteq \{ 1,\ldots,\ps \}$ and $\{ j_{p+1},\ldots,j_{p+q}
\} \subseteq \{ \ps+1,\ldots,\ps+\qs \}$ 
(a term variable is a parameter
variable iff its shape is a parameter shape variable).

\[
\cardinF(u_{p+1},\ldots,u_n,\us_{\ps+1},\ldots,\us_{\ns}) = \psi_1 \land \cdots \land \psi_d
\]
where each $\psi_i$ is a cardinality constraint of the form
\[
   |\phi(u_{j_1},\ldots,u_{j_l})|_{\us} = k
\]
or
\[
   |\phi(u_{j_1},\ldots,u_{j_l})|_{\us} \geq k
\]
where $\{ j_1,\ldots,j_l \} \subseteq \{ p+1,\ldots,n \}$
and the conjunct
$\termShape{u_{j_d}}=\us$ occurs in $\termHomF$ for $1 \leq
d \leq l$.
We require each structural base formula to satisfy the following
conditions:
\begin{enumerate}

\item[P0)] the graph associated with shape base formula 
      \[
         \exists  \us_1,\ldots,\us_{\ns}.\
            \shapeBaseF(\us_1,\ldots,\us_{\ns},\xs_1,\ldots,\xs_{\ms})
      \]
      is acyclic;
%% \item[P0)] $\shapeBaseF$ does not violate the occur-check:
%%   $\lnot (\us \deterplus{\shapeBaseF} \us)$ for every shape
%%   variable $\us$ occurring in $\shapeBaseF$;

\item[P1)] congruence closure property for $\shapeBaseF$ subformula:
      there are no two distinct variables $\us_i$ and
      $\us_j$ such that both $\us_i =
      f(\us_{l_1},\ldots,\us_{l_k})$ and $\us_j =
      f(\us_{l_1},\ldots,\us_{l_k})$ occur as conjuncts in
      formula $\shapeBaseF$;

\item[P2)] congruence closure property for $\termBaseF$ subformula:
      there are no two distinct variables $u_i$ and $u_j$
      such that both $u_i = f(u_{l_1},\ldots,u_{l_k})$ and $u_j =
      f(u_{l_1},\ldots,u_{l_k})$ occur as conjuncts in formula
      $\termBaseF$;

\item[P3)] homomorphism property of $\termShapeF$:
      for every non-parameter term variable $u$ such that $u =
      f(u_{i_1},\ldots,u_{i_k})$ occurs in $\termBaseF$, if
      conjunct $\termShape{u}=\us$ occurs in $\termHomF$,
      then for some shape variables
      $\us_{j_1},\ldots,\us_{j_k}$ term $\us =
      \fs(\us_{j_1},\ldots,\us_{j_k})$ occurs in $\shapeBaseF$
      where $\fs=\shapified{f}$ and for every $r$ where $1 \leq
      r \leq k$, conjunct $\termShape{u_{i_r}}=\us_{j_r}$
      occurs in $\termHomF$.

\end{enumerate}
\end{definition}
\end{raggedright}
Note that the validity of the occur check for term variables
follows from P0) and P3).  Another immediate consequence of
Definition~\ref{zzz:def:nonrecStructuralBase} is the
following Proposition~\ref{zzz:prop:nonrecStructexBaseBase}.

\begin{proposition}[Quantification of Str.\ Base Form.]
\label{zzz:prop:nonrecStructexBaseBase}
If $\beta$ is a structural base formula and $x$ a free shape
or term variable in $\beta$, then there exists a base structural formula
$\beta_1$ equivalent to $\exists x.\beta$.
\end{proposition}

We proceed to show that a quantifier-free formula can be
written as a disjunction of base formulas, and a base
formula can be written as a quantifier-free formula.

\subsection{Conversion to Base Formulas}

Conversion from a quantifier-free formula to the structural
base formula is given by
Proposition~\ref{zzz:prop:nonrecStructexBaseBase}.  The
proof of
Proposition~\ref{zzz:prop:nonrecStructqFreeToBaseDisj} is
analogous to the proof of
Proposition~\ref{prop:StructqFreeToBaseDisj} but uses
of~(\ref{zzz:eqn:decompose}) instead
of~(\ref{eqn:anyDecompose}).
\begin{proposition}[Quantifier-Free to Structural Base]
\label{zzz:prop:nonrecStructqFreeToBaseDisj}
Every well-defined quantifier-free formula $\phi$ is equivalent
on $\ThPe$ to $\boolTrue$, $\boolFalse$, or some disjunction
of structural base formulas.
\end{proposition}

\subsection{Conversion to Quantifier-Free Formulas}

The conversion from structural base formulas to
quantifier-free formulas is similar to the case of two
constant symbols in Section~\ref{sec:twoConstQE}, but
requires the use of Feferman-Vaught technique.

\begin{definition} \label{zzz:def:nonrecStructDeterminations}
  The set $\determinations$ of variable determinations of a
  structural base formula $\beta$ is the least set $S$ of pairs
  $\tu{\ustar,\tstar}$ where $\ustar$ is an internal term
  or shape variable and $\tstar$ is a term over the
  free variables of $\beta$, such such that:

\begin{raggedright}
\begin{enumerate} % \compr
  
\item if $\xstar=\ustar$ occurs in $\termBaseF$
  or $\shapeBaseF$, then $\tu{\ustar,\xstar} \in S$;
  
\item if $\tu{\ustar,\tstar} \in S$ and $\ustar =
  \fstar(\ustar_1,\ldots,\ustar_k)$ occurs in $\shapeBaseF$ or
  $\termBaseF$ then
  $\{\tu{\ustar_1,\fstar_1(\tstar)},\ldots,\tu{\ustar_k,\fstar_k(\tstar)}\}
  \subseteq S$;
  
\item if
  $\{\tu{\ustar_1,\fstar_1(\tstar)},\ldots,\tu{\ustar_k,\fstar_k(\tstar)}\}
  \subseteq S$ and $\ustar = \fstar(\ustar_1,\ldots,\ustar_k)$ occurs in
  $\shapeBaseF$ or $\termBaseF$ then $\tu{\ustar,\tstar} \in
  S$;
  
\item if $\tu{u,t} \in S$ and $\termShape{u}=\us$ occurs in $\termHomF$
      then $\tu{\us,\termShape{t}} \in S$.

      % !! unfortunate confusion of object and meta level possible
      
\end{enumerate}
\end{raggedright}
\end{definition}

\begin{definition} \label{zzz:def:determinedVar}
  An internal variable $\ustar$ is \emph{determined} if
  $\tu{\ustar,\tstar} \in \determinations$ for some term
  $\ts$.  An internal variable is \emph{undetermined} if it
  is not determined.
\end{definition}

\begin{lemma} \label{zzz:lemma:nonrecDeterminationWorks}
  Let $\beta$ be a structural base formula with matrix $\beta_0$ and
  let $\determinations$ be the determinations of $\beta$.  If
  $\tu{\ustar,\tstar} \in S$ then $\models \beta_0 \implies \ustar=\tstar$.
\end{lemma}
%% \begin{proof}
%% By induction, using Definition~\ref{zzz:def:nonrecStructDeterminations}.
%% \end{proof}

\begin{corollary} \label{zzz:cor:nonrecDeterminedToQfree}
  Let $\beta$ be a structural base formula such that every
  internal variable is determined.  Then $\beta$ is equivalent
  to a well-defined quantifier-free formula.
\end{corollary}
\begin{proof}
  By Lemma~\ref{zzz:lemma:nonrecDeterminationWorks}
  using
  \begin{equation} \label{zzz:eqn:replacement}
    \exists x. x = t \land \phi(x)  \iff   \phi(t)
  \end{equation}
\end{proof}

\begin{lemma} \label{zzz:lemma:nonrecStrucElimTermNonparam}
Let $u$ be an undetermined non-parameter
term variable in a
structural base formula $\beta$ such that $u$ is a source i.e.\
no conjunct of the form
\[
   u' = f(u_1,\ldots,u,\ldots,u_k)
\]
occurs in $\termBaseF$.  Let $\beta'$ be the result of removing
$u$ and conjuncts containing $u$ from $\beta$.  Then $\beta$ is
equivalent to $\beta'$.
\end{lemma}
\begin{proof}
  The conjunct containing $u$ in $\termHomF$ is a
  consequence of the remaining conjuncts, so we drop it.
  We then apply~(\ref{zzz:eqn:replacement}).
\end{proof}

\begin{corollary} \label{zzz:cor:nonrecStrucElimNonparam}
  Every base formula is equivalent to a base formula without
  undetermined non-parameter term variables.
\end{corollary}
\begin{proof}
  If a structural base formula has an undetermined
  non-parameter term variable, then it has an undetermined
  non-parameter term variable that is a source.
  Repeatedly apply
  Lemma~\ref{zzz:lemma:nonrecStrucElimTermNonparam} to eliminate
  all undetermined non-parameter term variables.
\end{proof}

The following Lemma~\ref{zzz:lemma:productSubstructure} is a
consequence of the fact that terms of a fixed shape $s$ form
a substructure of $\ThP$ isomorphic to the finite power
$\ThC^m$ where $m = |\leaves{s}|$ and follows from
Feferman-Vaught theorem in Section~\ref{sec:fefermanVaught}.
\begin{lemma} \label{zzz:lemma:productSubstructure}
  Let
  \begin{equation}
    \begin{array}{l}
      \alpha \ \equiv \ \exists u.\ 
      \begin{array}[t]{l}
        \termShape{u}=\us \ \land \mnl
        \termShape{u_1}=\us \ \land \ldots \land \ \termShape{u_k}=\us \ \land \mnl
        \psi_1 \ \land \ldots \land \ \psi_p
      \end{array}
    \end{array}
  \end{equation}
  where each $\psi_i$ is a cardinality constraint of the form
  $|\phi|_{\us}=k$ or $|\phi|_{\us} \geq k$ where all free variables
  of $\phi$ are among $u,u_1,\ldots,u_k$.  Then there exists formula
  $\psi$ such that $\psi$ is a disjunction of conjunctions of
  cardinality constraints $|\phi'|=k$ and $|\phi'|\geq k$ where
  the free variables in each $\phi'$ are among $u_1,\ldots,u_k$ and
  formula $\alpha$ is equivalent on $\ThPe$ to $\alpha'$ where
  \begin{equation}
    \begin{array}{l}
      \alpha' \ \equiv \ 
      \termShape{u_1}=\us \ \land \ldots \land \ \termShape{u_k}=\us \ \land \
      \psi
    \end{array}
  \end{equation}
\end{lemma}

\begin{proposition}[Struct.\ Base to Quantifier-Free]
\label{zzz:prop:nonrecStructbaseToSel}
Every structural base formula $\beta$ is equivalent on $\ThPe$ to some
well-defined quantifier-free formula $\phi$.
\end{proposition}
\begin{proofsketch}
  By Corollary~\ref{zzz:cor:nonrecStrucElimNonparam} we may
  assume that $\beta$ has no undetermined non-parameter term
  variables.  By Corollary~\ref{zzz:cor:nonrecDeterminedToQfree}
  we are done if there are no undetermined variables, so it
  suffices to eliminate undetermined parameter term
  variables and undetermined shape variables.
  
  Let $u$ be an undetermined parameter term variable. $u$
  does not occur in $\termBaseF$ because it cannot have a
  successor or a predecessors in the graph associated with
  term base formula.
%% $\lnot (u
%%   \deterplusSome u')$ for all $u'$, and $\lnot (u''
%%   \deterplusSome u)$ for all $u''$ since there are no
%%   undetermined non-parameter term variables.
  Therefore, $u'$ occurs only in $\termHomF$ and $\cardinF$.
  Let $\us$ be the shape variable such that
  $\us=\termShape{u}$ occurs in $\termHomF$.  Let
  $\psi_1,\ldots,\psi_p$ be all conjuncts of $\cardinF$ that contain
  $u$.
  
  Each $\psi_i$ is of the form $|\phi|_{\us} \geq k_i$ or $|\phi|_{\us} =
  k_i$ and for each variable $u'$ free in $\phi$ the conjunct
  $\termShape{u}=\us$ occurs in $\termHomF$.
  The base formula can therefore be
  written in form
  \[
    \beta_1 \ \equiv \ \exists x_1,\ldots,x_e,\xs_1,\ldots,\xs_f.\ \phi \land \alpha
  \]
  where $\alpha$ has the form as in
  Lemma~\ref{zzz:lemma:productSubstructure}.  Applying
  Lemma~\ref{zzz:lemma:productSubstructure} we eliminate $u$ and
  obtain $\psi = \bigvee_{i=1}^w \alpha_i$ where and each $\alpha_i$ is a
  conjunction of cardinality constraints.  Base formula
  $\beta_1$ is thus equivalent to the disjunction $\bigvee_{i=1}^w
  \beta_{1,i}$ where each $\beta_{1,i}$ is a base formula
  \[
  \beta_{1,i} \ \equiv \ \exists x_1,\ldots,x_e,\xs_1,\ldots,\xs_f.\ \phi \land \phi_{1,i}
  \]
  By repeating this process we eliminate all undetermined
  parameter term variables from a base formula.  Each of the
  resulting base formulas contains no undetermined term
  variables.

  It remains to eliminate undetermined shape variables.
  This process is similar to term algebra quantifier
  elimination; the key ingredient is
  Lemma~\ref{lemma:baseSat}, which relies on the fact that
  undetermined parameter variables may take on infinitely
  many values.  We therefore ensure that undetermined
  parameter shape variables are not constrained by term and
  parameter variables through conjuncts outside
  $\shapeBaseF$.  
  
  Consider an undetermined parameter shape variable $\us$.
  $\us$ does not occur in $\termHomF$, because all term
  variables are determined and a conjunct $\us =
  \termShape{u}$ would imply that $\us$ is determined as
  well.  $\us$ can thus occur only in $\cardinF$ within some
  cardinality constraint $|\phi|_{\us}=k$ or $|\phi|_{\us}\geq k$.
  Moreover, formula $\phi$ in each such cardinality constraint
  is closed: otherwise $\phi$ would contain some free variable
  $u$, by definition of base formula $u$ would have to be a
  parameter variable, all parameter term variables are
  determined, so $\us$ would be determined as well.  Let
  $\us$ denote some shape $s$.  Because $\phi$ is a closed
  formula, $|\phi|$ is equal to $0$ if
  $\interpretS{\phi}{\ThC}=\boolFalse$ and to the shape size
  $m=|\leaves{s}|$ if $\interpretS{\phi}{\ThC}=\boolTrue$.
  (The fact that closed formulas reduce to the constraints
  on domain size appears in \cite[Theorem 3.36, Page
  13]{Mostowski52DirectProductsTheories}.)  After
  eliminating constraints equivalent to $0=k$ and $0 \geq k$,
  we obtain a conjunction of simple linear constraints of the
  form $m = k$ and $m \geq k$.  These constraints specify a
  finite or infinite set $S \subseteq \{ 0,1,\ldots \}$ of possible sizes
  $m$.  Let $A = \{ s \mid \, |\leaves{s}| \in S \}$.  If the set
  $S$ is infinite then it contains an infinite interval of
  form $\{ m_0, m_0 + 1,\ldots \}$ so the set $A$ is infinite.  If
  $\FreeSig$ contains a unary constructor and $S$ is
  nonempty, then $A$ is infinite.  If $\FreeSig$ contains no
  unary constructors and $S$ is finite then $A$ is finite
  and the cardinality constraints containing $\us$ are
  equivalent to
  $
     \bigvee_{i=1}^p \us = \ts_i
  $
  where $A = \{ \ts_1,\ldots,\ts_p \}$.  We therefore generate a
  disjunction of base formulas $\beta_1,\ldots,\beta_p$ where $\beta_i$
  results from $\beta$ by replacing cardinality constraints
  containing $\us$ with with $\us=\ts_i$.  We convert each
  $\beta_i$ to a disjunction of base formulas by labelling
  subterms of $t_i$ with internal shape variables and doing
  case analysis on the equality between new internal shape
  variables to ensure the invariants of a base formula, as
  in the proof of~\ref{zzz:prop:nonrecStructqFreeToBaseDisj}.
  By repeating this process for all shape variables $\us$
  where the set $S$ is finite, we obtain base formulas where
  the set $A$ is infinite for every undetermined parameter
  shape variable $\us$.  We may then eliminate all
  undetermined parameter and non-parameter shape variables
  along with the conjuncts that contain them.  The result is
  an equivalent formula by Lemma~\ref{lemma:baseSat}.
  
  All variables in each of the resulting base formulas are
  determined.  By
  Corollary~\ref{zzz:cor:nonrecDeterminedToQfree} each formula
  can be written as a quantifier-free formula, and the
  resulting disjunction is a quantifier-free formula.
\end{proofsketch}

%%% Local Variables: 
%%% mode: latex
%%% TeX-master: "main"
%%% End: 

  \subsection{One-Relation-Symbol Variance}
\label{sec:oneVariance}

So far we have assumed that all constructors are covariant.
In this section we describe the changes needed to extend the
result to the case when the constructors have arbitrary
variance with respect to some distinguished binary relation
denoted $\order$.

\begin{definition}
  If $\phi$ is a first-order formula in the language $\lanC$ the
  \emph{contravariant version} of $\phi$, denoted $\cver{\phi}$,
  is defined by induction on the structure of formula by:
  \begin{equation}
    \begin{array}{rcl}
      \cver{(r(t_1,\ldots,t_k))} &=& r(t_1,\ldots,t_k), \mbox{ if } 
           r \in \lanC \setminus \{ \order \} \mnl
      \cver{(t_1 \order t_2)} &=& t_2 \order t_1 \mnl
      \cver{(\phi_1 \land \phi_2)} &=& \cver{\phi_1} \land \cver{\phi_2} \mnl
      \cver{(\phi_1 \lor \phi_2)} &=& \cver{\phi_1} \land \cver{\phi_2} \mnl
      \cver{(\lnot \phi)} &=& \lnot \cver{\phi} \mnl
      \cver{(\exists t. \phi)} &=& \exists t. \cver{\phi} \mnl
      \cver{(\forall t. \phi)} &=& \forall t. \cver{\phi}
    \end{array}
  \end{equation}
\end{definition}
Define $\ThCi$ to have the same domain and same
interpretation of operations and relations $r \in \lanC \setminus \{
\order \}$ but where
\begin{equation}
   \interpretS{\order}{\ThCi} = (\interpretS{\order}{\ThC})^{-1}
\end{equation}
We clearly have for every formula $\phi$ and every valuation $\sigma$:
\begin{equation}
  \interpretS{\cver{\phi}}{\ThC} =
  \interpretS{\phi}{\ThCi}
\end{equation}
If $l \in \leaves{s}$ is a leaf $l = \tu{f^1,i^1} \ldots \tu{f^n,i^n}$, define
$\variance(l)$ as the product of integers
\begin{equation}
  \prod_{j=1}^n \variance(f^j,i^j)
\end{equation}
We generalize~(\ref{zzz:eqn:leafsetCardinDef}) to
\begin{equation} \label{zzz:eqn:contraLeafsetCardinDef}
  \begin{array}{l}
    \interpretS{\phi(x_1,\ldots,x_k)}{\ThPe}(t_1,\ldots,t_k) = \mnl
    \qquad \{ l \mid  \interpretS{\phi(x_1,\ldots,x_k)}{\ThC'}(t_1[l],\ldots,t_k[l]) \}
  \end{array}
\end{equation}
where $\ThC'$ denotes $\ThC$ for $\variance{l}=1$ and
$\ThCi$ for $\variance{l}-1$.  Hence, isomorhism between
terms of some fixed shape $s$ with $|\leaves{s}|=m$ 
and $\ThC^m$ breaks, but there is still an isomorphism with
$\ThC^{P(s)} \times (\ThCi)^{N(s)}$ where 
\begin{equation} \label{eqn:PN}
\begin{array}{rcl}
  P(s) &=& |\{ l \in \leaves{s} \mid \variance(l) = 1 \}| \mnl
  N(s) &=& |\{ l \in \leaves{s} \mid \variance(l) = -1 \}|
\end{array}
\end{equation}
Because of this isomorphism,
Lemma~\ref{zzz:lemma:productSubstructure} still holds and we may
still use Feferman-Vaught theorem from Section~\ref{sec:fefermanVaught}.
% \cite{FefermanVaught59FirstOrderPropertiesProductsAlgebraicSystems,
% Mostowski52DirectProductsTheories}.

Equation (\ref{zzz:eqn:decompose}) generalizes to:
\begin{equation} \label{zzz:eqn:contraDecompose}
  \begin{array}{l}
    |\interpretS{\phi(x_1,\ldots,x_k)}{\ThPe}(f(t_{11},\ldots,t_{1l}),\ldots,f(t_{k1},\ldots,t_{kl}))|
    \mnl
    \quad = \sum_{i=1}^l |\interpretS{\wvar{\phi}{\variance(f,l)}
      (x_1,\ldots,x_k)}{\ThPe}(t_{1i},\ldots,t_{ki})|
  \end{array}
\end{equation}
The only change in the proof of
Proposition~\ref{zzz:prop:nonrecStructqFreeToBaseDisj} is
the use of~(\ref{zzz:eqn:contraDecompose}) instead
of~(\ref{zzz:eqn:decompose}).  Most of the proof of
Proposition~\ref{zzz:prop:nonrecStructbaseToSel} remains
unchanged as well; the only additional difficulty is
eliminating constraints of the form $|\phi|_{\us}=k$ and
$|\phi|_{\us} \geq k$ where $\us$ is a parameter shape variable
and $\phi$ is a closed formula.  Lemma~\ref{zzz:lemma:contraSets}
below addresses this problem.

We say that an algorithm $g$ \emph{finitely computes} some function
$f : A \to 2^B$ where $B$ is an infinite set iff $g$ is a function
from $A$ to the set $\FinPow{B} \cup \{ \infty \}$ where
$\FinPow{B}$ is the set of finite subsets of set $B$, $\infty$ is
a fresh symbol, and
\begin{equation}
  g(a) = \left\{ \begin{array}{rl}
      f(a), & \mbox{ if } f(a) \in \FinPow{B} \mnl
      \infty,    & \mbox{ if } f(a) \notin \FinPow{B}
    \end{array} \right.
\end{equation}
\begin{lemma} \label{zzz:lemma:contraSets}
  There exists an algorithm that, given a shape variable $\us$ and
  a conjunction $\psi \equiv \bigwedge_{i=1}^n \psi_i$
  of cardinality constraints where each $\psi_i$ is of form
  $|\phi_i|_{\us} = k_i$ or $|\phi_i|_{\us} \geq k_i$ for some closed formula $\phi_i$,
  finitely computes the set
  \begin{equation}
    A = \{ s \mid \interpretS{\psi}{\ThP} [\us \mapsto s] \}
  \end{equation}
  of shapes which satisfy $\psi$ in $\ThP$.
\end{lemma}
\begin{proofsketch}
  Let $\phi$ be a closed formula in language $\lanC$.  Compute
  $\interpretS{\phi}{\ThC}$ and $\interpretS{\cver{\phi}}{\ThC}$
  and then replace $|\phi|_s$ with one of the expressions
  $P(s)+N(s)$, $P(s)$, $N(s)$, $0$ according to the following
  table.
\begin{equation}
  \begin{array}{cc|c}
    \interpretS{\phi}{\ThC} & \interpretS{\cver{\phi}}{\ThC} & |\phi|_s = \\ \hline
    \boolTrue & \boolTrue & P(s) + N(s) \mnl
    \boolTrue & \boolFalse & P(s) \mnl
    \boolFalse & \boolTrue & N(s) \mnl
    \boolFalse & \boolFalse & 0
  \end{array}
\end{equation}
  The constraints of the form $N(s) + P(s)=k$ and $N(s) + P(s) =
  k$ can be expressed as propositional combinations of
  constraints of the form $N(s) = k$, $P(s)=k$, $P(s) \geq k$ and
  $N(s) \geq k$.  Therefore, $\psi$ can be written as a
  propositional combination of these four kinds of
  constraints and each conjunction $C(s)$ can further be
  assumed to have one of the forms:
  \begin{enumerate}
  \item[F1)] $\constr{k_P}{k_N}(s) \ \equiv \ P(s) = k_P \land N(s) = k_N$;
  \item[F2)] $\constr{k_P}{k^{+}_N}(s) \ \equiv \ P(s) = k_P \land N(s) \geq k_N$;
  \item[F3)] $\constr{k^{+}_P}{k_N}(s) \ \equiv \ P(s) \geq k_P \land N(s) = k_N$;
  \item[F4)] $\constr{k^{+}_P}{k^{+}_N}(s) \ \equiv \ P(s) \geq k_P \land N(s) \geq k_N$.
  \end{enumerate}
  Let $A = \{ s \in \DomShapes \mid C(s) \}$.  To compute $A$ when
  $\FreeSig$ contains unary constructors, we first restrict
  $\FreeSig$ to the language $\FreeSig'$ with no unary
  constructors, and compute the set $A' \subseteq A$ using language
  $\FreeSig'$.  If $A'$ is empty, so is $A$, otherwise $A$
  is infinite.  Assume that $\FreeSig$ contains no unary
  constructors.  Assume further $\FreeSig$ contains at least
  one binary constructor and at lest one constructor is
  contravariant in some argument. 
  Let
  \begin{equation*}
    S = \{ \tu{P(s),N(s)} \mid s \in A \}
  \end{equation*}
  Because $P(s)+N(s) = |\leaves{s}|$ and there are only
  finitely many shapes of any given size (every constructor
  is of arity at least two), it suffices to finitely compute
  $S$.  $S$ can be given an alternative characterization as
  follows.  If $f \in \FreeSig$, $\ar(f)=k$, $f$ is covariant
  in $l$ arguments and contravariant in $k{-}l$ arguments
  define
  \begin{equation}
    \begin{array}{l}
      \interpretS{f}{\sizeS}(\tu{p_1,n_1},\ldots,\tu{p_k,n_k}) = \mnl
      \quad \tu{\sum_{i=1}^l p_i + \sum_{i=l+1}^k n_i, 
        \sum_{i=1}^l n_i + \sum_{i=l+1}^k p_i}
    \end{array}
  \end{equation}
  Let $U$ be the subset of $\{ \tu{p,n} \mid p,n \geq
  0\}$ generated from element $\tu{1,0}$ using operations
  $\interpretS{f}{\sizeS}$ for $f \in \FreeSig$.
  Then
  \begin{equation}
    S = \{ \tu{p,n} \in U \mid c(p,n) \}
  \end{equation}
  where $c(p,n)$ is the linear constraint corresponding to the
  constraint $C(s)$.

  Let $C(s)=\constr{k_P}{k_N}(s)$.  Then $S \subseteq \{ \tu{p,n} \mid
  p+n=k_P+k_N\}$. $S$ is therefore a subset of a finite set and
  is easily computable, which solves case F1).
  
  Let $C(s) = \constr{k^{+}_P}{k^{+}_N}(s)$.  Because
  $\FreeSig$ contains a binary constructor, $S$ contains
  pairs $\tu{p,n}$ with arbitrarily large $p{+}n$, so either
  the $p$ components or $n$ component of elements of $S$
  grows unboundedly.  Because $\FreeSig$ contains a
  constructor $f$ contravariant in some argument, we can
  define using $f$ an operation $o$ acting as a constructor
  covariant in at least one argument and contravariant in at
  least one argument.  Using operation on tuples whose one
  component grows unboundedly yields tuples whose both
  components grow unboundedly.  Therefore, $S$ is infinite,
  which solves case F4).
  
  Finally, consider the case $C(s) =
  \constr{k_P}{k^{+}_N}(s)$ (this will solve the case $C(s)
  = \constr{k^{+}_P}{k_N}(s)$ as well).  Observe that
  \begin{equation}
    \constr{k_P}{k^{+}_N}(s) = \constr{k_P}{0^+}(s) \land
    \bigwedge_{i=0}^{k_P-1} \lnot \constr{k_P}{i}(s)
  \end{equation}
  Because the set $S$ for each $\constr{k_P}{i}(s)$ is finite,
  it suffices to finitely compute $S$ for
  $\constr{k_P}{0^+}(s)$.  In that case
  \begin{equation}
    S = \{ \tu{p,n} \in U \mid p = k_P \}
  \end{equation}
  Let
  \begin{equation}
    \begin{array}{rcl}
      S_i &=& \{ \tu{p,n} \in U \mid p = i \} \mnl
      T_i &=& \{ \tu{p,n} \in U \mid n = i \}
    \end{array}
  \end{equation}
  To finitely compute $S$, finitely compute the sets $S_i$
  and $T_i$ for $0 \leq i \leq k_P$.  The algorithm starts with
  all sets $S_i$ and $T_i$ empty and keeps adding elements
  according to operations $\interpretS{f}{\sizeS}$.
  
  Assume that $S_0,T_0,\ldots,S_{i-1},T_{i-1}$ are finitely
  computed.  The computation of $S_i$ and $T_i$ proceeds as
  follows.  Let $f \in\FreeSig$ be a constructor of arity $k$
  with $l$ covariant arguments.  For $S_i$ we consider all
  solutions of the equation
  \begin{equation}
    p_1 + \cdots + p_l + n_{l+1} + \ldots + n_k = i
  \end{equation}
  for nonnegative integers $p_1,\ldots,p_l,n_{l+1},\ldots,n_k$.  First
  consider solution solutions where no variable is equal to
  $i$.  If for one of the solutions, one of the sets
  $S_{p_1},\ldots,S_{p_l}$ is infinite, then $S_i$ is infinite,
  otherwise add to $S_i$ all elements $\tu{i,n}$ where
  \begin{equation} \label{zzz:eqn:forS}
    n = n_1 + \cdots + n_l + p_{l+1} + \ldots + p_k
  \end{equation}
  If $n \leq k_P$ then also add the same elements $\tu{i,n}$ to
  $T_n$.  Next, proceed analogously with $T_i$, considering
  solutions of
  \begin{equation} \label{zzz:eqn:forT}
    n_1 + \cdots + n_l + p_{l+1} + \ldots + p_k = i
  \end{equation}
  If at this point $S_i$ is not infinite and not empty, then
  also consider the solutions of~(\ref{zzz:eqn:forS}) where
  $p_j = i$ for some $j$.  If such solution exists, then
  mark $S_i$ as infinite.  Proceed analogously with $T_i$.
  Finally, if both $S_i$ and $T_i$ are still finite but
  there exists a solution for $S_i$ where $n_{l+j}=i$ for
  some $j$ \emph{and} exists a solution for $T_j$ where
  $p_{l+d}=i$ for some $d$, then mark both $S_i$ and $T_i$
  as infinite.  This completes the sketch of one step of the
  computation.  (This step also applies to $S_0$ and $T_0$;
  we initially assume that $\tu{1,0} \in T_0$.)
\end{proofsketch}

\begin{example}
  Let us apply this algorithm to the special case where
  $\FreeSig = \{ f, g \}$ and
  \[
      \variance(g) = \tu{1,1}
  \]
  \[
      \variance(f) = \tu{-1,1}
  \]
  Let us see what the set $S$ looks like.  If $\tu{x,y} \in S$
  define $k \tu{x,y} = \tu{k x, k y}$ as in a vector space.
  
  First, $\tu{1,0} \in S$ because of $\cs$.  Next $\tu{1,1} \in
  S$ because of $\fs$ and $\tu{2,0} \in S$ because of $\gs$.

  More generally, we have the following composition rule:
  If $\tu{p_1,n_1}, \tu{p_2,n_2}$ then
  \[
       \tu{p_1+p_2, n_1+n_2} \in S
  \]
  because of $\gs$, and
  \[
       \tu{n_1+p_2, p_1+n_2} \in S
  \]
  because of $\fs$.
  
  Using $\gs$ we obtain all pairs $\tu{p,0}$ for $p \geq 1$.
  Using $\fs$ once on those we obtain $\tu{1,n}$ for $n \geq
  0$.  Adding these we additionally obtain $\tu{p,n}$ for $p
  \geq 2$ and $n \geq 0$.  Hence we have all pairs $\tu{p,n}$ for
  $p \geq 1$ and $n \geq 0$ and those are the only ones that can
  be obtained.  Thus,
  \[
      S = \{ \tu{p,n} \mid p \geq 1 \land n \geq 0 \}
  \]
  As expected, the case F1) yields a finite and the case F4)
  an infinite set.  The case F2) for $k_P=0$ is an empty
  set, otherwise it is an infinite set.  The case F3) always
  yields an infinite set.  This solves the problem for two
  constructors $f,g$.
\end{example}

Lemma~\ref{zzz:lemma:contraSets} allows to carry our the proof
of Proposition~\ref{zzz:prop:nonrecStructbaseToSel} so we obtain
our main result for finite $C$.

\begin{theorem}[Term Power Quant.\ Elimination]
  \label{zzz:thm:nonrecTwoConstQElim}
  There exists an algorithm that for a given
  well-defined formula $\phi$
  produces a quantifier-free formula $\phi'$ that is equivalent to $\phi$
  on $\ThPe$.
\end{theorem}

\begin{corollary}[Decidability of Structural Subtyping]
  Let $\ThC$ be a structure with a finite carrier and $\ThP$ a
  $\FreeSig$-term-power of $\ThC$.  Then the first-order
  theory of $\ThP$ is decidable.
\end{corollary}

%%% Local Variables: 
%%% mode: latex
%%% TeX-master: "main"
%%% End: 

\section{Term-Powers of Decidable Theories}
\label{sec:nonrec}

In this section we extend the result of
Section~\ref{sec:nconst} on decidability of term-powers of a
base structure $\ThC$ to allow $\ThC$ to be an arbitrary
decidable theory, even if the carrier $C$ is infinite.  

To keep a finite language in the case when $C$ is infinite,
we introduce a predicate $\IsPrim$ that allows testing
whether $t \in C$ for a term $t \in \DomP$.

In structural base formulas, we now distinguish between 1)
composed variables, denoting elements $t \in \DomP$ for which
$\Is{f}(t)$ holds for some constructor $f \in \FreeSig$, and
2) primitive variables, denoting elements $t \in \DomP$ for
which $\IsPrim(t)$ holds.

Another generalization compared to Section~\ref{sec:nconst}
is the use of a syntactically richer language for term power
algebras; to some extent this richer language can be viewed
as syntactic sugar and can be simplified away.

The generalization to infinitely many primitive types and
the generalization to a richer language are orthogonal.

For most of the section we focus on covariant constructors,
Section~\ref{sec:variance} discusses a generalized notion of
variance.

\begin{figure}
\newcommand{\tit}[1]{\multicolumn{3}{l}{{\mbox{\sl #1:}}}\mnl}
\newcommand{\btit}[1]{\multicolumn{3}{c}{{\mbox{\bf #1}}}\mnl}
\newcommand{\nextTablePart}{\\[2ex] \hline \\}
\begin{equation*}
\begin{array}{rcl}
\btit{lifted relations $r'$ for $r \in \lanC$}
r' &::& \termSort^k \to \boolSort \mnl
\btit{term algebra on terms}
\tit{constructors, $f \in \FreeSig$}
f &::& \termSort^k \to \termSort \mnl
\tit{constructor test, $f \in \FreeSig$}
\Is{f} &::& \termSort \to \boolSort \mnl
\tit{selectors, $f \in \FreeSig$}
f_i &::& \termSort \to \termSort \mnl
\end{array}
\end{equation*}
\caption{Basic Operations of $\Sigma$-term-power Structure
\label{fig:termPowerOps}}
\end{figure}

As in Section~\ref{sec:fefermanVaught} let $\ThC = \tu{C,R}$
be a decidable structure where $C$ is a non-empty set and
$R$ is a set of relations interpreting some relational
language $\lanC$, such that each $r \in R$ is a relation of
arity $\ar(r)$ on set $C$, i.e. $r \subseteq C^{\ar(r)}$.  We assume
that $R$ contains a binary relation symbol $\req \in R$,
interpreted as equality on the set $C$.

Operations and relations of the $\Sigma$-term-power structure are
summarized in Figure~\ref{fig:termPowerOps}.  We will show
the decidability of the first-order theory of the structure
with these operations.

In the special case when $C=\{ a,b \}$ and
\[
    r = \{ \tu{a,a}, \tu{a,b}, \tu{b,b} \}
\]
we obtain the theory in Section~\ref{sec:twoconst}.  When $R
= \{r\}$ where $r$ is a partial order on types, we obtain the
theory of structural subtyping of non-recursive covariant
types.  For arbitrary relational structure $\ThC$, if $f \in
\FreeSig$ for $\ar(f)=k$ we obtain a structure that properly
contains the $k$-th strong power of structure $\ThC$, in the
terminology of \cite{Mostowski52DirectProductsTheories}.

The structure of this section follows
Sections~\ref{sec:twoconst}.  We also associate a boolean
algebra of sets with each term $t$.  However, in this case,
the elements of the associated boolean algebra are sets of
occurrences of the constants that satisfy the given
first-order formula interpreted over $C$.  The occurrences
of constants within the terms of a given shape correspond to
the indices of the product structure in
Section~\ref{sec:fefermanVaught}.  We call these occurrences
{\em leaves}, because they can be represented as leaves of
the tree corresponding to a term.

\subsection{Product Theory of Terms of a Given Shape}

In this section we define the notions shape and leafset, and
state some properties that we use in the sequel.

Let
\[
    \FreeSigConst = \{ \cs \} \cup \{ \fs \mid f \in \FreeSig \}
\]
be a set of function symbols such that $\cs$ is a fresh
constant symbol with $\ar(\cs)=0$ and $\fs$ are fresh
distinct constant symbols with $\ar(\fs)=\ar(f)$ for each
$f \in
\FreeSig$.  Let $\shapifiedSym : \FreeSigExt \to
\FreeSigConst$ be defined by
\[\begin{array}{rcll}
\shapified{x} &=& \cs, & \mbox{ if } x \in C \mnl
\shapified{f} &=& \fs, & \mbox{ if } f \in \FreeSig
\end{array}\]
Let $\FTConst$ be the set of ground terms with signature
$\FreeSigConst$ and $\FTExt$ the set of ground terms of signature $\FreeSigExt$.

Define function $\termShapeF :: \FTExt \to \FTConst$ mapping each term 
to its shape by
\[
   \termShape{f(t_1,\ldots,t_n)} =
      \shapified{f}(\termShape{t_1},\ldots,\termShape{t_n})
\]
for each $f \in \FreeSigExt$.  Define $t_1 \sim t_2$ iff $\termShape{t_1}=\termShape{t_2}$.

Let $t$ be a term or shape and $t'$ the tree representing
$t$ as in Section~\ref{sec:termsAsTrees}.  If $p$ is a path
such that $t'(p)$ is defined and denotes a constant,
we write $t[p]$ to denote $t'(p)$ and call $p$ a {\em leaf}.
Note that $t[p]$ is defined iff $\termShape{t}[p]$ is
defined.  On the set of equivalent terms leaves act as
indices of Section~\ref{sec:fefermanVaught}.  If $s$ is a
shape, let $\leaves{s}$ denote the set of all leaves defined
on shape $s$.

Generalizing $\termContentF$ of Section~\ref{sec:baEquivTerms}, 
define function $\termContentF : \FTExt \to C^{*}$ by:
\[
\begin{array}{rcll}
   \termContent{c} &=& \cs, \mbox{if $c \in C$} \mnl
   \termContent{f(t_1,\ldots,t_k)} &=& \termContent{t_1} \cdot \ldots \cdot \termContent{t_k}
\end{array}
\]
Define $\delta(t)=\tu{\termShape{t},\termContent{t}}$
and
\[
   B = \{ \tu{s,w} \mid s \in \FTConst, w \in C^{*}, \termLength{s}=\seqLength{w} \}
\]
If all constructors $f \in \FreeSig$ are covariant then $\delta$ is
a bijection between $\FTExt$ and $B$.  Let
\[
   B(s_0) = \{ \tu{s,w} \in B \mid s = s_0 \}
\]
For a fixed $s_0$, the set $B(s_0)$ is isomorphic to the
power structure $C^n$ where $n = \termLength{s}$.

For each shape $s$ we introduce operations from
Section~\ref{sec:fefermanVaught}.  To distinguish the sets
of positions belonging to different shapes, we tag each set
of positions $L$ with a shape $s$.  We call the pair
$\tu{s,L}$ a \emph{leafset}.  The interpretation of each
relation $r \in \lanC$ is the leafset:
\[
    \interpret{r_s}(t_1,\ldots,t_k) = 
       \tu{s,\{ p \mid \interpretS{r}{C}(t_1[p],\ldots,t_k[p]) \}}
\]
We let $\slandI{s}$, $\slorI{s}$, $\slnotI{s}$,
$\strueI{s}$, $\sfalseI{s}$ stand for intersection, union,
complement, full set and empty set in the algebra of subsets
of the set $\leaves{s}$.  We also introduce $\sexistsI{s}$
as the union of a family of subsets indexed by a term of
shape $s$ and $\sforallI{s}$ as the intersection of a
family of subsets indexed by a term.

We use constructor-selector language for the term algebra on
terms.  We introduce constructor-selector language on shapes
by generalizing operations in Section~\ref{sec:baEquivTerms}
in a natural way.  In addition, we introduce a
constructor-selector language on leafsets.
For each $f \in \FreeSig$ we introduce a constructor
symbol $\fl$ on leafsets and define
\[
    \leafified{f} = \fl
\]
Constructors $\fl$ act on leafsets as follows.  If $L_i \subseteq
\leaves{s_i}$ for $1 \leq i \leq k$ define
\[
\fl(\tu{s_1,L_1},\ldots,\tu{s_k,L_k}) = \tu{s,L}
\]
where $s = \fs(s_1,\ldots,s_k)$, and $L \subseteq \leaves{s}$ is given by
\[
    L = (\{ 1 \} \cdot L_1) \ \cup \cdots \cup \ (\{ k \} \cdot L_k)
\]
(Here we define
$A \cdot B = \{ a \cdot b \mid a \in A \land b \in B \}$.)

We define selector functions on leafsets as follows.
If $s = \fs(s_1,\ldots,s_k)$ and $L \subseteq \leaves{s}$, then
$\fl_i(\tu{s,L}) = \tu{s_i,L_i}$ where
$L_i \subseteq \leaves{s_i}$ is defined by
\[
    L_i = \{ w \mid w \cdot i \in A \}
\]
Equivalently, we require that
\[
   \fl_i(\fl(\tu{s_1,L_1},\ldots,\tu{s_n,L_n})) = \tu{s_i,L_i}
\]
We can now express relations $r'$ in Figure~\ref{fig:termPowerOps}
using the fact:
\begin{equation} \label{eqn:expressingPrimes}
\begin{array}{l}
   r'(t_1,\ldots,t_k)  \iff \mnl
\begin{array}[t]{l}
  \termShape{t_2}=\termShape{t_1} \ \land \ldots \land \ \termShape{t_k}=\termShape{t_1}
  \ \land \mnl 
  r_{\termShape{t_1}}(t_1,\ldots,t_k) = \strueI{\termShape{t_1}}
\end{array}
\end{array}
\end{equation}

To handle an infinite number of elements of the base
structure $\ThC$, we do not introduce into the language
constants for every element of $C$ as in
Section~\ref{sec:nconst}.  Instead, we introduce the
predicate $\IsPrim :: \termSort \to \boolSort$ called
\emph{primitive-term test} that checks whether a term is a
constant:
\begin{equation*}
  \IsPrim(x) = (x \in C)
\end{equation*}
and the predicate $\IsPrimL :: \leafsetSort \to \boolSort$ called
\emph{primitive-leafset test}:
\begin{equation*}
  \IsPrimL(\tu{s,L}) = (s = \cs)
\end{equation*}
Instead of the rule~(\ref{eqn:IsfPartit}), we have
for $f, g \in \FreeSig \cup \{ \PRIM \}$:
\begin{equation} \label{eqn:IsfPartitMany}
\begin{array}{l}
  \forall x.\ \bigvee\limits_{f \in \FreeSig \cup \{ \PRIM \}} \Is{f}(x) \mnlb
  \forall x.\ \lnot (\Is{f}(x) \land \Is{g}(x)), \qquad \mbox{ for $f \not\equiv g$}
\end{array}
\end{equation}
Analogous rules hold for term algebra of leafsets:
\begin{equation} \label{eqn:IsfPartitManyL}
\begin{array}{l}
  \forall x.\ \bigvee\limits_{f \in \FreeSig \cup  \{ \PRIM \}} \Is{\fl}(x) \mnlb
  \forall x.\ \lnot (\Is{\fl}(x) \land \Is{\gl}(x)), \qquad \mbox{ for $f \not\equiv g$}
\end{array}
\end{equation}
Term algebra of shapes satisfies the original
rules~(\ref{eqn:IsfPartit}) of term algebra.

\subsection{A Logic for Term-Power Algebras}
\label{sec:logicTermPower}

\begin{figure*}
\newcommand{\tit}[1]{\multicolumn{3}{l}{{\mbox{\sl #1:}}}\mnl}
\newcommand{\btit}[1]{\multicolumn{3}{c}{{\mbox{\bf #1}}}\mnl}
\newcommand{\nextTablePart}{\\[2ex] \hline \\}
\[
\begin{array}{c|c}
\begin{array}{rcl}
\btit{per-shape product structure}
%--------------------------------------------------
\tit{inner formula relations for $r \in \lanC$}
   \sr{\_} &::& \shapeSort \times \termSort^k \to \leafsetSort \mnl
%--------------------------------------------------
\tit{inner logical connectives}
   \slandI{\_}, \slorI{\_}  %, \impliesI 
     &::& \shapeSort \times \leafsetSort \times \leafsetSort \to \leafsetSort \mnl
   \slnotI{\_} &::& \leafsetSort \to \leafsetSort \mnl
   \strueI{\_}, \sfalseI{\_} &::& \leafsetSort \mnl
%--------------------------------------------------
\tit{inner formula quantifiers}
  \sexistsI{\_}, \sforallI{\_} &::&
   \shapeSort \times (\termSort \to \leafsetSort) \to \leafsetSort \mnl
%--------------------------------------------------
\tit{leafset equality}
  \leafsetEq &::& 
     \leafsetSort \times \leafsetSort \to \boolSort \mnl
%--------------------------------------------------
\tit{leafset cardinality constraints, $k \geq 0$}
  |\_|_{\_} \geq k, \ |\_|_{\_} = k &::& 
     \shapeSort \times \leafsetSort \to \boolSort \mnl
%--------------------------------------------------
\tit{leafset quantifiers}
  \existsl, \foralll &::&
     (\leafsetSort \to \boolSort) \to \boolSort \mnl
%--------------------------------------------------
\tit{term equality}
  \termEq &::& \termSort \times \termSort \to \boolSort \mnl
%--------------------------------------------------
\tit{term quantifiers}
  \exists, \forall &::&
     (\termSort \to \boolSort) \to \boolSort \mnl
%--------------------------------------------------
\tit{shape equality}
  \shapeEq &::& \shapeSort \times \shapeSort \to \boolSort \mnl
%--------------------------------------------------
\tit{shape quantifiers}
  \existsS, \forallS &::&
     (\shapeSort \to \boolSort) \to \boolSort \mnl
%--------------------------------------------------
\tit{logical connectives}
   \land_, \lor &::& \boolSort \times \boolSort \to \boolSort \mnl
   \lnot &::& \boolSort \to \boolSort \mnl
   \boolTrue, \boolFalse, \boolUndef &::& \boolSort \mnl
\end{array}
%--------------------------------------------------
%--------------------------------------------------
& \begin{array}{rcl}
\btit{term algebra on terms}
\tit{constructors, $f \in \FreeSig$}
f &::& \termSort^k \to \termSort \mnl
\tit{constructor test, $f \in \FreeSig$}
\Is{f} &::& \termSort \to \boolSort \mnl
\tit{primitive-term test}
\IsPrim &::& \termSort \to \boolSort \mnl
\tit{selectors, $f \in \FreeSig$}
f_i &::& \termSort \to \termSort \mnl
\tit{term shape}
\termShapeF &::& \termSort \to \shapeSort \nextTablePart
%--------------------------------------------------
%--------------------------------------------------
\btit{term algebra on leafsets}
\tit{constructors, $f \in \FreeSig$}
\fl &::& \leafsetSort^k \to \leafsetSort \mnl
\tit{constructor test, $f \in \FreeSig$}
\Isfl &::& \leafsetSort \to \boolSort \mnl
\tit{primitive-leafset test}
\IsPrimL &::& \leafsetSort \to \boolSort \mnl
\tit{selectors, $f \in \FreeSig$}
\fl_i &::& \leafsetSort \to \leafsetSort \mnl
\tit{leafset shape}
\leafsetShapeF &::& \leafsetSort \to \shapeSort \nextTablePart
%--------------------------------------------------
%--------------------------------------------------
\btit{term algebra on shapes}
\tit{constructors, $f \in \FreeSigConst$}
\fs &::& \shapeSort^k \to \shapeSort \mnl
\tit{constructor test, $f \in \FreeSigConst$}
\Isfs &::& \shapeSort \to \boolSort \mnl
\tit{selectors, $f \in \FreeSig$}
\fs_i &::& \shapeSort \to \shapeSort \mnl
\end{array}
\end{array}
\]
\caption{Operations and relations in structure $\ThP$
\label{fig:nonrecLogic}}
\end{figure*}

To show the decidability of the first-order theory
of the structure $\termPower$ with operations in
Figure~\ref{fig:termPowerOps}, we show decidability for a
richer structure.  Figure~\ref{fig:nonrecLogic} shows the
operations and relations of this richer structure.

The structure has four sorts: $\boolSort$ representing truth
values, $\termSort$ representing terms, $\shapeSort$
representing shapes, and $\leafsetSort$ representing sets of
leaves within a given shape.  The structure can be seen as
as a combination of the operations of
Figure~\ref{fig:twoSortedSig} and
Figure~\ref{fig:productStructureLogic}.

%% As in Section \ref{sec:twoConstLogic}, our logic includes
%% sorts $\termSort$, $\shapeSort$, and $\boolSort$.  
For each
relation symbol $r \in R$ we define a relation symbol
$\termize{r}$ of sort $\shapeSort \times \termSort^k \to \boolSort$
acting on terms of the same shape.  While in Section
\ref{sec:twoConstLogic} we associate a boolean algebra with
the terms of same shape, in this section we associate a
cylindric algebra \cite{HenkinETAL71CylindricAlgebrasPartI}
with terms of the same shape.  This is a particularly simple
cylindric algebra resulting from lifting first-order logic
on the base structure $\ThC$ so that elements are replaced
by terms of a given shape (which are isomorphic to functions
from leaves to elements), and boolean values are replaced by
sets of leaves (isomorphic to functions from leaves to
booleans).  In both cases, operations on the set $X$
are lifted to operations on the set $\leaves{s} \to X$.
Syntactically, we introduce a copy of all
propositional connectives and quantifiers: $\slandI{\_}$,
$\slorI{\_}$, $\slnotI{\_}$, $\strueI{\_}$, $\sfalseI{\_}$.
Like boolean algebra operations in Figure~\ref{fig:twoSortedSig},
these syntactic
constructs in Figure~\ref{fig:nonrecLogic}
take an additional shape argument, because
term-power algebra contains one copy of a strong power
$\ThC^n$ of base structure for each shape.  We call formulas
built using the operations of the cylindric algebra 
{\em inner formulas}.

For each operation in Figure~\ref{fig:productStructureLogic}
there is an operation in Figure~\ref{fig:nonrecLogic},
potentially taking a shape as an additional argument (for
operations used to build inner formulas).  The logic further
contains term algebra operations on terms, leafsets, and
shapes.

We use undecorated identifiers (e.g.\ $u$) to denote
variables of $\termSort$ sort, variables with superscript $S$
to denote shape variables (e.g.\ $\us$) and variables with
superscript $L$ to denote leafset variables (e.g.\ $\ul$).

\begin{figure*}
\newcommand{\tit}[1]{\multicolumn{3}{l|}{{\mbox{\sl #1:}}} & \ \mnl}
\newcommand{\btit}[1]{\multicolumn{3}{c|}{{\mbox{\bf #1}}} &\mnl}
\newcommand{\btitT}[1]{\multicolumn{3}{c}{{\mbox{\bf #1}}}\mnl}
\newcommand{\nextTablePart}{\\[2ex] \hline \\}
\[
\begin{array}{c}
\begin{array}{rcl}
\btitT{interpretation of sorts}
\interpret{\termSort} &=& \FTExt \mnl
\interpret{\shapeSort} &=& \FTConst \mnl
\interpret{\leafsetSort} &=& \{ \tu{s,L} \mid L \subseteq \leaves{s} \} \mnl
\interpret{\boolSort} &=& \{ \boolTrue, \boolFalse, \boolUndef \} \mnl
\end{array}
\\
\\
\begin{array}{rcl|l}
\multicolumn{3}{c|}{\bf semantics} & 
\mbox{\bf well-definedness} \\ \hline
%\btit{per-shape product structure}
%--------------------------------------------------
\tit{inner formula relations for $r \in \lanC$}
\interpret{r}(s,t_1,\ldots,t_k) &=&
  \tu{s, \{ l \mid \interpretS{r}{C}(t_1[l],\ldots,t_k[l]) \}} &
\termShape{t_1}=s \land \ldots \land \termShape{t_k}=s \mnl
%--------------------------------------------------
\tit{inner logical connectives}
\interpret{\landI}(s,\tu{s_1,L_1},\tu{s_2,L_2}) &=&
    \tu{s,L_1 \cap L_2} & s_1 = s \land s_2 = s \mnl
\interpret{\lorI}(s,\tu{s_1,L_1},\tu{s_2,L_2}) &=&
    \tu{s,L_1 \cup L_2} & s_1 = s \land s_2 = s \mnl
\interpret{\lnotI}(s,\tu{s_1,L_1}) &=&
    \tu{s,\leaves{s} \setminus L_1} & s_1 = s \mnl
\interpret{\trueI}(s) &=&
    \tu{s,\leaves{s}} \mnl
\interpret{\falseI}(s) &=&
    \tu{s,\emptyset} \mnl
%--------------------------------------------------
\tit{inner formula quantifiers, for
$h : \interpret{\termSort} \to \interpret{\leafsetSort}$}
\interpret{\existsI}(s,h) &=& 
\tu{s,
\bigcup \{ L \mid \exists t \in \interpret{\termSort}.\ 
               \termShape{t} = s \land h(t) = \tu{s,L} \}
} &
\forall t \in \interpret{\termSort}.\ \leafsetShape{h(t)} = s \mnl
\interpret{\forallI}(s,h) &=& 
\tu{
\bigcap \{ L \mid \exists t \in \interpret{\termSort}.\ 
               \termShape{t} = s \land h(t) = \tu{s,L} \},
} &
\forall t \in \interpret{\termSort}.\ \leafsetShape{h(t)} = s \mnl
%--------------------------------------------------
\tit{leafset equality}
\interpret{\leafsetEq}(\tu{s_1,L_1},\tu{s_2,L_2}) &=&
  s_1 = s_2 \land L_1 = L_2 \mnl
%--------------------------------------------------
\tit{leafset cardinality constraints}
  \interpret{|\tu{s_1,L_1}|_s \geq k} &=&
  (|L_1| \geq k) &
  s_1 = s \mnl
  \interpret{|\tu{s_1,L_1}|_s = k} &=&
  (|L_1| = k) &
  s_1 = s \mnl
%--------------------------------------------------
%% \tit{logical connectives}
%%    \land_, \lor &::& \boolSort \times \boolSort \to \boolSort \mnl
%%    \lnot &::& \boolSort \to \boolSort \mnl
%%    \boolTrue, \boolFalse &::& \boolSort \mnl
%--------------------------------------------------
\tit{leafset quantifiers, for 
$h : \interpret{\leafsetSort} \to \interpret{\boolSort}$}
\interpret{\existsl} h &=&
  \exists \tu{s,t} \in \interpret{\leafsetSort}.\ h(\tu{s,t}) \mnl
\interpret{\foralll} h &=&
  \forall \tu{s,t} \in \interpret{\leafsetSort}.\ h(\tu{s,t}) \mnl
%--------------------------------------------------
\tit{term equality}
\interpret{=}(t_1,t_2) &=& 
  (t_1 = t_2) \mnl
%--------------------------------------------------
\tit{term quantifiers, for
$h : \interpret{\termSort} \to \interpret{\boolSort}$}
\interpret{\exists} h &=&
  \exists t \in \interpret{\termSort}.\ h(t) \mnl
\interpret{\forall} h &=&
  \forall t \in \interpret{\termSort}.\ h(t) \mnl
%--------------------------------------------------
\tit{shape equality}
\interpret{\shapeEq}(\ts_1,\ts_2) &=& 
  (\ts_1 = \ts_2) \mnl
%--------------------------------------------------
\tit{shape quantifiers, for
$h : \interpret{\shapeSort} \to \interpret{\boolSort}$}
\interpret{\existsS} h &=&
  \exists t \in \interpret{\shapeSort}.\ h(t) \mnl
\interpret{\forallS} h &=&
  \forall t \in \interpret{\shapeSort}.\ h(t) \mnl
\end{array}
\end{array}
\]
\caption{Semantics for Logic of Term-Power Algebra (Part I)
\label{fig:termPowerSemanticsOne}}
\end{figure*}

\begin{figure*}
\newcommand{\tit}[1]{\multicolumn{3}{l|}{{\mbox{\sl #1:}}} & \ \mnl}
\newcommand{\btit}[1]{\multicolumn{3}{c|}{{\mbox{\bf #1}}} &\mnl}
\newcommand{\btitT}[1]{\multicolumn{3}{c}{{\mbox{\bf #1}}}\mnl}
\newcommand{\nextTablePart}{\\[2ex] \hline \\}
\[
\begin{array}{rcl|l}
\multicolumn{3}{c|}{\bf semantics} & 
\mbox{\bf well-definedness} \\ \hline
\btit{term algebra on terms}
\tit{constructors, $f \in \FreeSig$}
\interpret{f}(t_1,\ldots,t_k) &=& 
  f(t_1,\ldots,t_k) \mnl
\tit{constructor test, $f \in \FreeSig$}
\interpret{\Is{f}}(t) &=&
  \exists t_1,\ldots,t_k.\ t = f(t_1,\ldots,t_k) \mnl
\tit{primitive-term test}
\interpret{\IsPrim}(t) &=&
  (t \in C) \mnl
\tit{selectors, $f \in \FreeSig$}
\interpret{f_i}(t) &=&
  \epsilon t_i.\ t=f(t_1,\ldots,t_i,\ldots,t_k) &
  \interpret{\Is{f}}(t) \mnl
\tit{term shape}
\interpret{\termShape{f(t_1,\ldots,t_n)}} &=&
  \shapified{f}(\termShape{t_1},\ldots,\termShape{t_n}) \mnl
%--------------------------------------------------
%--------------------------------------------------
\btit{term algebra on leafsets}
\tit{constructors, $f \in \FreeSig$}
\interpret{\fl}(\tu{s_1,L_1},\ldots,\tu{s_k,L_k}) &=&
  \tu{f(s_1,\ldots,s_k),(\{ 1 \} \cdot L_1) \ \cup \cdots \cup \ (\{ k \} \cdot L_k)} \mnl
\tit{constructor test, $f \in \FreeSig$}
\interpret{\Isfl}(\tu{s,L}) &=&
  \exists s_1,L_1,\ldots,s_k,L_k.\ \tu{s,L} = \interpret{\fl}(\tu{s_1,L_1},\ldots,\tu{s_k,L_k}) \mnl
\tit{primive-leafset test}
\interpret{\IsPrimL}(\tu{s,L}) &=&
  (s = \cs)
  \mnl
\tit{selectors, $f \in \FreeSig$}
\interpret{\fl_i}(\tu{s,L}) &=& 
  \epsilon \tu{s_i,L_i}.\ \tu{s,L} = 
      \interpret{\fl}(\tu{s_1,L_1},\ldots,\tu{s_i,L_i},\ldots,\tu{s_k,L_k}) &
  \interpret{\Isfl}(\tu{s,L}) \mnl
\tit{leafset shape}
\interpret{\leafsetShapeF}(\tu{s,L}) &=& s \mnl
%--------------------------------------------------
%--------------------------------------------------
\btit{term algebra on shapes}
\tit{constructors, $f \in \FreeSig$}
\interpret{\fs}(s_1,\ldots,s_k) &=&
  \fs(s_1,\ldots,s_k) \mnl
\tit{constructor test, $f \in \FreeSigConst$}
\interpret{\Isfs}(s) &=&
  \exists s_1,\ldots,s_k.\ s = \fs(s_1,\ldots,s_k) \mnl
\tit{selectors, $f \in \FreeSig$}
\interpret{\fs_i}(s) &=&
  \epsilon s_i.\ s = \fs(s_1,\ldots,s_i,\ldots,s_k) &
  \interpret{\Isfs}(s) \mnl
\end{array}
\]
\caption{Semantics for Logic of Term-Power Algebra (Part II)
\label{fig:termPowerSemanticsTwo}}
\end{figure*}

Figures~\ref{fig:termPowerSemanticsOne}
and~\ref{fig:termPowerSemanticsTwo} show the semantics of
logic in Figure~\ref{fig:nonrecLogic}.  The first row
specifies semantics of operations in the case when all
arguments are defined and are in the domain of the
operation.  The domain of each operation is in the second
column, it is omitted if it is equal to the entire domain
resulting from interpreting the sort of the operation.  All
operations except for plain logical operations and
quantifiers over the $\boolSort$ domain are strict.  Logical
operations and quantifiers over the $\boolSort$ domain are
defined as in the three-valued logic of
Section~\ref{sec:partFun}.

We remark that values of $\leafsetSort$ act as terms with two
constants in Figure~\ref{fig:twoSortedSig}.  In fact, if the
base structure $\ThC$ has only two constants then the formula
$x=a$ and its
propositional combinations are sufficient to express all
facts about $\ThC$, so in that case there is no need to
distinguish between terms and leafsets.

%% To make it easier to check well-formedness in the logic, we
%% let leafset variables range over pairs $\tu{s,L}$ where $s$
%% is a shape and $L$ is a set of leaves applicable to $s$.
%% Operations on sets are well-defined iff the shapes of
%% arguments match appropriately.

\subsection{Some Properties of Term-Power Structure}

In this section we establish some further properties of the
term-power structure, including the homomorphism properties
between the term algebra of terms and the term algebra of
leafsets.  We also argue that it suffices to consider a
restricted class of formulas called \emph{simple formulas}.

Recall that $\req \in R$ is the equality relation on $C$.
Given $\req$, we can express the equality between terms by:
\begin{equation}  \label{eqn:expressingEquality}
\begin{array}{rcl}
  t_1 = t_2 & \iff & {\req}'(t_1,t_2) \mnl
         & \iff  &
 \termShape{t_2} = \termShape{t_1} \land \req(t_1,t_2) = \strueI{\termShape{t_1}}
\end{array}
\end{equation}

We define the notion of a $\us$-term as in
Definition~\ref{def:usterms} except that we use different
symbols for boolean algebra operations.
\begin{definition}[$\us$-terms] \label{def:nonrecUsterms}
  Let $\us \in \SVar$ be a shape variable.  The set of
  $\us$-terms $\usterm{\us}$ is the least set such that:
  \begin{enumerate}
  \item $\ul \in \usterm{\us}$ for every leafset variable $\ul$;
  \item $\sfalseI{\us}, \strueI{\us} \in \usterm{\us}$;
  \item if $\tl_1,\tl_2 \in \usterm{\us}$, then also
    \[\begin{array}{l}
      \tl_1 \slandI{\us} \tl_2 \in \usterm{\us}, \mnl
      \tl_1 \slorI{\us} \tl_2 \in \usterm{\us}, \mbox{ and} \mnl
      \slnotI{\us} \tl_1 \in \usterm{\us}
    \end{array}\]
  \end{enumerate}
\end{definition}
If $\ts$ is a term of shape sort, the notion of $\ts$-inner
formula is defined as follows.
\begin{definition}[$\us$-inner formula]
  Let $\us \in \SVar$ be a shape variable.  The set of
  $\us$-inner formulas $\usinner{\us}$ is the least set such
  that:
  \begin{enumerate}
  \item if $u_1,\ldots,u_k$ are term variables and $r \in \lanC$ such
    that $\ar(r)=k$, then
    \[
    r_{\us}(u_1,\ldots,u_k) \in \usinner{\us}
    \]
    
  \item $\sfalseI{\us}, \strueI{\us} \in \usinner{\us}$
    
  \item if $\phi_1,\phi_2 \in \usinner{\us}$ then also
    \[\begin{array}{l}
      \phi_1 \slandI{\us} \phi_2 \in \usinner{\us} \mnl
      \phi_1 \slorI{\us} \phi_2 \in \usinner{\us} \mnl
      \slnotI{\us} \phi_1 \in \usinner{\us}
    \end{array}\]
    
  \item if $\phi \in \usinner{\us}$ and $u$ is a term variable
    that does not occur in $\us$, then also
    \[\begin{array}{l}
      \sexistsI{\us} u. \phi \ \in\ \usinner{\us} \mnl
      \sforallI{\us} u. \phi \ \in\ \usinner{\us}
    \end{array}\]
    
  \end{enumerate}
  If $\phi \in \usinner{\us}$ and $u_1,\ldots,u_n$ is the set of free
  term variables of $\phi$, we write $\phi(\us,u_1,\ldots,u_n)$ for
  $\phi$.  Furthermore, if $\ts$ is a term of $\shapeSort$ sort
  and $t_1,\ldots,t_n$ terms of $\termSort$ sort, we write
  $\phi(\ts,t_1,\ldots,t_n)$ for 
  \[
      \phi[\us:=\ts,u_1:=t_1,\ldots,u_n:=t_n]
  \]
  where we assume that variables bound by $\sexistsI{\_}$ and
  $\sforallI{\_}$ are renamed to avoid the capture of
  variables that are free in $\ts,t_1,\ldots,t_n$.  
  
  We call $\phi(\ts,t_1,\ldots,t_n)$ an \emph{instance} of the
  $\us$-inner formula $\phi(\us,u_1,\ldots,u_n)$.
  
  If $\phi(\us,u_1,\ldots,u_n)$ is an inner formula, we abbreviate
  it by writing $[\phi'(u_1,\ldots,u_n)]_{\us}$ where $\phi'$ results
  from $\phi(\us,u_1,\ldots,u_n)$ by omitting the shape argument
  $\us$ from the operations occurring in $\phi(\us,u_1,\ldots,u_n)$.
  Similarly, we write $[\phi'(t_1,\ldots,t_n)]_{\ts}$ for
  $\phi(\ts,t_1,\ldots,t_n)$.
\end{definition}

According to the semantics in
Figure~\ref{fig:termPowerSemanticsOne}, $\termShapeF$ is a
homomorphism from the term algebra of terms to the term
algebra of shapes.  In addition, $\leafsetShapeF$ is a
homomorphism from the term algebra of leafsets to the term
algebra of shapes.

We also have the following important property.  Let $r \in
\lanC$ be a relation symbol of arity $n$, let $f \in \FreeSig$
be a function symbol of arity $k$, and let
\[
   \termShape{t_{1j}} = \ldots = \termShape{t_{nj}} = s_j
\]
for $1 \leq j \leq k$.  If $\fs = \shapified{f}$, $\fl =
\leafified{f}$, and $s = \fs(s_1,\ldots,s_k)$ then
\begin{equation} \label{eqn:leafsetTermHomRel}
\begin{array}{l}
   r_s(f(t_{11},\ldots,t_{1k}),\ldots,f(t_{n1},\ldots,t_{nk})) = \mnl
\qquad  \fl(r_{s_1}(t_{11},\ldots,t_{n1}),\ldots,r_{s_k}(t_{1k},\ldots,t_{nk}))
\end{array}
\end{equation}
Furthermore, if $\leafsetShape{l_j} =
\leafsetShape{l'_j}=s_j$ for $1 \leq j \leq k$ and $s=\fs(s_1,\ldots,s_k)$
then
\begin{equation} \label{eqn:leafsetTermHomPropagate}
\begin{array}{l}
% ---------------- \land ----------------------
\fl(l_1,\ldots,l_k) \slandI{s} \fl(l'_1,\ldots,l'_k) \leafsetEq \mnl
\qquad 
\fl(l_1 \slandI{s_1} l'_1,\ldots,l_k \slandI{s_k} l'_k) \mnl
% ---------------- \lor ----------------------
\fl(l_1,\ldots,l_k) \slorI{s} \fl(l'_1,\ldots,l'_k) \leafsetEq \mnl
\qquad 
\fl(l_1 \slandI{s_1} l'_1,\ldots,l_k \slandI{s_k} l'_k) \mnl
% ---------------- \lnot ----------------------
\slnotI{s} \fl(l_1,\ldots,l_k) \leafsetEq \mnl
\qquad
\fl(\slnotI{s_1} l_1,\ldots,\slnotI{s_k} l_k) \mnl
% ---------------- \exists ----------------------
\sexistsI{s} t. \fl(h_1(t),\ldots,h_k(t)) \leafsetEq \mnl
\qquad
\fl(\sexistsI{s_1} t. h_1(t),\ldots,\sexistsI{s_k} t. h_k(t)) \mnl
% ---------------- \forall ----------------------
\sforallI{s} t. \fl(h_1(t),\ldots,h_k(t)) \leafsetEq \mnl
\qquad
\fl(\sforallI{s_1} t. h_1(t),\ldots,\sforallI{s_k} t. h_k(t))
\end{array}
\end{equation}
From these properties by induction we conclude that
if $\phi(\us,u_1,\ldots,u_n)$ is an inner formula, then
\begin{equation} \label{eqn:nonrecDecompose}
\begin{array}{l}
   \phi(s,f(t_{11},\ldots,t_{1k}),\ldots,f(t_{n1},\ldots,t_{nk})) = \mnl
\qquad  \fl(\phi(s_1,t_{11},\ldots,t_{n1}),\ldots,\phi(s_k,t_{1k},\ldots,t_{nk}))
\end{array}
\end{equation}
Let $\phi(\us,u_1,\ldots,u_n)$ be an inner formula and let
$\phi'(u_1,\ldots,u_n)$ be a first-order formula that results from replacing
operations $\slandI{s},\slorI{s},\slnotI{s}$,
$\sforallI{s},\sexistsI{s}$ by $\land,\lor,\lnot$, $\forall,\exists$.  Interpreting
$\phi'(u_1,\ldots,u_n)$ over the structure $\ThC$ yields a relation
$\rho' \subseteq C^n$.  If
\[
    \termShape{t_1} = \ldots = \termShape{t_k} = s
\]
then
\[
    \interpret{\phi}(s,t_1,\ldots,t_k) = \tu{s,\{ l \mid \rho'(t_1[l],\ldots,t_k[l]) \}}
\]

The following Definition~\ref{def:nonrecSimpleFormulas}
introduces a more restricted set of formulas than the set of
formulas permitted by sort declarations in
Figure~\ref{fig:nonrecLogic}.  We call this restricted set
of formulas \emph{simple formulas}.  One of the main
properties of simple formulas compared to arbitrary formulas
is that simple formulas allow the use of operations
$\sexistsI{\_}, \sforallI{\_}$, and relations $\sr{\_}$, $r \in
\lanC$ only within instances of $\us$-inner formulas.

\begin{definition}
  A simple operation is any operation or relation in
  Figure~\ref{fig:nonrecLogic} except for operations
  $\sexistsI{\_}, \sforallI{\_}$, and relations $\sr{\_}$ for $r \in
  \lanC$.
\end{definition}

\begin{definition}[Simple Formulas] \label{def:nonrecSimpleFormulas}
  The set of simple formulas is the least set that satisfies
  the following.
  \begin{enumerate}
  \item if $\phi(\us,u_1,\ldots,u_n)$ is a an inner formula,
    $\ts$ a term of $\shapeSort$
    sort, $t_1,\ldots,t_n$ terms of $\termSort$ sort
    and $\ul$ is a leafset variable, then 
    \[
         \ul \leafsetEq \phi(\ts,t_1,\ldots,t_n)
    \]
    is a simple formula.
  \item applying simple operations to simple formulas
    yields simple formulas.    
  \end{enumerate}
\end{definition}

\begin{example}
A formula
\begin{equation} \label{eqn:nonsimpleExa}
  \ul \leafsetEq \sexistsI{\us_1} u.\ r_{\us_2}(u,u)
\end{equation}
is not a simple formula for $\us_1 \not\equiv \us_2$.  Formula
\[\begin{array}{l}
  (\us_1 = \us_2
   \ \land \ 
   \ul \leafsetEq \sexistsI{\us_1} u.\ r_{\us_1}(u,u)) \ \lor  \mnl
  (\us_1 \neq \us_2 \land \boolUndef)
\end{array}\]
is a simple formula equivalent to
formula~(\ref{eqn:nonsimpleExa}).  We abbreviate
$\sexistsI{\us_1} u.\ r_{\us_1}(u,u)$
as $[\existsI u.\ r(u,u)]_{\us_1}$.
\end{example}

Lemma~\ref{lemma:nonrecSimplification} shows that for every
formula in the logic of Figure~\ref{fig:nonrecLogic} there
exists an equivalent simple formula.  Note that even simple
formulas are sufficient to express the relations of
structural subtyping.  A reader not interested in the
decidability of the more general logic of
Figure~\ref{fig:nonrecLogic} may therefore ignore
Lemma~\ref{lemma:nonrecSimplification}.

\begin{lemma}[Formula Simplification] \label{lemma:nonrecSimplification}
  For every well-defined formula in the logic of
  Figure~\ref{fig:nonrecLogic} there exists an equivalent
  well-defined simple formula.
\end{lemma}
\begin{proofsketch}
  According to the definition of simple formula, we need to
  ensure that every occurrence of quantifiers
  $\sforallI{\_},\sexistsI{\_}$ and relations $\sr{\_}$ is an
  occurrence in some inner-formula instance
  $\phi(\ts,t_1,\ldots,t_n)$.  Each occurrence $\sr{\ts}(t_1,\ldots,t_n)$
  is an inner formula instance by itself, so the main
  difficulty is fitting the quantifiers $\sforallI{\_}$ and
  $\sexistsI{\_}$ into inner formulas.  
  
  Let us examine the syntactic structure of formulas of
  logic in Figure~\ref{fig:nonrecLogic}.  This syntactic
  structure is determined by sort declarations.  Each
  expression of $\leafsetSort$ is formed starting from
  \begin{enumerate}
  \item relations $r \in \lanC$;
  \item $\leafsetSort$ variables;
  \item $\strueI{\_}, \sfalseI{\_}$
  \end{enumerate}
  using operations $\slandI{\_}$, $\slorI{\_}$, $\slnotI{\_}$, 
  $\sforallI{\_}$,
  $\sexistsI{\_}$, as well as $\fl$ and $\fl_i$.  
  The
  $\leafsetSort$ expressions can be used in a formula in the
  following ways (in addition to constructing new
  $\leafsetSort$ expressions):
  \begin{enumerate}
  \item to compare for equality using $\leafsetEq$;
  \item to test for the top-level constructor using $\Isfl$;
  \item to form leafset cardinality constraints;
  \item to form a shape using $\leafsetShapeF$.
  \end{enumerate}
  
  Because the top-level sort of a formula is $\boolSort$,
  every term $\tl_0$ of sort leafset occurs within some
  formula $\tl_1 \leafsetEq \tl_2$ or $\Isfl(\tl)$,
  $|\tl|_{\ts} = k$, $|\tl|_{\ts} \geq k$ or as part of some
  term $\leafsetShape{\tl}$.  We can replace $\Isfl(\tl)$
  with
  \begin{equation*}
    \exists \ul.\ \ul \leafsetEq \tl \lsland \Isfl(\ul)
  \end{equation*}
  according to Lemma~\ref{lemma:generalUnnesting}, so we
  need not consider that case.  We can similarly eliminate
  non-variable leafset terms from cardinality constraints.
  If a leafset term $\tl$ occurs in an expression
  $\leafsetShape{\tl}$, we consider the smallest atomic
  formula $\psi(\leafsetShape{\tl})$ enclosing
  $\leafsetShape{\tl}$, and replace $\psi(\tl)$ with
  \begin{equation*}
    \exists \ul.\ \ul \leafsetEq \tl \lsland \psi(\ul)
  \end{equation*}
  This transformation is valid by
  Lemma~\ref{lemma:generalUnnesting} because $\psi$ and
  $\leafsetShapeF$ are strict.  

  We further assume that in every atomic formula $\tl_1
  \leafsetEq \tl_2$, the term $\tl_1$ is a leafset variable.
  
  Suppose that a term $\tl$ in a formula $\ul \leafsetEq
  \tl$ is not an instance of an inner formula.
  Then there are two possibilities.
  \begin{enumerate}
  \item There are some occurrences of leafset term algebra
    operations $\fl$, $\fl_i$ or leafset variables $\ul_1$
    in $\tl$.  Here by ``occurrence'' in $\tl$ we mean
    occurrence that is reachable without going through a
    shape argument or a relation, but only through
    operations $\sforallI{\_},\sexistsI{\_}$, $\slandI{\_},
    \slorI{\_}, \slnotI{\_}$.  For example, we ignore the
    occurrences of $\fl$, $\fl_i$ within terms $\ts$ that
    occur in $\slandI{\ts}$.
  
  \item not all shape arguments in
    $\sforallI{\_},\sexistsI{\_}$, $\slandI{\_}, \slorI{\_},
    \slnotI{\_}$, $\strueI{\_}, \sfalseI{\_}$, $\sr{\_}$
    occurring in $\tl$ are syntactically identical.
  \end{enumerate}
  
  We eliminate the first possibility by propagating leafset
  term algebra operations $\fl$, $\fl_i$ inwards until they
  reach expressions of form $\rl{\ts}(t_1,\ldots,t_n)$, applying
  the equations~(\ref{eqn:leafsetTermHomPropagate}) from
  left to right.  We then convert $\fl$, $\fl_i$ operations
  of term algebra of leafsets into operations of the term
  algebra of terms applying (\ref{eqn:leafsetTermHomRel})
  from right to left.
  
  To eliminate the second possibility, let $\ts_1,\ldots,\ts_n$
  be the occurrences (reachable through $\strueI{\_},
  \sfalseI{\_}$, $\slandI{\_}$, $\slorI{\_}$, $\slnotI{\_}$,
  $\sforallI{\_}$, $\sexistsI{\_}$) in term $\tl$ of the shape
  arguments of operations $\strueI{\_}, \sfalseI{\_}$,
  $\slandI{\_}$, $\slorI{\_}$, $\slnotI{\_}$, $\sforallI{\_}$,
  $\sexistsI{\_}$.  Then replace
  \begin{equation*}
    \ul = \tl(\ts_1,\ldots\ts_n)
  \end{equation*}
  with
  \newcommand{\ALL}{\forall^{\m{CL}}}
  \begin{equation*}
    \begin{array}{l}
      (\existsS \us.
      \begin{array}[t]{l}
        \ALL_1 (\us \shapeEq \ts_1) \lsland \ldots \lsland
        \ALL_n (\us \shapeEq \ts_n) \lsland \mnl
        \ul \leafsetEq \tl(\us,\ldots,\us)) \ \lor
      \end{array} \mnl
      (\boolUndef \ \land \ \bigvee_{1 \leq i < j \leq n} \ts_i \neq \ts_j)
    \end{array}      
  \end{equation*}
  Here $\ALL_i$ denotes universal quantification
  $\forall u_{i,1},\ldots,u_{i,n_i}$ where $u_{i,1},\ldots,u_{i,n_i}$ is a list of those term
  variables occurring in $\ts_i$ that are bound by some quantifier
  $\sexistsI{\_},\sforallI{\_}$ within $\tl$.
\end{proofsketch}
 
\subsection{Quantifier Elimination}

In this section we give a quantifier elimination procedure
for the term-power structure.  The procedure of this section
is applicable whenever $\ThC$ is a structure with a
decidable first-order theory.

Definition~\ref{def:nonrecStructuralBase} below
generalizes the notion of structural base formula
of Definition~\ref{def:structuralBase},
Section~\ref{sec:twoConstQE}.  There are two main
differences between Definition~\ref{def:structuralBase}
and the present Definition~\ref{def:nonrecStructuralBase}.

The first difference is the presence of three (instead of two)
base formulas: shape base, leafset base, and term base.
This difference is a consequence of the distinction between
leafsets and terms and is needed whenever base structure
$\ThC$ has more than two elements.  There is a homomorphism
formula relating leafset base formula to shape base formula
and a homomorphism formula relating term base formula to
shape base formula.  Furthermore, some of the leafset
variables are determined by term variables using inner
formula maps, which establishes the relationship between
term base formula and leafset base formula.  Cardinality
constraints now apply to leafset variables.

The second difference is the distinction between composed
and primitive non-parameter leafset and term variables.  A
composed non-parameter variable denotes a leafset or a term
whose shape $s$ has property $\Isfs(s)$ for some $f \in
\FreeSig$.  A primitive non-parameter variable denotes a
leafset or a term whose shape is $\cs$ and has property
$\IsPrim$ or $\IsPrimL$.  The purpose of this distinction is
to allow cardinality constraints and inner formula maps not
only on parameter variables, but also on primitive
non-parameter variables, which is useful when the base
structure $\ThC$ is decidable but infinite.

\begin{definition}[Structural Base Formula] \ \\
\label{def:nonrecStructuralBase}
A {\em structural base formula} with:
\begin{itemize}
\item free term variables $x_1,\ldots,x_m$;
\item internal composed non-parameter term variables $u_1,\ldots,u_r$;
\item internal primitive non-parameter term variables $u_{r+1},\ldots,u_p$;
\item internal parameter term variables $u_{p+1},\ldots,u_{p+q}$;
\item free leafset variables $\xl_1,\ldots,\xl_{\ml}$;
\item internal composed non-parameter leafset variables $\ul_1,\ldots,\ul_{\irl}$;
\item internal primitive non-parameter leafset
      variables $\ul_{\irl+1},\ldots,\ul_{\pl}$;
\item internal parameter leafset variables $u_{\pl+1},\ldots,u_{\pl+\ql}$;
\item free shape variables $\xs_1,\ldots,\xs_{\ms}$;
\item internal non-parameter shape variables $\us_1,\ldots,\us_{\ps}$;
\item internal parameter shape variables $\us_{\ps},\ldots,\us_{\ps+\qs}$
\end{itemize}
is a formula of form:
\[\begin{array}{l}
  \exists  u_1,\ldots,u_n,\ul_1,\ldots,\ul_{\nl}, \us_1,\ldots,\us_{\ns}. \mnl
\begin{array}[t]{l}
    \shapeBaseF(\us_1,\ldots,\us_{\ns},\xs_1,\ldots,\xs_{\ms}) \ \land \mnl
    \leafsetBaseF(\ul_1,\ldots,\ul_{\nl},\xl_1,\ldots,\xl_{\ml}) \ \land \mnl
    \leafsetHomF(\ul_1,\ldots,\ul_{\nl},\us_1,\ldots,\us_{\ns}) \ \land \mnl
    \termBaseF(u_1,\ldots,u_n,x_1,\ldots,x_m) \ \land \mnl
    \termHomF(u_1,\ldots,u_n,\us_1,\ldots,\us_{\ns}) \ \land \mnl
    \cardinF(\ul_{\irl+1},\ldots,\ul_{\nl},\us_{\ps+1},\ldots,\us_{\ns}) \ \land \mnl
    \innerMapF(u_{r+1},\ldots,u_n,\ul_{\irl+1},\ldots,\ul_{\nl},\us_{\ps+1},\ldots,\us_{\ns})
\end{array}
\end{array}\]
where $n=p+q$, $\nl = \pl + \ql$, $\ns = \ps + \qs$, and
formulas $\shapeBaseF$, $\leafsetBaseF$, $\termBaseF$,
$\leafsetHomF$, $\termHomF$, $\cardinF$, $\innerMapF$ are
defined as follows.
\[\begin{array}{l}
\shapeBaseF(\us_1,\ldots,\us_{\ns},\xs_1,\ldots,\xs_{\ms}) = \mnl
\qquad \bigwedge\limits_{i=1}^{\ps} \us_i = t_i(\us_1,\ldots,\us_{\ns}) \ \land\
       \bigwedge\limits_{i=1}^{\ms} \xs_i = \us_{j_i} \mnl
\qquad \land \ \distinctF(\us_1,\ldots,\us_n)
\end{array}\]
where each $t_i$ is a shape term of form $\fs(\us_{i_1},\ldots,\us_{i_k})$ for
some $f \in \FreeSigConst$, $k=\ar(f)$, and 
$j : \{ 1,\ldots,\ms \} \to \{ 1,\ldots,\ns\}$ is a function mapping
indices of free shape variables to indices of internal shape variables.

\[\begin{array}{l}
\leafsetBaseF(\ul_1,\ldots,\ul_{\nl},\xl_1,\ldots,\xl_{\ml}) = \mnl
 \qquad \begin{array}[t]{cl}
        \bigwedge\limits_{i=1}^{\irl} & \ul_i = t_i(\ul_1,\ldots,\ul_{\nl}) \ \land \mnl
        \bigwedge\limits_{i=\irl+1}^{\pl} & \IsPrimL(\ul_i) \ \land \mnl
        \bigwedge\limits_{i=1}^{\ml} & \xl_i = \ul_{j_i}
       \end{array}
\end{array}\]
where each $t_i$ is a term of form $f(u_{i_1},\ldots,u_{i_k})$
for some $f \in \FreeSig$, $k = \ar(f)$, and $j : \{ 1,\ldots,\ml \} \to
\{ 1,\ldots,\nl\}$ is a function mapping indices of free leafset
variables to indices of internal leafset variables.

\[\begin{array}{ll}
\leafsetHomF(\ul_1,\ldots,\ul_{\nl},\us_1,\ldots,\us_{\ns}) = &
  \bigwedge\limits_{i=1}^{\nl} \leafsetShape{\ul_i} = \us_{j_i}
\end{array}\]
where $j : \{ 1,\ldots,\nl \} \to \{ 1,\ldots,\ns\}$ is some function such
that $\{ j_1,\ldots,j_p \} \subseteq \{ 1,\ldots,\ps \}$ and $\{ j_{\pl+1},\ldots,j_{\pl+\ql}
\} \subseteq \{ \ps+1,\ldots,\ps+\qs \}$ 
(a leafset variable is a parameter
variable iff its shape is a parameter shape variable).

\[\begin{array}{l}
\termBaseF(u_1,\ldots,u_n,x_1,\ldots,x_m) = \mnl
 \qquad \begin{array}[t]{cl}
        \bigwedge\limits_{i=1}^r & u_i = t_i(u_1,\ldots,u_n) \ \land \mnl
        \bigwedge\limits_{i=r+1}^p & \IsPrim(u_i) \ \land \mnl
        \bigwedge\limits_{i=1}^m & x_i = u_{j_i}
      \end{array}
\end{array}\]
where each $t_i$ is a term of form $f(u_{i_1},\ldots,u_{i_k})$ for
some $f \in \FreeSig$, $k = \ar(f)$, and 
$j : \{ 1,\ldots,m \} \to \{ 1,\ldots,n\}$ is a function mapping
indices of free term variables to indices of internal term variables.

\[\begin{array}{ll}
\termHomF(u_1,\ldots,u_n,\us_1,\ldots,\us_{\ns}) = &
  \bigwedge\limits_{i=1}^n \termShape{u_i} = \us_{j_i}
\end{array}\]
where $j : \{ 1,\ldots,n \} \to \{ 1,\ldots,\ns\}$ is some function such
that $\{ j_1,\ldots,j_p \} \subseteq \{ 1,\ldots,\ps \}$ and $\{ j_{p+1},\ldots,j_{p+q}
\} \subseteq \{ \ps+1,\ldots,\ps+\qs \}$ 
(a term variable is a parameter
variable iff its shape is a parameter shape variable).

\[
\cardinF(\ul_{\irl+1},\ldots,\ul_{\nl},\us_{\ps+1},\ldots,\us_{\ns}) = \psi_1 \land \cdots \land \psi_d
\]
where each $\psi_i$ is of form
\[
   |\tl(\ul_{\irl+1},\ldots,\ul_{\nl})|_{\us}=k
\]
or
\[
   |\tl(\ul_{\irl+1},\ldots,\ul_{\nl})|_{\us}\geq k
\]
for some $\us$-term
$\tl(\ul_{\irl+1},\ldots,\ul_{\nl})$ that contains no variables other
than some of the variables $\ul_{\irl+1},\ldots,\ul_{\nl}$,
and the following
condition holds:
\begin{equation} \label{eqn:nonrecWellDefinednessCondition}
\begin{minipage}{2.5in}
If a variable $\ul_j$ for $\irl+1 \leq j \leq \nl$ occurs in the term
$\tl(\ul_{\irl+1},\ldots,\ul_{\nl})$,
then $\leafsetShape{\ul_j}=\us$
occurs in formula $\leafsetHomF$.
\end{minipage}
\end{equation}

\begin{equation*}
  \begin{array}{l}
    \innerMapF(u_{r+1},\ldots,u_n,\ul_{\irl+1},\ldots,\ul_{\nl},\us_{\ps+1},\ldots,\us_{\ns}) = \mnl
    \qquad \eta_1 \land \cdots \land \eta_e
  \end{array}
\end{equation*}
where each $\eta_i$ is of form
\[
   \ul_j = \phi^I(\us,u_{i_1},\ldots,u_{i_k})
\]
for some inner formula $\phi^I(\us,u_{i_1},\ldots,u_{i_k}) \in
\usinner{\us}$ where $\rl+1 \leq j \leq \nl$ i.e.\ $\ul_j$ is a
primitive non-parameter leafset variable or parameter leafset
variable, $\{u_{i_1},\ldots,u_{i_k}\} \subseteq \{ u_{r+1},\ldots,u_n \}$ are
primitive non-parameter term variables and parameter
variables, the conjunct $\leafsetShape{\ul}=\us$ occurs in
$\leafsetHomF$, and the following condition holds:
\begin{equation} \label{eqn:nonrecInnerWellDefinednessCondition}
\begin{minipage}{2.5in}
$\termShape{u_{i_j}}=\us$ occurs in formula $\termHomF$
for every $j$ where $1 \leq j \leq k$.
\end{minipage}
\end{equation}
We require each structural base formula to satisfy the following
conditions:
\begin{enumerate}

\item[P0)] the graph associated with shape base formula 
      \[
         \exists  \us_1,\ldots,\us_{\ns}.\
            \shapeBaseF(\us_1,\ldots,\us_{\ns},\xs_1,\ldots,\xs_{\ms})
      \]
      is acyclic (compare to Definition~\ref{def:baseFormula});

\item[P1)] congruence closure property for $\shapeBaseF$ subformula:
      there are no two distinct variables $\us_i$ and
      $\us_j$ such that both $\us_i =
      f(\us_{l_1},\ldots,\us_{l_k})$ and $\us_j =
      f(\us_{l_1},\ldots,\us_{l_k})$ occur as conjuncts in
      formula $\shapeBaseF$;

\item[P2)] congruence closure property for $\leafsetBaseF$ subformula:
      there are no two distinct variables $\ul_i$ and $\ul_j$
      such that both $\ul_i = \fl(\ul_{l_1},\ldots,\ul_{l_k})$ and $\ul_j =
      \fl(\ul_{l_1},\ldots,\ul_{l_k})$ occur as conjuncts in formula
      $\leafsetBaseF$;

\item[P3)] congruence closure property for $\termBaseF$ subformula:
      there are no two distinct variables $u_i$ and $u_j$
      such that both $u_i = f(u_{l_1},\ldots,u_{l_k})$ and $u_j =
      f(u_{l_1},\ldots,u_{l_k})$ occur as conjuncts in formula
      $\termBaseF$;

\item[P4)] homomorphism property of $\leafsetShapeF$:
      for every non-parameter leafset variable $\ul$ such that 
      $\ul = \fl(\ul_{i_1},\ldots,\ul_{i_k})$ occurs in $\leafsetBaseF$, if
      conjunct $\leafsetShape{\ul}=\us$ occurs in $\leafsetHomF$,
      then for some shape variables
      $\us_{j_1},\ldots,\us_{j_k}$ term $\us =
      \fs(\us_{j_1},\ldots,\us_{j_k})$ occurs in $\shapeBaseF$
      where $\fs=\shapified{f}$ and for every $r$ where $1 \leq
      r \leq k$, conjunct $\leafsetShape{u_{i_r}}=\us_{j_r}$
      occurs in $\leafsetHomF$.

\item[P5)] homomorphism property of $\termShapeF$:
      for every non-parameter term variable $u$ such that $u =
      f(u_{i_1},\ldots,u_{i_k})$ occurs in $\termBaseF$, if
      conjunct $\termShape{u}=\us$ occurs in $\termHomF$,
      then for some shape variables
      $\us_{j_1},\ldots,\us_{j_k}$ term $\us =
      \fs(\us_{j_1},\ldots,\us_{j_k})$ occurs in $\shapeBaseF$
      where $\fs=\shapified{f}$ and for every $r$ where $1 \leq
      r \leq k$, conjunct $\termShape{u_{i_r}}=\us_{j_r}$
      occurs in $\termHomF$.

\end{enumerate}
\end{definition}

As in Section~\ref{sec:qeTA} and
Section~\ref{sec:twoConstQE} we proceed to show that each
quantifier-free formula can be written as a disjunction of
base formulas and each base formula can be written as a
quantifier-free formula.  We first give a small example to
illustrate how the techniques of
Section~\ref{sec:twoConstQE} extend to the more general case
of $\Sigma$-term-power.

\begin{example}
  We solve one subproblem from
  Example~\ref{exa:twoConstMain} using the language of
  term-power algebras.

  Consider the formula
  \begin{equation} \label{eqn:nonrecExa}
    \begin{array}{l}
      \exists v.\ g(v,z) \leq g(z,v) \ \land \ \Is{g}(v) \ \land \ \Is{g}(w) \ \land \mnl
      \qquad \lnot (g_1(w) \leq g_1(v))
    \end{array}
  \end{equation}
  Formula~(\ref{eqn:nonrecExa}) is in the language of
  Figure~\ref{fig:termPowerOps}, with $\leq$ a binary lifted relation.  
  After converting~(\ref{eqn:nonrecExa})
  into the language of Figure~\ref{fig:nonrecLogic} we obtain
  as one of the possible cases formula:
  \begin{equation}
    \begin{array}{l}
      \exists v. 
      \begin{array}[t]{l}
        [g(v,z) \leqc g(z,v)]_{\termShape{g(z,v)}} \leafsetEq
          \strueI{\termShape{g(z,v)}} \ \land \mnl
        \termShape{g(z,v)} \shapeEq \termShape{g(z,v)} \ \land \mnl
        \Is{g}(v) \ \land \ \Is{g}(w) \ \land \mnl
        [g_1(w) \leqc g_1(v)]_{\termShape{g_1(w)}} \not\leafsetEq
          \strueI{\termShape{g_1(w)}} \ \land \mnl
        \termShape{g_1(v)} \shapeEq \termShape{g_1(w)}
      \end{array}
    \end{array}
  \end{equation}
  where $\leqc$ is the subtyping relation on the base
  structure $C$ so that ${\leq} = {\leqc'}$.  We next transform
  the formula into unnested form, obtaining:
  \begin{equation} \label{eqn:nonrecExaUnnested}
    \begin{array}{l}
      \exists v, u_{vz},u_{zv},u_{w1},u_{v1}.\ \existsl \ul_{vz}, \ul_{w1}.\
      \existsS \us_{vz}, \us_{w1}. \mnl
      \quad
      \begin{array}[t]{l}
        u_{vz} = g(v,z) \ \land \ u_{zv} = g(z,v) \ \land \mnl
        u_{w1} = g_1(w) \ \land \ u_{v1} = g_1(v) \ \land \mnl
        \us_{vz} \shapeEq \termShape{u_{vz}} \ \land \
        \us_{w1} \shapeEq \termShape{u_{w1}} \ \land \mnl
        \termShape{u_{zv}} \shapeEq \us_{vz} \ \land \
        \termShape{u_{v1}} \shapeEq \us_{w1} \ \land \mnl
        \Is{g}(v) \ \land \ \Is{g}(w) \ \land \mnl
        \ul_{vz} \leafsetEq [u_{vz} \leqc u_{zv}]_{\us_{vz}} \mnl
        |\lnot \ul_{vz}|_{\us_{vz}} = 0 \ \land \mnl
        \ul_{w1} \leafsetEq [u_{w1} \leqc u_{v1}]_{\us_{w1}} \ \land \mnl
        |\lnot \ul_{w1}| \geq 1
      \end{array}      
    \end{array}  
  \end{equation}
  We next transform~(\ref{eqn:nonrecExaUnnested}) into
  disjunction of base formulas.
  % !! more blah on how we get here, see twoconst
  A typical base formula is:
  \begin{equation} \label{eqn:nonrecExaBase}
    \begin{array}{l}
      \exists u_{vz}, u_{zv}, u_v, u_z, u_w, u_{v1}, u_{v2}, u_{z1}, u_{z2}, u_{w1}, u_{w2}. \mnl
      \existsl \ul_{vz}, \ul_v, \ul_z, \ul_{v1}, \ul_{v2}, 
               \ul_{z1}, \ul_{z2}, \ul_{w1}. \mnl
      \existsS \us_{vz}, \us_w, \us_{w1}, \us_{w2}. \mnl
      \begin{array}{l}
        \shapeBaseF_1 \ \land \mnl
        \leafsetBaseF_1 \ \land \ \leafsetHomF_1 \ \land \mnl
        \termBaseF_1 \ \land \ \termHomF_1 \ \land \mnl
        \cardinF_1 \ \land \ \innerMapF_1
      \end{array}
    \end{array}
  \end{equation}
\[
\begin{array}{l}
\shapeBaseF_1 = 
\begin{array}[t]{l}
\us_{vz} = \gs(\us_w,\us_w) \land \us_w = \gs(\us_{w1},\us_{w2}) \ \land \mnl
\distinctF(\us_{vz},\us_w,\us_{w1},\us_{w2})
\end{array}
\end{array}
\]
\[
\begin{array}{l}
\leafsetBaseF_1 =
\begin{array}[t]{l}
\ul_{vz} = \gl(\ul_v,\ul_z) \ \land \mnl
\ul_v = \gl(\ul_{v1},\ul_{v2}) \ \land
\ul_z = \gl(\ul_{z1},\ul_{z2})
\end{array}
\end{array}
\]
\[
\leafsetHomF_1 = \mnl
\begin{array}[t]{l}
\leafsetShape{\ul_{vz}} = \us_{vz} \ \land \mnl
\leafsetShape{\ul_v} = \us_w \land
\leafsetShape{\ul_z} = \us_w \ \land \mnl
\leafsetShape{\ul_{v1}} = \us_{w1} \land
\leafsetShape{\ul_{v2}} = \us_{w2} \ \land \mnl
\leafsetShape{\ul_{z1}} = \us_{w1} \land
\leafsetShape{\ul_{z2}} = \us_{w2} \ \land \mnl
\leafsetShape{\ul_{w1}} = \us_{w1}
\end{array}
\]
\[
\begin{array}{l}
\termBaseF_1 =
\begin{array}[t]{l}
u_{vz} = g(u_v,u_z) \land u_{zv} = g(u_z,u_v) \ \land \mnl
u_v = g(u_{v1},u_{v2}) \land u_z = g(u_{z1},u_{z2}) \ \land \mnl
u_w = g(u_{w1},u_{w2}) \ \land \mnl
z = u_z \land w = u_w
\end{array}
\end{array}
\]
\[
\begin{array}{l}
\termHomF_1 = \mnl
\begin{array}[t]{l}
\termShape{u_{vz}}=\us_{vz} \land
\termShape{u_{zv}}=\us_{vz} \ \land \mnl
\termShape{u_v}=\us_w \land 
\termShape{u_z}=\us_w \land 
\termShape{u_w}=\us_w \ \land \mnl
\termShape{u_{v1}}=\us_{w1} \land 
\termShape{u_{z1}}=\us_{w1} \land 
\termShape{u_{w1}}=\us_{w1} \ \land \mnl
\termShape{u_{v2}}=\us_{w2} \land 
\termShape{u_{z2}}=\us_{w2} \land 
\termShape{u_{w2}}=\us_{w2}
\end{array}
\end{array}
\]
\[
\begin{array}{l}
\innerMapF_1 = \mnl
\begin{array}[t]{l}
  \ul_{v1} \leafsetEq [u_{v1} \leqc u_{z1}]_{\us_{w1}} \ \land 
  \ul_{z1} \leafsetEq [u_{z1} \leqc u_{v1}]_{\us_{w1}} \ \land \mnl
  \ul_{v2} \leafsetEq [u_{v2} \leqc u_{z2}]_{\us_{w2}} \ \land 
  \ul_{z2} \leafsetEq [u_{z2} \leqc u_{v2}]_{\us_{w2}} \ \land \mnl
  \ul_{w1} \leafsetEq [u_{w1} \leqc u_{v1}]_{\us_{w1}}
\end{array}
\end{array}
\]
\[
\begin{array}{l}
\cardinF_1 = 
\begin{array}[t]{l}
  |\lnot \ul_{v1}|_{\us_{w1}} = 0 \ \land \ |\lnot \ul_{z1}|_{\us_{w1}} = 0 \ \land \mnl
  |\lnot \ul_{v2}|_{\us_{w2}} = 0 \ \land \ |\lnot \ul_{z2}|_{\us_{w2}} = 0 \ \land \mnl
  |\lnot \ul_{w1}|_{\us_{w1}}| \geq 1
% |u_{v1} \cap u^c_{z1}|_{\us_{w1}} = 0 \ \land\ |u_{z1} \cap u^c_{v1}|_{\us_{w1}} = 0 \ \land \mnl
% |u_{v2} \cap u^c_{z2}|_{\us_{w2}} = 0 \ \land\ |u_{z2} \cap u^c_{v2}|_{\us_{w2}} = 0 \ \land \mnl
% |u_{w1} \cap u^c_{v1}|_{\us_{w1}} \geq 1
\end{array}
\end{array}
\]
We next show how to transform the base
formula~(\ref{eqn:nonrecExaBase}) into quantifier-free form.

We substitute away non-parameter term variables
$u_{vz},u_{zv},u_v$ and non-parameter leafset variables
$\ul_{vz},\ul_v,\ul_z$, because the homomorphism constraints
they participate in may be derived from the remaining
conjuncts.  We next eliminate parameter term variables
$u_{v1}, u_{v2}$ and parameter leafset variables
$\ul_{v1},\ul_{v2},\ul_{z1},\ul_{z2},\ul_{w1}$.  Grouping
the conjuncts in $\cardinF_1$ and $\innerMapF_1$ by their
shape, we may extract the subformulas $\psi_1$ and $\psi_2$
of~(\ref{eqn:nonrecExaBase}).
\[
\begin{array}{l}
\psi_1 \ \equiv \mnl
\begin{array}[t]{l}
  \exists u_{v1}. \existsl \ul_{v1},\ul_{z1},\ul_{w1}. \mnl
  \begin{array}{l}
    \termShape{u_{v1}} \shapeEq \us_{w1} \ \land \
    \termShape{u_{z1}} \shapeEq \us_{w1} \ \land \
    \termShape{u_{w1}} \shapeEq \us_{w1} \ \land \mnl
    \leafsetShape{\ul_{v1}} \shapeEq \us_{w1} \ \land \
    \leafsetShape{\ul_{z1}} \shapeEq \us_{w1} \ \land \mnl
    \leafsetShape{\ul_{w1}} \shapeEq \us_{w1} \ \land \mnl
    \ul_{v1} \leafsetEq [u_{v1} \leqc u_{z1}]_{\us_{w1}} \ \land 
    \ul_{z1} \leafsetEq [u_{z1} \leqc u_{v1}]_{\us_{w1}} \ \land \mnl
    \ul_{w1} \leafsetEq [u_{w1} \leqc u_{v1}]_{\us_{w1}} \ \land \mnl
    |\lnot \ul_{v1}|_{\us_{w1}} = 0 \ \land \ |\lnot \ul_{z1}|_{\us_{w1}} = 0 \ \land \mnl
    |\lnot \ul_{w1}|_{\us_{w1}}| \geq 1    
  \end{array}
\end{array}
\end{array}
\]
and
\[
\begin{array}{l}
\psi_2 \ \equiv \mnl
\begin{array}[t]{l}
  \exists u_{v2}. \existsl \ul_{v2},\ul_{z2}. \mnl
  \begin{array}{l}
    \termShape{u_{v2}} \shapeEq \us_{w2} \ \land \
    \termShape{u_{z2}} \shapeEq \us_{w2} \ \land \mnl
    \termShape{\ul_{v2}} \shapeEq \us_{w2} \ \land \
    \termShape{\ul_{z2}} \shapeEq \us_{w2} \ \land \mnl
    \ul_{v2} \leafsetEq [u_{v2} \leqc u_{z2}]_{\us_{w2}} \ \land 
    \ul_{z2} \leafsetEq [u_{z2} \leqc u_{v2}]_{\us_{w2}} \ \land \mnl
    |\lnot \ul_{v2}|_{\us_{w2}} = 0 \ \land \ |\lnot \ul_{z2}|_{\us_{w2}} = 0
  \end{array}
\end{array}
\end{array}
\]
Formula $\psi_1$ expresses a fact in a structure isomorphic to
the power $\ThC^n$ where $n$ is the number of leaves in the
shape denoted by $\us_{w1}$.  Similarly, $\psi_2$ expresses a
fact in a product structure $\ThC^m$ where $m$ is the number
of leaves in the shape denoted by $\us_{w2}$.  We can
therefore use the technique of Feferman-Vaught technique
(Section~\ref{sec:fefermanVaught}) to eliminate the
quantifiers from formulas $\psi_1$ and $\psi_2$.  According to
Example~\ref{exa:feferman}, $\psi_1$ is equivalent to:
\begin{equation*}
  \begin{array}{l}
    \existsl \ul_0,\ul_4.\mnl
    \begin{array}{l}
      \ul_0 \leafsetEq [\existsI t.\ t \leqc u_{z1} \ \landI \ 
                                   u_{z1} \leqc t \ \landI \
                                   u_{w1} \leqc t]_{\us_{w1}} \ \land \mnl
      \ul_4 \leafsetEq [\existsI t.\ t \leqc u_{z1} \ \landI \ 
                                   u_{z1} \leqc t \ \landI \
                                   \lnotI u_{w1} \leqc t]_{\us_{w1}} \ \land \mnl
      |\ul_4|_{\us_{w1}} \geq 1 \ \land \ 
      |\lnotI \ul_0 \ \landI \ \lnotI \ul_4|_{\us_{w1}} = 0
    \end{array}
  \end{array}
\end{equation*}
We similarly apply Feferman-Vaught construction to $\psi_2$ and
obtain the result $\boolTrue$.  We may now substitute
the results of quantifier elimination in
$\psi_1$ and $\psi_2$.  The resulting formula is:
\begin{equation*}
    \begin{array}{l}
      \exists u_{vz}, u_{zv}, u_v, u_z, u_w, u_{v1}, u_{v2}, u_{z1}, u_{z2}, u_{w1}, u_{w2}. \mnl
      \existsl \ul_{vz}, \ul_v, \ul_z, \ul_{v1}, \ul_{v2}, 
               \ul_{z1}, \ul_{z2}, \ul_{w1}. \mnl
      \existsS \us_{vz}, \us_w, \us_{w1}, \us_{w2}. \mnl
      \begin{array}{l}
        \shapeBaseF_1 \ \land \mnl
        \leafsetHomF_2 \ \land \mnl
        \termBaseF_2 \ \land \ \termHomF_2 \ \land \mnl
        \cardinF_1 \ \land \ \innerMapF_1
      \end{array}
    \end{array}
\end{equation*}
where
\[
\leafsetHomF_2 = \mnl
\begin{array}[t]{l}
\leafsetShape{\ul_0} = \us_{w1} \ \land \
\leafsetShape{\ul_4} = \us_{w1}
\end{array}
\]
\[
\begin{array}{l}
\termBaseF_2 =
\begin{array}[t]{l}
u_z = g(u_{z1},u_{z2}) \ \land \ 
u_w = g(u_{w1},u_{w2}) \ \land \mnl
z = u_z \land w = u_w
\end{array}
\end{array}
\]
\[
\begin{array}{l}
\innerMapF_2 = \mnl
\begin{array}[t]{l}
      \ul_0 \leafsetEq [\existsI t.\ t \leqc u_{z1} \ \landI \ 
                                   u_{z1} \leqc t \ \landI \
                                   u_{w1} \leqc t]_{\us_{w1}} \ \land \mnl
      \ul_4 \leafsetEq [\existsI t.\ t \leqc u_{z1} \ \landI \ 
                                   u_{z1} \leqc t \ \landI \
                                   \lnotI u_{w1} \leqc t]_{\us_{w1}} \ \land \mnl
\end{array}
\end{array}
\]
\[
\begin{array}{l}
\cardinF_2 = 
\begin{array}[t]{l}
      |\ul_4|_{\us_{w1}} \geq 1 \ \land \ 
      |\lnotI \ul_0 \ \landI \ \lnotI \ul_4|_{\us_{w1}} = 0
\end{array}
\end{array}
\]
In the resulting formula all variables are expressible in
terms of free variables, so we can write the formula without
quantifiers $\exists,\forall,\existsl,\foralll$.
% !! FIX: should I express \us_{vz} ?! or drop it?

%% !! FIX
%% Finally, note that all transformations we have performed are
%% valid regardless of the cardinality of the structure $C$ and
%% the way $\leqc$ is defined on $C$.  Our decision procedure
%% is in fact uniform with respect to the details of $C$.
% say about lattice vs. non-lattice
\end{example}

The following Proposition~\ref{prop:nonrecStructexBaseBase}
is analogous to Proposition~\ref{prop:StructexBaseBase}; the
proof is straightforward.
\begin{proposition}[Quantification of Struct.\ Base]
\label{prop:nonrecStructexBaseBase}
If $\beta$ is a structural base formula and $x$ a free shape, leafset,
or term variable in $\beta$, then there exists a base structural formula
$\beta_1$ equivalent to $\exists x.\beta$.
\end{proposition}
The following
Proposition~\ref{prop:nonrecStructqFreeToBaseDisj}
corresponds to Proposition~\ref{prop:StructqFreeToBaseDisj}.
\begin{proposition}[Quantifier-Free to Structural Base]
\label{prop:nonrecStructqFreeToBaseDisj}
% NOTE: I am requiring here that the formula is simple
% to avoid talking about the resulting form that 
% Lemma~\ref{lemma:nonrecSimplification} would produce.
Let $\phi$ be a well-defined simple formula without
quantifiers $\existsl, \foralll$, $\exists, \forall$, $\existsS,
\forallS$.  Then $\phi$ can be written as $\boolTrue$,
$\boolFalse$, or a disjunction of structural base formulas.
\end{proposition}
\begin{proofsketch} 
  The overall idea of the transformation to base formula is
  similar to the transformation in the proof of
  Proposition~\ref{prop:StructqFreeToBaseDisj}.
  Additional complexity is due to inner formulas.
  However, note that an inner formula $\phi(\us,u_1,\ldots,u_n)$ is
  well-defined iff $\delta(\us,u_1,\ldots,u_n)$ holds where
  \[
    \delta(\us,u_1,\ldots,u_n) \ \equiv \ 
   \termShape{u_1}=\us \ \land \ldots \land \ \termShape{u_n}=\us
  \]
  Hence, each formula $\phi(\us,u_1,\ldots,u_n)$ can be
  treated as a partial operation $p$ of sort
  \[
     \shapeSort \times \termSort^n \to \leafsetSort
  \]
  and the domain given by
  \[
     D_p = \tu{\tu{\us,u_1,\ldots,u_n},\delta(\us,u_1,\ldots,u_n)}
  \]
  This means that we may
  apply~Proposition~\ref{prop:unnestedDNFpartial} and
  convert formula to disjunction existentially quantified
  well-defined conjunctions of literals in one of the
  following forms:
  \begin{enumerate}
  \item equality with inner formulas: 
    $\ul_0 \leafsetEq \phi(\us,u_1,\ldots,u_n)$ where
    $\phi(\us,u_1,\ldots,u_n)$ is a $\us$-inner formula;
  \item formulas of leafset boolean algebra:
    \[\begin{array}{l}
    \ul_0 \leafsetEq \ul_1 \slandI{\us} \ul_2 \mnl
    \ul_0 \leafsetEq \ul_1 \slorI{\us} \ul_2 \mnl
    \ul_0 \leafsetEq \slnotI{\us} \ul_1 \mnl 
    \ul_0 \leafsetEq \strueI{\us} \mnl
    \ul_0 \leafsetEq \sfalseI{\us}
    \end{array}\]
  \item formulas of term algebra of terms:
    \[\begin{array}{l}
      u_1 = u_2,\ u_1 \neq u_2 \mnl
      u_0 = f(u_1,\ldots,u_n) \mnl
      u = f_i(u_0) \mnl
      \Is{f}(u_0),\ \lnot \Is{f}(u_0) \mnl
      \termShape{u} = \us
    \end{array}\]
  \item formulas of term algebra of leafsets:
    \[\begin{array}{l}
      \ul_1 \leafsetEq \ul_2,\ \ul_1 \not \leafsetEq \ul_2 \mnl
      \ul_0 \leafsetEq \fl(\ul_1,\ldots,\ul_n) \mnl
      \ul \leafsetEq \fl_i(\ul_0) \mnl
      \Isfl(\ul_0),\ \lnot \Isfl(\ul_0) \mnl
      \leafsetShape{\ul} \leafsetEq \us      
    \end{array}\]
  \item formulas of term algebra of shapes:
    \[\begin{array}{l}
      \us_1 \shapeEq \us_2,\ \us_1 \not \shapeEq \us_2 \mnl
      \us_0 \shapeEq \fs(\us_1,\ldots,\us_n) \mnl
      \us \shapeEq \fs_i(\us_0) \mnl
      \Isfs(\us_0),\ \lnot \Isfs(\us_0) \mnl
    \end{array}\]    
  \end{enumerate}

  We next describe transformation of each existentially
  quantified conjunction.  In the sequel, whenever we
  perform case analysis and generate a disjunction of
  conjunctions, existential quantifiers propagate to the
  conjunctions, so we keep working with a existentially
  quantified conjunction.  The existentially quantified
  variables will become internal variables of a structural
  base formula.

  Analogously to the proof of
  Proposition~\ref{prop:qFreeToBaseDisj}, we use
  (\ref{eqn:IsfPartitMany}), (\ref{eqn:IsfPartitManyL}),
  (\ref{eqn:IsfPartit}) to eliminate literals $\lnot
  \Is{f}(u_0)$, $\lnot \Isfl{\fl}(\ul_0)$, $\lnot \Isgs(\us_0)$.
  
  As in the proof of
  Proposition~\ref{prop:StructqFreeToBaseDisj}, we replace
  formulas of leafset boolean algebra by cardinality
  constraints, similarly to Figure~\ref{fig:baUnnestElim}.
  
  We next convert formulas of term algebra of terms into a
  base formula, formulas of term algebra of leafsets into a
  base formula, and formulas of term algebra of shapes into
  a base formula.

  We simultaneously make sure that every term or leafset variable
  has an associated associated shape variable,
  introducing new shape variables if needed.

  We also ensure homomorphism
  requirements by replacing internal variables when we entail
  their equality.  

  Another condition we ensure is that
  parameter term variables map to parameter shape variables,
  and non-parameter term variables to non-parameter shape
  variables; we do this by performing expansion of term and
  shape variables.  

  We perform expansion of shape variables as
  in Section~\ref{sec:qeBoolalg}.  Expansion of term and 
  variables
  is even simpler because there is no need to do case analysis
  on equality of term variable with other variables.
  
  We eliminate disequality between term variables
  using~(\ref{eqn:expressingEquality}).  We eliminate
  disequalities between leafset variables as in
  Example~\ref{exa:disEq}, by converting each disequality
  into a cardinality constraint.  Elimination of
  disequalities might violate previously established
  homomorphism invariants, so we may need to reestablish
  these invariants by repeating the previously described
  steps.  The overall process terminates because we never
  introduce new inequalities between term or leafset
  variables.
  
  As a final step, we convert all cardinality constraints into
  constraints on parameter term variables,
  using~(\ref{eqn:nonrecDecompose}).
  
  In the case when the shape of cardinality constraint is
  $\cs$, we cannot apply~(\ref{eqn:nonrecDecompose}).
  However, in this case, unlike
  Proposition~\ref{prop:StructqFreeToBaseDisj}, we do not do
  case analysis on all possible constant leafsets (this is
  not even possible in general).  This is because
  Definition~\ref{def:nonrecStructuralBase}, unlike
  Definition~\ref{def:structuralBase} implies no need to
  further decompose cardinality constraints in that case,
  because we allow primitive non-parameter leafset
  variables.
  
  This completes our sketch of transforming a
  quantifier-free formula into disjunction of structural
  base formulas.
\end{proofsketch}

%% !! FIX: modify the two constant case to work with determined
% instead of covered?
We introduce the notion of determined variables in
structural base formula generalizing
Definition~\ref{def:coveredVarDef} and
Definition~\ref{def:structCovering}.  

For brevity, we write $\ustar$ for internal shape, term, or
leafset variables, similarly $\xstar$ for a free variable,
$\tstar$ for a term and $\fstar$ for a shape, term, or
leafset term algebra constructor and $\fstar_i$ for a shape,
term, or leafset term algebra selector.
\begin{definition} \label{def:nonrecStructDeterminations}
  The set $\determinations$ of variable determinations of a
  structural base formula $\beta$ is the least set $S$ of pairs
  $\tu{\ustar,\tstar}$ where $\ustar$ is an internal term,
  leafset, or shape variable and $\tstar$ is a term over the
  free variables of $\beta$, such such that:
\begin{enumerate}
  
\item if $\xstar=\ustar$ occurs in $\termBaseF$,
  $\leafsetBaseF$, or $\shapeBaseF$, then
  $\tu{\ustar,\xstar} \in S$;

\item if $\tu{\ustar,\tstar} \in S$ and $\ustar =
  \fstar(\ustar_1,\ldots,\ustar_k)$ occurs in $\shapeBaseF$,
  $\termBaseF$, or $\leafsetBaseF$ then
  $\{\tu{\ustar_1,\fstar_1(\tstar)},\ldots,\tu{\ustar_k,\fstar_k(\tstar)}\}
  \subseteq S$;
  
\item if
  $\{\tu{\ustar_1,\fstar_1(\tstar)},\ldots,\tu{\ustar_k,\fstar_k(\tstar)}\}
  \subseteq S$ and $\ustar = \fstar(\ustar_1,\ldots,\ustar_k)$ occurs in
  $\shapeBaseF$, $\termBaseF$, or $\leafsetBaseF$ then
  $\tu{\ustar,\tstar} \in S$;
  
\item if $\tu{u,t} \in S$ and $\termShape{u}=\us$ occurs in $\termHomF$
      then $\tu{\us,\termShape{t}} \in S$;

\item if $\tu{\ul,\tl} \in S$ and $\leafsetShape{\ul}=\us$ occurs 
  in $\leafsetHomF$ then $\tu{\us,\leafsetShape{\tl}} \in S$;
  
\item if $\ul = \phi(\us,u_1,\ldots,u_n)$ occurs in $\innerMapF$
  where $\phi(\us,u_1,\ldots,u_n)$ is an inner formula and $\{
  \tu{\us,\ts}, \tu{u_1,t_1}, \ldots, \tu{u_n,t_n} \} \subseteq S$, then
  $\tu{\ul,\phi(\ts,t_1,\ldots,t_n)} \in S$.   (In the special case
  when $\phi$ contains no free term variables, if
  $\tu{\us,\ts} \in S$ then $\tu{\ul,\phi(\us)} \in S$.

      % !! unfortunate confusion of object and meta level possible
      
\end{enumerate}
\end{definition}

\begin{definition}
  An internal variable $\ustar$ is \emph{determined} if
  $\tu{\ustar,\tstar} \in \determinations$ for some term
  $\ts$.  An internal variable is \emph{undetermined} if it
  is not determined.
\end{definition}

\begin{lemma} \label{lemma:nonrecDeterminationWorks}
  Let $\beta$ be a structural base formula with matrix $\beta_0$ and
  let $\determinations$ be the determinations of $\beta$.  If
  $\tu{\ustar,\tstar} \in S$ then $\models \beta_0 \implies \ustar=\tstar$.
\end{lemma}
\begin{proof}
By induction, using Definition~\ref{def:nonrecStructDeterminations}.
\end{proof}

\begin{corollary} \label{cor:nonrecDeterminedToQfree}
  Let $\beta$ be a structural base formula such that every
  internal variable is determined.  Then $\beta$ is equivalent
  to a well-defined formula without quantifiers $\existsl,
  \foralll$, $\exists, \forall$, $\existsS, \forallS$.
\end{corollary}
\begin{proof}
  By Lemma~\ref{lemma:nonrecDeterminationWorks}
  using~(\ref{eqn:replacement}).
\end{proof}

\begin{lemma} \label{lemma:nonrecStrucElimTermNonparam}
Let $u$ be an undetermined composed non-parameter
term variable in a
structural base formula $\beta$ such that $u$ is a source i.e.\
no conjunct of form
\[
   u' = f(u_1,\ldots,u,\ldots,u_k)
\]
occurs in $\termBaseF$.  Let $\beta'$ be the result of dropping
$u$ from $\beta$.  Then $\beta$ is equivalent to $\beta'$.
\end{lemma}
\begin{proof}
  Because $u$ is a composed non-parameter term variable, it
  does not occur in $\innerMapF$, so it only occurs in
  $\termBaseF$ and $\termHomF$.  The conjunct containing $u$
  in $\termHomF$ is a consequence of the remaining
  conjuncts, so it may be dropped.  After that,
  applying~(\ref{eqn:replacement}) yields a structural base
  formula $\beta'$ not containing $u$, where $\beta'$ is equivalent
  to $\beta$.
\end{proof}

\begin{lemma} \label{lemma:nonrecStrucElimLeafsetNonparam}
Let $\ul$ be an undetermined composed non-parameter 
leafset variable in a
structural base formula $\beta$ such that $\ul$ is a source i.e.\
no conjunct of form
\[
   {\ul}' = \fl(\ul_1,\ldots,\ul,\ldots,\ul_k)
\]
occurs in $\leafsetBaseF$.  Let $\beta'$ be the result of dropping
$\ul$ from $\beta$.  Then $\beta$ is equivalent to $\beta'$.
\end{lemma}
\begin{proof}
  Because $\ul$ is a composed non-parameter term variable,
  it does not occur in $\innerMapF$ or $\cardinF$, so it
  only occurs in $\leafsetBaseF$ and $\leafsetHomF$.  The
  conjunct containing $\ul$ in $\leafsetHomF$ is a
  consequence of the remaining conjuncts, so it may be
  dropped.  After that, applying~(\ref{eqn:replacement})
  yields a structural base formula $\beta'$ not containing
  $\ul$, where $\beta'$ is equivalent to $\beta$.
\end{proof}

\begin{corollary} \label{cor:nonrecStrucElimNonparam}
  Every base formula is equivalent to a base formula without
  undetermined composed non-parameter term variables and
  without undetermined composed non-parameter leafset
  variables.
\end{corollary}
\begin{proof}
  If a structural base formula has an undetermined composed
  non-parameter term variable, then it has an undetermined
  composed non-parameter term variable that is a source,
  similarly for leafset variables.  By repeated application
  of Lemma~\ref{lemma:nonrecStrucElimTermNonparam} and
  Lemma~\ref{lemma:nonrecStrucElimLeafsetNonparam} we
  eliminate all undetermined non-parameter term and leafset
  variables.
\end{proof}

The following Proposition~\ref{prop:nonrecStructbaseToSel}
corresponds to Proposition~\ref{prop:StructbaseToSel} and
Proposition~\ref{zzz:prop:nonrecStructbaseToSel}.
\begin{proposition}[Struct.\ Base to Quantifier-Free]
\label{prop:nonrecStructbaseToSel}
Every structural base formula $\beta$ is equivalent to a
well-defined simple formula $\phi$ without quantifiers
$\existsl, \foralll$, $\exists, \forall$, $\existsS, \forallS$.
\end{proposition}
\begin{proofsketch}
  By Corollary~\ref{cor:nonrecStrucElimNonparam} we may
  assume that $\beta$ has no undetermined composed non-parameter
  term and leafset variables.  By
  Corollary~\ref{cor:nonrecDeterminedToQfree} we are done if
  there are no undetermined variables, so it suffices to
  eliminate:
  \begin{enumerate}
  \item undetermined parameter term variables,
  \item undetermined primitive non-parameter term variables,
  \item undetermined parameter leafset variables,
  \item undetermined primitive non-parameter leafset variables, and
  \item undetermined shape variables.
  \end{enumerate}
  If $u$ is an undetermined parameter term variable or a
  primitive non-parameter term variable, then $u$ does not
  occur in $\termBaseF$, so it occurs only in $\termHomF$
  and $\innerMapF$.  If $\ul$ is an undetermined parameter
  leafset variable or a primitive non-parameter leafset
  variable then $\ul$ does not occur in $\leafsetBaseF$, so
  it occurs only in $\leafsetHomF$, $\innerMapF$, and
  $\cardinF$.

  For a undetermined term or leafset variable of shape $\us$
  such that there is an uncovered parameter or primitive
  non-parameter term or leafset variable with shape $\us$,
  consider all conjuncts $\gamma_i$ in $\innerMapF$ of form
  \[
  \ul_j = \phi(\us,u_{i_1},\ldots,u_{i_k})
  \]
  and all conjuncts $\delta_i$ from $\cardinF$ of form:  
  \[
  |\tl(\ul_{\irl+1},\ldots,\ul_{\nl})|_{\us}=k
  \]
  or
  \[
  |\tl(\ul_{\irl+1},\ldots,\ul_{\nl})|_{\us}\geq k
  \]
  Together with formulas from $\termHomF$ and $\leafsetHomF$
  that contain term and leafset variables free in formulas
  $\gamma_i$ and $\delta_i$, these conjuncts form a formula $\eta$ which
  expresses a relation in the substructure of term-power
  algebra which (because constructors are covariant) is
  isomorphic to a term-power of $\ThC$.  We therefore use
  Feferman-Vaught theorem from
  Section~\ref{sec:fefermanVaught} to eliminate all term and
  parameter variables from $\eta$.  By repeating this process
  we eliminate all undetermined parameter and leafset
  variables.
  
  It remains to eliminate undetermined shape variables.
  This process is similar to term algebra quantifier
  elimination in Section~\ref{sec:qeTA}.  An essential part
  of construction in Section~\ref{sec:qeTA} is
  Lemma~\ref{lemma:baseSat}, which relies on the fact that
  undetermined parameter variables may take on infinitely
  many values.  We therefore ensure that undetermined
  parameter shape variables are not constrained by term and
  parameter variables through conjuncts outside
  $\shapeBaseF$.  An undetermined parameter shape variable
  $\us$ does not occur in $\termHomF$ or $\leafsetHomF$
  because there are no parameter term and leafset variables,
  so $\us$ can occur only in $\innerMapF$ and $\cardinF$.
  
  However, because undetermined parameter and leafset
  variables are eliminated from the formula, if $\us$ is a
  parameter shape variable then exactly one of these two
  cases holds:
  \begin{enumerate}
  \item there are some conjuncts in $\innerMapF$ and
    $\cardinF$ that contain $\us$ and contain some determined
    term and leafset variables, in this case $\us$ is
    determined, or
  \item there are no conjuncts in $\innerMapF$ containing $\us$
    and $\cardinF$ contains only domain cardinality constraints
    of form $|1|_{\us} = k$ and $|1|_{\us} \geq k$.
  \end{enumerate}
  Hence, if $\us$ is a shape variable it remains to
  eliminate the constraints of form $|1|_{\us} = k$ and
  $|1|_{\us} \geq k$.  We eliminate these constraints as in the
  proof of Proposition~\ref{zzz:prop:nonrecStructbaseToSel}.
  
  In the resulting formula all variables are determined.  By
  Corollary~\ref{cor:nonrecDeterminedToQfree} the formula
  can be written as a formula without quantifiers $\existsl,
  \foralll$, $\exists, \forall$, $\existsS, \forallS$.
\end{proofsketch}

The following is the main result of this paper.

\begin{theorem}[Term Power Quant.\ Elimination]
  \label{thm:nonrecTwoConstQElim}
  There exist algorithms $A$, $B$ such that for a given
  formula $\phi$ in the language of
  Figure~\ref{fig:nonrecLogic}:
\begin{enumerate}
\item[a)] $A$ produces a quantifier-free formula $\phi'$ in
  selector language
\item[b)] $B$ produces a disjunction $\phi'$ of structural base
  formulas
\end{enumerate}
\end{theorem}

We also explicitly state the following corollary.
\begin{corollary}
  Let $\ThC$ be a structure with decidable first-order
  theory.  Then the set of true sentences in the logic of
  Figure~\ref{fig:nonrecLogic} interpreted in the structure
  $\ThP$ according to
  Figures~\ref{fig:termPowerSemanticsOne} and
  \ref{fig:termPowerSemanticsTwo} is decidable.
\end{corollary}

\subsection{Handling Contravariant Constructors}
\label{sec:variance}

In this section we discuss the decidability of the
$\FreeSig$-term-power structure for a decidable theory
$\ThC$ when some of the function symbols $f \in \FreeSig$ are
contravariant.  We then suggest a generalization of the
notion of variance to multiple relations and to relations
with arity greater than two.

The modifications needed to accommodate contravariance with
respect to some distinguished relation symbol $\leq \in R$ for
the case of infinite $C$ are analogous to the modifications
in Section~\ref{sec:oneVariance}.  We this obtain a
quantifier elimination procedure for any decidable theory
$\ThC$ in the presence of contravariant constructors.

\begin{theorem}[Decidability of Structural Subtyping]
  Let $\ThC$ be a decidable structure and $\ThP$ a
  $\FreeSig$-term-power of $\ThC$.  Then the first-order
  theory of $\ThP$ is decidable.
\end{theorem}

In the rest of this section we consider a generalization
that allows defining variance for every relation
symbol $r \in R$ of any arity, and not just the relation
symbol $\leq \in R$.

For a given relation symbol $r \in R$, function symbol $f \in
\FreeSig$, with $k = \ar(f)$, and integer $i$ where $1 \leq i \leq
k$, let $P_r(f,i)$ denote a permutation of the set $\{1,\ldots,k\}$
that specifies the variance of the $i$-th argument of $f$
with respect to the relation $r$.  For example, if $r$ is a
binary relation then $P_r(f,i)$ is the identity permutation
$\{\tu{1,1}\tu{2,2}\}$ if $i$-th argument of $f$ is covariant,
or a the transpose permutation $\{\tu{1,2},\tu{2,1}\}$ if
$i$-th argument of $f$ is contravariant.

If $l \in \leaves{s}$ is a leaf $l = \tu{f^1,i^1} \ldots
\tu{f^n,i^n}$, define the permutation $\variance(l)$ as the
composition of permutations:
\begin{equation*}
  \variance(l) = P_r(f^n,i^n) \circ \cdots \circ P_r(f^1,i^1)
\end{equation*}
Then define
$\interpret{r}$ by
\begin{equation*}
  \begin{array}{ll}
     \interpret{r}(s,t_1,\ldots,t_k) = \mnl 
  \langle s, 
  \{ l \mid
     \begin{array}[t]{l}
       \interpretS{r}{C}(t_{p_1}[l],\ldots,t_{p_k}[l]) \ \land \mnl
       \tu{p_1,\ldots,p_k} = \variance(l)
     \end{array} \mnl
  \qquad  \} \rangle
   \end{array}
\end{equation*} 
We generalize~(\ref{eqn:PN}) by defining
\begin{equation*}
  N_{\pi}(s) = |\{ l \in \leaves{s} \mid \variance(l) = \pi \}|
\end{equation*}

As in Section~\ref{sec:oneVariance}, we can transform the
constraints $|1|_{\us} = k$ and $|1|_{\us} \geq k$ on each
parameter shape variable into a conjunction of constraints
of form:
\begin{equation*}
  N_\pi(\us) = k
\end{equation*}
or
\begin{equation*}
  N_\pi(\us) \geq k
\end{equation*}
  
{\bf A problem on nonnegative integers.}  To solve the
problem of variance with any number of relation symbols of
any arity, it suffices to solve the following problem on
sets of tuples of non-negative integers.

Let $\Nat = \{0,1,2,\ldots\}$.  Consider the structure $\St =
\Nat^d$ for some $d \geq 2$ and let $D = \{1,2,\ldots,d\}$.  If $p$ is
a permutation on $D$, let $M_p$ denote an operation $\St \to
\St$ defined by
\begin{equation*}
  M_p(x_1,\ldots,x_d) = (x_{p_1},\ldots,x_{p_d})
\end{equation*}
If $\tu{x_1,\ldots,x_d}, \tu{y_1,\ldots,y_d} \in \St$ define
\begin{equation*}
  \tu{x_1,\ldots,x_d} + \tu{y_1,\ldots,y_d} = \tu{x_1+y_1,\ldots,x_d+y_d}
\end{equation*}
Consider a finite set of operations $f : \St^k \to \St$
where each operation $f$ is determined by $k$
permutations $p_1^f,\ldots,p_k^f$ in the following way:
\begin{equation*}
  f(t_1,\ldots,t_k) = M_{p_1^f}(t_1) + \ldots + M_{p_k^f}(t_k)
\end{equation*}
Hence, each operation $f$ of arity $k$ is given by a
permutation which specifies how to exchange the order of
arguments in the tuple.  After permuting the arguments the
tuples are summed up.

Given a finite set $F$ of operations $f$, let $S$ be
the set generated by operations in $F$ starting from the
element $(1,0,\ldots,0) \in \St$.  Let $C(n_1,\ldots,n_d)$ be a
conjunction of simple linear constraints of the forms
\begin{equation*}
  n_i = a_i
\end{equation*}
and
\begin{equation*}
  n_i \geq a_i
\end{equation*}
Consider the set
\begin{equation*}
  A_C = \{ (n_1,\ldots,n_d) \in S | C(n_1,\ldots,n_d) \}
\end{equation*}
The problem is: For given set of operations $F$, is there an
algorithm that given $C(n_1,\ldots,n_d)$ finitely computes the set $A_C$.

\noindent
{\bf End of a problem on nonnegative integers.}
  
We conjecture that the technique of
Lemma~\ref{zzz:lemma:contraSets} can be generalized to yield
a solution to the problem on nonnegative integers and thus
establish the decidability for the notion of variance with
respect to any number of relations with any number of
arguments.

\subsection{A Note on Element Selection}

We make a brief note related to the choice of the language
for making statements in term-power algebras.  In
Section~\ref{sec:nconst} we avoided the use of leafset
variables by substituting them into cardinality constraints.
In this section we use a cylindric algebra of leafsets.

An apparently even more flexible alternative is to allow the
element selection operation
\begin{equation*}
  \select :: \termSort \times \leafSort \to \elemSort
\end{equation*}
where $\elemSort$ is a new sort, interpreted over the set
$C$, and $\leafSort$ is a sort interpreted over the set of
pairs of a shape and a leaf.  Instead of the formula
\begin{equation*}
\sr{\us}(t_1,\ldots,t_n) \leafsetEq \strueI{\us}
\end{equation*}
we would then write
\begin{equation*}
\forall l.\
\sr{\us}(\select(t_1,l),\ldots,\select(t_n,l)) \leafsetEq \strueI{\us}
\end{equation*}
Using $\select$ operation we can define update relation:
\begin{equation*}
  \begin{array}{l}
  \update(t_1,l_0,e,t_2) \ \equiv \mnl
  \begin{array}[t]{l}
    \forall l.
    \begin{array}[t]{l}
      ((l = l_0 \ \land \ \select(t_2,l)=e) \ \lor \mnl
      (l \neq l_0 \ \land \ \select(t_2,l)=\select(t_1,l)))
    \end{array}
  \end{array}
\end{array}
\end{equation*}
The resulting language is at least as expressive as the
language in Figure~\ref{fig:twoSortedSig}.  This language is
interesting because it allows reasoning about updates to
leaves of a tree of fixed shape, thus generalizing the
theory of updatable arrays
\cite{McCarthyPainter67CorrectnessCompilerArithmeticExpressions}
to the theory of trees with update operations, which would
be useful for program verification.  We did not choose this
more expressive language in this report for the following
reason.

If the base structure $\ThC$ has a finite domain $C$, then
for certain reasonable choice of the relations interpreting
$\lanC$ it is possible to express statements of this
extended language in the logic of
Figure~\ref{fig:nonrecLogic}.  The idea is to assume a
partial order on the elements of $C$ with a minimal element,
and use terms $t$ with exactly one leaf non-minimal to model
the leaves.

On the other hand, in the more interesting case when $C$ is
infinite, we can easily obtain undecidable theories in the
presence of selection operation.  Namely, the selection
operation allows terms to be used as finite sets of elements
of $C$.  The term-power therefore increases the expressiveness
from the first-order theory to the weak monadic second-order
theory, which allows quantification over finite sets of
objects.  Weak monadic theory allows in particular inductive
definitions.  If theory of structure $C$ is decidable, weak
monadic theory might therefore still be undecidable, as an
example we might take the term algebra itself, whose weak
monadic theory would allow defining subterm relation,
yielding an undecidable theory \cite[Page
508]{Venkataraman87DecidabilityExistentialTermAlgebras}.

%%% Local Variables: 
%%% mode: latex
%%% TeX-master: "main"
%%% End: 
% LocalWords:  kl arity leafset iff leafsets definedness formedness Feferman CL
% LocalWords:  Vaught subformula unnested blah twoconst subformulas vz vs Quant
% LocalWords:  Struct disequality disequalities Contravariant contravariant
% LocalWords:  contravariance updatable subterm

\section{Some Connections with MSOL}
\label{sec:connectionWithMSOL}

This section explores some relationships between the theory
of structural subtyping and monadic second-order logic
(MSOL) interpreted over tree-like structures.  We present it
as a series of remarks that are potentially useful for
understanding the first-order theory of structural subtyping
of recursive types, see
\cite{MuellerNiehren00OrderingConstraintsFeatureTreesMSOL,
  MuellerETAL01OrderingConstraintsFeatureTrees} for similar
results in the context of the theory of feature trees.

In Section~\ref{sec:msolEmbedding} we exhibit an embedding
of MSOL of {\em infinite} binary tree into the first-order
theory of structural subtyping of {\em recursive types} with
two constant symbols $a$,$b$ and one covariant binary
function symbol $f$.  MSOL of infinite binary tree is
decidable.  Although the embedding does not give an answer
to the decidability of the structural subtyping of recursive
types, it does show that the problem is at least as
difficult as decidability of MSOL over infinite trees.  We
therefore expect that, if the theory of structural subtyping
of recursive types is decidable, the decidability proof will
likely either use decidability of MSOL over infinite trees,
or use directly techniques similar to those of
\cite{GurevichHarrington82TreesAutomataGames,
  Walukiewicz96MonadicSecondOrderLogicTreeLikeStructures}.

In Section~\ref{sec:embeddedTree} we use the embedding in
Section~\ref{sec:msolEmbedding} to argue the decidability of
formulas of the first-order theory of structural subtyping
of recursive types where variables range over terms of
certain fixed infinite shape $s_e$.

In Section~\ref{sec:termsIntoTerms} we present an encoding of
all terms using terms of shape $s_e$.  We argue that the
main obstacle in using this encoding to show the
decidability of the first-order theory of structural
subtyping recursive types is inability to define the set of
all prefix-closed terms of the shape $s_e$.

In Section~\ref{sec:fixedRational} we generalize the
decidability result of Section~\ref{sec:embeddedTree} by
allowing different variables to range over different
constant shapes.

In Section~\ref{sec:feature} we illustrate some of the
difficulties in reducing first-order theory of structural
subtyping to MSOL over tree-like structures.  We show that
if we use a certain form of infinite feature trees instead
of infinite terms, the decidability follows.

In Section~\ref{sec:structuralUndec} we point out that
monadic second-order logic with prefix-closed sets is
undecidable, which follows from
\cite{SuETAL02FirstOrderTheorySubtypingConstraints}.  This
fact indicates that if we hope to show the decidability of
structural subtyping of recursive types, it is essential to
maintain the incomparability of types of different shape.

\subsection{Structural Subtyping Recursive Types} 
\label{sec:msolEmbedding}

In this section we define the problem of structural
subtyping of \emph{recursive} types.  We then give an
embedding of MSOL of the infinite binary tree into the
first-order theory of structural subtyping of infinite terms
over the signature $\FreeSig = \{ a, b, g \}$ with the partial
order $\leq$.

We define MSOL over infinite binary tree \cite[Page
317]{BoergerETAL97ClassicalDecisionProblem} as the structure
$\msolTwoVar = \tu{\{0,1\}^{*}, \succZero, \succOne}$.  The
domain of the structure is the set $\{0,1\}^{*}$ of all finite
strings over the alphabet $\{0,1\}$.  We denote first-order
variables by lowercase letters such as $x,y,z$.  First-order
variables range over finite words $w \in \{0,1\}^{*}$.  We
denote second-order variables by uppercase letters such as
$X,Y,Z$.  Second-order variables range over finite and
infinite subsets $S \subseteq \{0,1\}^{*}$.  The only relational
symbol is equality, with the standard interpretation.  There
are two function symbols, denoting the appending of the
symbol $0$ and the appending of the symbol $1$ to a word:
\[\begin{array}{rcl}
\succZero\, w &=& w \cdot 0 \mnl
\succOne\, w &=& w \cdot 1
\end{array}\]

For the purpose of embedding into the first-order theory of
structural subtyping, we consider a structure $\msolOneVar =
\tu{\{0,1\}^{*},\msolSubseteq,\setSuccZero,\setSuccOne}$
equivalent to $\msolTwoVar$.  We use the language of MSOL
without first-order variables to make statements within
$\msolOneVar$.  $\msolSubseteq$ is a binary relation
on sets denoting the subset relation:
\[
    Y_1 \msolSubseteq Y_2  \iff  \forall x.\ x \in Y_1 \implies x \in Y_2
\]
$\setSuccZero$ and $\setSuccOne$ are binary relations on
sets, $\setSuccZero, \setSuccOne \subseteq 2^{\{0,1\}^{*}} \times
2^{\{0,1\}^{*}}$, defined as follows:
\[\begin{array}{rcl}
\setSuccZero(Y_1,Y_2) &\iff& Y_2 = \{ w \cdot 0 \mid w \in Y_1 \} \mnl
\setSuccOne(Y_1,Y_2) &\iff& Y_2 = \{ w \cdot 1 \mid w \in Y_1 \} \mnl
\end{array}\]
The structure $\msolOneVar$ is similar to one
in~\cite{GurevichHarrington82TreesAutomataGames}; the
difference is that relations $\setSuccZero$ and
$\setSuccOne$ are true even for non-singleton sets.

Lemmas \ref{lemma:twoToOne} and \ref{lemma:oneToTwo} show
the expected equivalence of $\msolTwoVar$ and $\msolOneVar$.

\begin{lemma}[$\msolTwoVar$ expresses $\msolOneVar$]
\label{lemma:twoToOne}
Every relation on sets definable in $\msolOneVar$ is
definable in $\msolTwoVar$.
\end{lemma}
\begin{proof}
We express relations $\subseteq$, $\setSuccZero$, $\setSuccOne$ as
formulas in $\msolTwoVar$, as follows.  We express
$Y_1 \subseteq Y_2$ as % $\phi_1(Y_1,Y_2)$ where
\[
   % \phi_1(Y_1,Y_2) \ \equiv\ 
   \forall x.\ Y_1(x) \implies Y_2(x),
\]
$\setSuccZero(Y_1,Y_2)$ as % $\phi_2(Y_1,Y_2)$ where
\[
   % \phi_2(Y_1,Y_2) \ \equiv\ 
   \forall x. Y_2(x) \iff \exists y. y=\succZero(x),
\]
and $\setSuccOne(Y_1,Y_2)$ as % $\phi_3(Y_1,Y_2)$ where
\[
   % \phi_3(Y_1,Y_2) \ \equiv\ 
   \forall x. Y_2(x) \iff \exists y. y=\succOne(x).
\]
The statement follows by induction on the structure of
formulas.
\end{proof}

\noindent
Let $R \subseteq (2^{\{0,1\}^{*}})^k \times (\{0,1\}^{*})^n$ be relation of arity $k+n$.  Define
$R^{*} \subseteq (2^{\{0,1\}^{*}})^k \times (2^{\{0,1\}^{*}})^n$ by
\[\begin{array}{l}
    R^{*}(Y_1,\ldots,Y_k,X_1,\ldots,X_n) \ \equiv\ \mnl
\qquad
\begin{array}{rl}
     \exists x_1,\ldots,x_n. & X_1 = \{x_1\} \land \cdots \land X_n = \{x_n\} \ \land \mnl
                & R(Y_1,\ldots,Y_k,x_1,\ldots,x_n)
\end{array}
\end{array}\]

\begin{lemma}[$\msolOneVar$ expresses $\msolTwoVar$]
\label{lemma:oneToTwo}
If $R$ is definable in $\msolTwoVar$, then $R^{*}$ is
definable in $\msolOneVar$.
\end{lemma}
\begin{proofsketch}
Property of being an empty set is definable in $\msolOneVar$
by the formula
\[
   \phi_0(Y_1) \ \equiv\ \forall Y_2. Y_1 \msolSubseteq Y_2
\]
The relation $\msolSubset$ of being a proper subset is definable in
$\msolOneVar$ by formula
\[
   \phi_1(Y_1,Y_2) \ \equiv\ Y_1 \msolSubseteq Y_2 \land Y_1 \neq Y_2
\]
and the relation $\msolSubsetOne$ of having one element more
is definable by formula
\[
   \phi_2(Y_1,Y_2) \ \equiv\ Y_1 \msolSubset Y_2 \land 
                 \lnot \exists Z.\ Y_1 \msolSubset Z \land Z \msolSubset Y_2
\]
The property of being a singleton set can then be expressed by
formula
\[
   \phi_3(Y_1) \ \equiv\ \exists Y_0.\ \phi_0(Y_0) \land Y_0 \msolSubsetOne Y_1
\]
We define the relation on singletons corresponding to
$\succZero$ by
\[
   \phi_4(Y_1,Y_2) \ \equiv\ \phi_3(Y_1) \land \phi_3(Y_2) \land \setSuccZero(Y_1,Y_2)
\]
Similarly, the relation corresponding to $\succOne$ is defined
by
\[
   \phi_5(Y_1,Y_2) \ \equiv\ \phi_3(Y_1) \land \phi_3(Y_2) \land \setSuccOne(Y_1,Y_2)
\]
If $R$ is expressible by some formula $\psi$ in $\msolTwoVar$,
then $R$ is expressible by a formula in prenex normal form,
so suppose $\psi$ is of form
\[
    Q_1 V_1 \ldots Q_n V_n. \psi_0
\]
where $\psi_0$ is quantifier free.  We construct a formula $\psi'$
expressing $R^{*}$ in $\msolOneVar$.  We obtain the matrix
$\psi_0'$ of $\psi'$ by translating $\psi_0$ as follows.  If $x$ is a
first-order variable in $\psi_0$, we represent it with a
second-order variable $X$ denoting a singleton set.  We
replace membership relation $Y(x)$ with subset relation $X
\msolSubset Y$.  We replace $\succZero$ with $\phi_4$ and
$\succOne$ with $\phi_5$.  We construct $\psi'$ by adding
quantifiers to $\psi_0'$ as follows.  Second-order quantifiers
remain the same.  First-order quantifiers are relativized to
range over singleton sets: $\forall x. \psi_i$ becomes $\forall X. \phi_3(X)
\implies \psi_i'$ and $\exists x. \psi_i$ becomes $\exists X.\ \phi_3(X) \land
\psi'_i(X)$.
\end{proofsketch}

We can view $\msolOneVar$ as a first-order structure with
the domain $2^{\{0,1\}^{*}}$.  We show how to embed
$\msolOneVar$ into the first-order theory of structural
subtyping.

We define the first-order structure of structural subtyping
of \emph{recursive} types similarly to the corresponding
structure for non-recursive types in
Section~\ref{sec:twoconst}; the only difference is that the
domain contains both finite and \emph{infinite} terms.
Infinite terms correspond to infinite
trees~\cite{Courcelle83FundamentalPropertiesInfiniteTrees,
  Maher88CompleteAxiomatizationsAlgebrasTrees}.

We define infinite trees as follows.  We use alphabet
$\{l,r\}$ to denote paths in the tree.
A {\em tree domain} $D$ is a finite or infinite subset of the
set $\{l,r\}^{*}$ such that:
\begin{enumerate}
\item $D$ is prefix-closed: if $w \in \{l,r\}^{*}$, $x \in \{l,r\}$ then \\
     $w \cdot x \in D$ implies $w \in D$;
\item if $w \in D$ then
      exactly one of the following two properties hold:
      \begin{enumerate}
      \item $w$ is an interior node: $\{w \cdot l,w \cdot r\} \subseteq D$
      \item $w$ is a leaf: $\{w \cdot l,w \cdot r\} \cap D = \emptyset$.
      \end{enumerate}
\end{enumerate}
A {\em tree} with a tree domain $D$ is a total function $T$
from the set of leaves of $D$ to the set $\{a,b\}$.  

Note that the tree domain $D$ of a tree $T$ can be
reconstructed from $T$ as the prefix closure of the domain
of the graph of function $T$; we write $\treeDomain(T)$ for
the tree domain of tree $T$.

Two trees are equal if they are equal as functions.  Hence,
equal trees have equal function domains and equal tree
domains.

We say that $T_1 \leq T_2$ iff
$\treeDomain(T_1)=\treeDomain(T_2)$ and $T_1(w) \leq_0 T_2(w)$
for every word $w \in \treeDomain(T_1)$.  Here $\leq_0$ is the
relation $\{ \tu{a,a},\tu{a,b},\tu{b,b}\}$.

If $T_1$ and $T_2$ are trees, then $g(T_1,T_2)$ denotes
the tree $T$ such that
\[
   \treeDomain(T) = \{ l \cdot w \mid w \in T_1 \} \cup \{ r \cdot w \mid w \in T_2 \}
\]
\[ 
    T(l \cdot w) = T_1(w), \qquad \mbox{ if } w \in T_1
\]
\[
    T(r \cdot w) = T_2(w), \qquad \mbox{ if } w \in T_2
\]
Let $\InfTrees$ denote the set of all infinite trees.  The
structural subtyping structure is the structure $\InfTheory
= \tu{\InfTrees,g,a,b,\leq}$.  $\InfTheory$ is an infinite-term
counterpart to the structure $\binarySubtyping$ from
Section~\ref{sec:twoconst}.

Similarly to the case of finite terms, define the relation $\sim$ of
``being of the same shape'' in $\InfTheory$ by
\[
    t_1 \sim t_2  \ \equiv\ \exists t_0.\ t_0 \leq t_1 \land t_0 \leq t_2
\]
Observe that $t_1 \sim t_2$ iff $\treeDomain(t_1)=\treeDomain(t_2)$.
% ..., perhaps we should use the term ``domain'' instead of ``shape''
% even in quantifier elimination?

We next present an embedding $\msolEmbed$ of $\msolOneVar$
into $\InfTheory$.  The image of
the embedding $\msolEmbed$ are the infinite trees that are
in the same $\sim$-equivalence-class with the tree $t_e$.
We define $t_e$ as the unique solution of the equation:
\[
    t_e = g(g(t_e,t_e),a)
\]
% Why is the solution unique?  Does prefix-closure guarantee that?!
Trees in the $\sim$-equivalence class of $t_e$ have the tree
domain $D = \treeDomain(t_e)$ given by the regular
context-free grammar
\[
    \Dnt \to \epsilon \mid r \mid l \mid lr\Dnt \mid ll\Dnt
\]
whereas the leaves $L$ of $D$ are given by the context-free grammar
\[
    \Lnt \to \epsilon \mid r \mid lr \Lnt \mid ll \Lnt
\]
or the regular expression $(lr|ll)^{*}r$.  Let $h$ be the
homomorphism of words from $\{0,1\}^{*}$ to $\{l,r\}^{*}$ such that
\[\begin{array}{rcl}
  h(0) &=& ll \mnl
  h(1) &=& lr
\end{array}\]
If $w = a_1\ldots a_n$ is a word, then $w^R$ denotes the reverse
of the word, $w^R = a_n \ldots a_1$.

We define the embedding $\msolEmbed$ to map a set $Y \subseteq
\{0,1\}^{*}$ into the unique tree $t$ such that $t \sim t_e$ and
for every $w \in \{0,1\}^{*}$,
\begin{equation}
  w \in Y  \iff  T(h(w^R) \cdot r) = b
\end{equation}
Observe that $\msolEmbed(\emptyset)=t_e$.  Define formulas
$\termSuccZero(t_1,t_2)$ and $\termSuccOne(t_1,t_2)$ as
follows:
\[\begin{array}{rcl}
\termSuccZero(t_1,t_2) &\equiv& t_2 = g(g(t_1,t_e),t_e) \mnl
\termSuccOne(t_1,t_2) &\equiv& t_2 = g(g(t_e,t_1),t_e) \mnl
\end{array}\]
It is straightforward to show that $\msolEmbed$ is an
injection and that $\msolEmbed$ maps relation $\msolSubset$
into $\leq$, relation $\setSuccZero$ into $\termSuccZero$, and
relation $\setSuccOne$ into $\termSuccOne$.  Moreover, the
range of $\msolEmbed$ is the set of all terms $t$ such that
$\termShape{t}=s_e$ where $s_e = \termShape{t_e}$.

\subsection{A Decidable Substructure}
\label{sec:embeddedTree}

Section~\ref{sec:msolEmbedding} shows that terms of shape
$s_e$ form a substructure within $\InfTheory$ that is
isomorphic to $\msolOneVar$.  In this section we consider
the following converse problem.

Consider the formulas $\BoundedFormulas$ that, instead of
quantifiers $\exists,\forall$, contain bounded quantifiers
$\existsE,\forallE$ that range over the elements of the
set
\begin{equation*}
  T_e = \{ t \mid \termShape{t} = s_e \}
\end{equation*}
We show that the set of closed formulas from
$\BoundedFormulas$ that are true in $\InfTheory$ is
decidable.

Although the quantifiers are bounded, terms in this logic can
still denote elements of shape other than $s_e$.  For example,
the in the atomic formula
\begin{equation*}
  g(x_1,x_2) \leq g(x_3,g(g(x_4,x_5),b))
\end{equation*}
the term $g(x_1,x_2)$ denotes a term of the shape
$\gs(s_e,s_e)$.  First we show that all atomic formulas are
of one of the following forms:
\begin{enumerate}
\item $x_0 = g(g(x_1,x_2),a)$;
\item $x_0 = g(g(x_1,x_2),b)$;
\item $x_1 = x_2$;
\item $x_1 \leq x_2$.
\end{enumerate}
Consider an atomic formula $t_1 = t_2$.  The key idea is
that if $\termShape{t_1} \neq \termShape{t_2}$ then the formula $t_1=t_2$
is $\boolFalse$.

If none of the term $t_1$ and $t_2$ is a variable then one
of them is a constant or a constructor application.  If $t_1
\equiv g(t_{11},t_{12})$ then either $t_1=t_2$ is $\boolFalse$ or
$t_2 \equiv g(t_{21},t_{22})$ for some $t_{21},t_{22}$.  We may
therefore decompose $t_1=t_2$ into $t_{11}=t_{21}$ and
$t_{12}=t_{22}$.  By repeating this decomposition we arrive
at terms of form $t_1=t_2$ where both $t_1$ and $t_2$ are
constants or at the equality of form $x_0 = t(x_1,\ldots,x_n)$.
The equalities between the constants can be trivially
evaluated.  This leaves only terms of form $x_0 =
t(x_1,\ldots,x_n)$.  Let $\ts(\xs_1,\ldots,\xs_n)$ be a shape term
that results from replacing $a$ and $b$ with $\cs$ and
replacing $g$ with $\gs$ in $t$.  Because all variables
range over $T_e$, we conclude that $x_0 =
t(x_1,\ldots,x_n)$ can be true only if
\begin{equation*} \label{eqn:shapesConseq}
  s_e = \ts(s_e,\ldots,s_e)
\end{equation*}
If $t(x_1,\ldots,x_n) \in \{ a, b\}$ is then~(\ref{eqn:shapesConseq})
is false.  If $t(x_1,\ldots,x_n) \equiv x_1$, we obtain formula of the
desired form.  So assume $t(x_1,\ldots,x_n) \equiv g(t_{21},t_{22})$.
Then $\termShape{t_{21}} = \gs(s_e,s_e)$ and
$\termShape{t_{22}}=\cs$.  Therefore, $t_{21} \equiv
g(t_{211},t_{212})$ where either
$\termShape{t_{211}}=\termShape{t_{212}}=s_e$ or $t_1=t_2$
is false.  Similarly, either $t_{22} \in \{ a,b \}$ or $t_1=t_2$
is false.  Therefore, $t(x_1,\ldots,x_n) \equiv
g(g(t_{211},t_{212}),a)$, $t(x_1,\ldots,x_n) \equiv
g(g(t_{211},t_{212}),b)$, or $t_1=t_2$ is false.  If 
$t(x_1,\ldots,x_n) \equiv
g(g(t_{211},t_{212}),a)$ then we may replace the $t_1=t_2$ with
the formula
\begin{equation*}
  \existsE y_1,y_2.\ x_0 = g(g(y_1,y_2),a) \ \land \ y_1 = t_{211} \ \land \ y_2 = t_{212}
\end{equation*}
and similarly in the other case.  By continuing this process
by the induction on the structure of the term $t(x_1,\ldots,x_n)$
we either conclude that $t_1=t_2$ is false, or we conclude
that $t_1=t_2$ is equivalent to a conjunction of formulas
of the desired form.

Conversion of atomic formula of form $t_1 \leq t_2$ is
analogous to the conversion of formulas $t_1 = t_2$.

To see the decidability it now suffices to convert the
formulas of the form $x_0 = g(g(x_1,x_2),a)$ and $x_0 =
g(g(x_1,x_2),b)$ into formulas $\termSuccZero(t_1,t_2)$ and
$\termSuccOne(t_1,t_2)$.  Expressibility of $x_0 =
g(g(x_1,x_2),a)$ follows from the fact that the following
relationship between $X_0,X_1,X_2$ is expressible in MSOL:
\begin{equation*}
  X_0 = \{ w \cdot 0 \mid w \in X_1 \} \cup \{ w \cdot 1 \mid w \in X_2 \}
\end{equation*}
Similarly, the expressibility of $x_0 = g(g(x_1,x_2),b)$
follows from the fact that
\begin{equation*}
  X_0 = \{ w \cdot 0 \mid w \in X_1 \} \cup \{ w \cdot 1 \mid w \in X_2 \} \cup \{ \epsilon \}
\end{equation*}
is expressible in MSOL.  We conclude that the set of closed
$\BoundedFormulas$ formulas that are true in $\InfTheory$ is
decidable.

\subsection{Embedding Terms into Terms}
\label{sec:termsIntoTerms}

We next give an embedding of the set of all terms into
$T_e$.  As in Section~\ref{sec:msolEmbedding} $t_e$ be the
unique solution of the equation $t_e = g(g(t_e,t_e),a)$ and
let
\begin{equation*}
  t_4(x_1,x_2,x_3,x_4) \equiv
    g(g(
        g(g(x_1,x_2),x_3), 
        t_e
        ),x_4)
\end{equation*}
Define
\begin{equation*}
  \begin{array}{l}
    t_a \equiv t_4(t_e,t_e,a,a) \mnl
    t_b \equiv t_4(t_e,t_e,a,b) \mnl
    t_g(x_1,x_2) \equiv t_4(x_1,x_2,b,b) \mnl
  \end{array}
\end{equation*}
Then define the homomorphism $\hT$ from the set of all terms
to the set $T_e$ by
\begin{equation*}
  \begin{array}{rcl}
    \hT(a) &=& t_a \mnl
    \hT(b) &=& t_b \mnl
    \hT(g(t_1,t_2)) &=& t_g(\hT(t_1),\hT(t_2))
  \end{array}
\end{equation*}
Then $\hT$ is embedding of the set of all terms into the
subset subset $T_e$ of all terms.  The term algebra
operations $a,b,g$ map to $t_a,t_b,t_g$ and $\leq$ maps to $\leq$.

Note that, if it were possible to define a predicate $P(t)$
such that
\begin{equation} \label{eqn:rangePredicate}
  P(x) \iff \exists y. \hT(y) = x
\end{equation}
then we could express all statements of $\InfTheory$ within
the $\BoundedFormulas$ subtheory, and therefore $\InfTheory$
would be decidable.

The fundamental problem with specifying $P(x)$ is not the
use of two bits to encode the three possible elements
$\{a,b,g\}$, but the constraint that if a term contains a
subterm of the form $t_4(t_1,t_2,a,a)$ or $t_4(t_1,t_2,a,b)$
at some even depth, then $t_1 \equiv t_2 \equiv t_e$.  Compared to the
relationships given by constructor $g$, this constraint
requires taking about successor relation at the opposite
side of the paths within a tree, see
Section~\ref{sec:structuralUndec}.

%% This should not be too surprising given \ref{sec:msolEmbedding}.
%% See also \cite{Courcelle83FundamentalPropertiesInfiniteTrees}.

%% Consider an existentially quantified conjunction that
%% specifies a shape as a unique rational tree $\us$:
%% \[
%%    \exists \us.  \phi(\us)
%% \]
%% Now consider a formula
%% \[
%%    \exists \us. \phi(\us) \ \land \ \psi(\us)
%% \]
%% where all quantification over terms in $\psi(\us)$ is bounded
%% to be over terms of $\us$ shape.  Then $\psi(\us)$ can be
%% translated to a formula in a MSOL of infinite tree and can
%% therefore be decided.

\subsection{Subtyping Trees of Known Shape}
\label{sec:fixedRational}

We next argue that if we allow the logic to have a copy of
bounded quantifiers $\exists_s,\forall_s$ for every \emph{constant}
shape $s$, we obtain a decidable theory.  To denote constant
shapes in a finite number of symbols we consider in addition
to term algebra symbols $\gs,\cs$ the expressions that yield
solutions of mutually recursive equations on shapes; the
details of the representation of types are not crucial for
our argument, see e.g.\ 
\cite{Courcelle83FundamentalPropertiesInfiniteTrees}

Consider a closed formula in such language.  Because every
variable has an associated constant shape, we can compute
the set of all shapes occurring in the formula.  This means
that all variables of the formula range over a finite known
set of shapes.  This allows us to define the predicate $P$
given by~(\ref{eqn:rangePredicate}) as a disjunction of
cases, one case for every shape.  Define $\hmin$, $\hmax$
functions that take a shape and produce a lower and upper
bound for terms of that shape:
\begin{equation*}
  \begin{array}{rcl}
    \hmin(\cs) &=& t_a \mnl
    \hmin(\gs(\ts_1,\ts_2)) &=& t_g(\hmin(\ts_1),\hmin(\ts_2)) \mnl
    \hmax(\cs) &=& t_b \mnl
    \hmax(\gs(\ts_1,\ts_2)) &=& t_g(\hmax(\ts_1),\hmax(\ts_2))
  \end{array}
\end{equation*}
If $s_1,\ldots,s_n$ is the list of shapes occurring in a formula,
we then define a predicate $P$ specific to that formula
by
\begin{equation*}
  P(t) = \bigvee_{i=1}^n (\hmin(s_i) \leq t \land t \leq \hmax(t))
\end{equation*}
We can therefore define $P(t)$ and use it to translate the
formula into a $\BoundedFormulas$ formula of the same truth
value.  Therefore, structural subtyping with quantification
bounded to constant shapes is decidable.

For decidability of the structural subtyping recursive types
it would be interesting to examine the decision procedure
for MSOL and determine whether there is some uniformity in
it that would allow us to handle even quantification over
shapes that are determined by variables.

\subsection{Recursive Feature Trees}
\label{sec:feature}

We next remark that certain notion of subtyping of recursive
feature trees is decidable.  By a feature tree we mean an
infinite tree built using a constructor which takes other
feature trees and an optional node label as an argument.  In
this section we consider the simple case of one binary
constructor $f$ and assume only one label denoted by $1$.
Hence, an empty feature tree is a feature tree, and if $t_1$
and $t_2$ are feature trees then so are $f^{\epsilon}(t_1,t_2)$ and
$f^{1}(t_1,t_2)$.  We represent an empty feature tree $e$ by
an infinite tree that has all features $\epsilon$.  We compare
feature trees as follows.  Let $\leq$ be defined on the
features $\{ \epsilon, 1 \}$ as the relation $\{ \tu{\epsilon,\epsilon}, \tu{\epsilon,1},
\tu{1,1} \}$.  Define $\leq$ on trees as the least relation such
that:
\begin{enumerate}
\item $e \leq t$ for all terms $t$;
\item $t_1 \leq t_1'$ and $t_2 \leq t_2'$ implies
\begin{equation*}
  f^{r_1}(t_1,t_2) \leq f^{r_2}(t_1',t_2')
\end{equation*}
for all $r_1,r_2 \in \{ \epsilon,1\}$ such that $r_1 \leq r_2$.
\end{enumerate}

The decidability of feature trees follows from
Section~\ref{sec:msolEmbedding} because of the isomorphism
$\hF$ between the set of terms $T_e$ and the set of feature
trees.  Here $\hF$ is defined by:
\begin{equation*}
  \begin{array}{rcl}
    \hF(e) &=& t_e \mnl
    \hF(f^{\epsilon}(t_1,t_2)) &=& g(\hF(t_1),\hF(t_2),a) \mnl
    \hF(f^{1}(t_1,t_2)) &=& g(\hF(t_1),\hF(t_2),b)
  \end{array}
\end{equation*}

The feature trees as we defined them have a limited feature
and node label alphabet.  This is not a fundamental problem.
Muchnik's theorem
\cite{Walukiewicz96MonadicSecondOrderLogicTreeLikeStructures}
gives the decidability of MSOL of trees over arbitrary
decidable structures.  It is reasonable to expect that the
decidability of MSOL over decidable structures yields a
generalization of the result of
Section~\ref{sec:msolEmbedding} and therefore the
decidability of feature trees with a richer vocabulary of
features.

The crucial property of our definition of feature trees is
that features can appear in any node of the tree.  Hence,
there are no prefix closure requirements on trees as in
Section~\ref{sec:termsIntoTerms}, which is responsible for
relatively simple reduction to MSOL.

\subsection{Reversed Binary Tree with Prefix-Closed Sets}
\label{sec:structuralUndec}

It is instructive to compare the difficulties our approach
faces in showing the decidability of structural subtyping of
recursive types with the difficulties reported in
\cite{SuETAL02FirstOrderTheorySubtypingConstraints}.  In
\cite[Section
5.3]{SuETAL02FirstOrderTheorySubtypingConstraints} the
authors remark that the difficulty with applying tree
automata is that the set $x=f(y,z)$ is not regular.  By
reversing the set of paths in a tree representing a term we
have shown in Section~\ref{sec:msolEmbedding} that the
relationship $x=f(y,z)$ becomes expressible.  However, the
difficulty now becomes specifying a set of words that
represents a valid term, because there is no immediate way
of stating that a set of words is prefix-closed.  If we add
an operation that allows expressing relationship at both
``ends'' of the words, we obtain a structure whose MSOL is
undecidable due to the following result \cite[Page
183]{Thomas90AutomataInfiniteObjects}.
\begin{theorem} \label{thm:fromBook}
  MSOL theory of the structure with two successor operations
  $w \cdot 0$ and $w \cdot 1$ and one inverse successor operation $0
  \cdot w$ is undecidable.
\end{theorem}
The case that is of interest of us is the dual to
Theorem~\ref{thm:fromBook} under the word-reversing
isomorphism: a structure with operations $0 \cdot w$, $1 \cdot w$,
$w \cdot 0$ has undecidable MSOL closed formulas.

Instead of expressing prefix-closure using operations $w \cdot
0$, $w \cdot 1$, let us consider MSOL over the structure that
contains only operations $0 \cdot w$ and $1 \cdot w$, but where all
second-order variables range over prefix-closed sets.  This
logic also turns out to be undecidable.

Let $\PCl$ be the set of prefix-closed sets.
For each word $w$, there exists the smallest $\PCl$
set containing $w$, namely the set $C(w)$ given by:
\[
   C(w) = \{ w' \mid w' \prec w \}
\]
Every subset of $C(w)$ in $\PCl$ is a of the form $C(w_1)$
for some word $w_1$.  Define $\PrefixSuccZero$ and
$\PrefixSuccOne$ on $\PCl$ by:
\begin{equation*}
  \begin{array}{rcl}
    \PrefixSuccZero(X_1,X_2) &=& \exists w.\ X_1 = C(w) \land X_2 = C(0 \cdot w) \mnl
    \PrefixSuccOne(X_1,X_2) &=& \exists w.\ X_1 = C(w) \land X_2 = C(1 \cdot w)
  \end{array}
\end{equation*}
Consider a monadic theory $\PrefT$ with relations
$\PrefixSuccZero$ and $\PrefixSuccOne$ where second-order
variables range over the subsets of $\PCl$.  It is easy to
see that $\PrefT$ corresponds to the first-order theory of
\emph{non-structural} subtyping of recursive types, with
subset relation $\subseteq$ corresponding to subtype relation $\leq$,
empty set corresponding to the least type $\bot$,
$\PrefixSuccZero(X_1,X_2)$ corresponding to $X_2 =
f(X_1,\bot)$, and $\PrefixSuccOne(X_1,X_2)$ corresponding to
$X_2 = f(\bot,X_2)$.  The first-order theory of non-structural
subtyping was shown undecidable in
\cite{SuETAL02FirstOrderTheorySubtypingConstraints}, so
$\PrefT$ is undecidable.  An interesting open problem is the
decidability of fragments of the first-order theory of
structural subtyping.  This problem translates directly to
the decidability of the fragments of $\PrefT$, a monadic
theory with prefix-closed sets, or, under the word-reversal
isomorphism, the decidability of fragments of the monadic
theory of two successor symbols with suffix-closed sets.

\nocite{Malcev71MetamathematicsAlgebraicSystems,
KozenETAL95EfficientRecursiveSubtyping,
Pottier01SimplifyingSubtypingConstraints,
GurevichHarrington82TreesAutomataGames,
Walukiewicz96MonadicSecondOrderLogicTreeLikeStructures,
BoergerETAL97ClassicalDecisionProblem,
Comon91Disunification, ComonLescanne89EquationalProblemsDisunification}

%BackofenTreinen98HowWinGameFeatures,
%MuellerETAL01OrderingConstraintsFeatureTrees,
%SeynhaeveETAL01GridStructuresUndecidableConstraint
%Treinen02FirstOrderTheoriesConcreteDomains
%Treinen97FeatureTreesArbitraryStructures

%%% Local Variables: 
%%% mode: latex
%%% TeX-master: "main"
%%% End: 
% LocalWords:  KozenETAL EfficientRecursiveSubtyping Pottier GurevichHarrington
% LocalWords:  SimplifyingSubtypingConstraints TreesAutomataGames Walukiewicz
% LocalWords:  MonadicSecondOrderLogicTreeLikeStructures BoergerETAL Comon
% LocalWords:  ClassicalDecisionProblem Disunification ComonLescanne Treinen
% LocalWords:  EquationalProblemsDisunification BackofenTreinen MuellerETAL
% LocalWords:  HowWinGameFeatures OrderingConstraintsFeatureTrees SeynhaeveETAL
% LocalWords:  GridStructuresUndecidableConstraint
% LocalWords:  FirstOrderTheoriesConcreteDomains
% LocalWords:  FeatureTreesArbitraryStructures

\section{Conclusion}

In this paper we presented a quantifier elimination
procedure for the first-order theory of structural subtyping
of non-recursive types.  Our proof uses quantifier
elimination.  Our decidability proof for the first-order
theory of structural subtyping clarifies the structure of
the theory of structural subtyping by introducing explicitly
the notion of \emph{shape} of a term.

We presented the proof in several stages with the hope of
making the paper more accessible and self-contained.  Our
result on the decidability of $\Sigma$-term-power is more general
than the decidability of structural subtyping non-recursive
types, because we allow even infinite decidable base
structures for primitive types.  We view this decidability
result as an interesting generalization of the decidability
for term algebras and decidability of products of decidable
theories.  This generalization is potentially useful in
theorem proving and program verification.

Of potential interest might be the study of axiomatizability
properties; the quantifier elimination approach is
appropriate for this purpose
\cite{Malcev71MetamathematicsAlgebraicSystems,
  Maher88CompleteAxiomatizationsAlgebrasTrees}, we did not
pay much attention to this because we view the language and
the mechanism for specifying the axioms of secondary
importance.

Our goal in describing quantifier elimination procedure was
to argue the decidability of the theory of structural
subtyping.  While it should be relatively easy to extract an
algorithm from our proofs, we did not give a formal
description of the decision procedure.  One possible
formulation of the decision procedure would be a
term-rewriting system such as
\cite{ComonLescanne89EquationalProblemsDisunification}; this
formulation is also appropriate for implementation within a
theorem prover.  Our approach eliminates quantifiers as
opposed to quantifier alternations.  For that purpose we
extended the language with partial functions.  The use of
Kleene logic for partial functions seems to preserve most of
the properties of two valued logic and appears to agree with
the way partial functions are used in informal mathematical
practice.  An alternative direction for proving decidability
of structural subtyping would be to use Ehrenfeucht-Fraisse
games \cite[Page 405]{Thomas97LanguagesAutomataLogic};
\cite{FerranteRackoff79ComputationalComplexityLogicalTheories}
uses techniques based on games to study both the
decidability and the computational complexity of theories.

The complexity of our the decidability for structural
subtyping non-recursive types is non-elementary and is a
consequence of the non-elementary complexity of the term
algebra, whose elements and operations are present in the
theory of structural subtyping.  Tools like MONA
\cite{KlarlundETAL00MONA} show that non-elementary
complexity does not necessarily make the implementation of a
decision procedure uninteresting.  An interesting property
of quantifier elimination is that it can be applied
partially to elimination an innermost quantifier from some
formula.  This property makes our decision procedure
applicable as part of an interactive theorem prover or a
subroutine of a more general decision procedure.

In this paper we have left open the decidability of
structural subtyping of \emph{recursive} types, giving only
a few remarks in Section~\ref{sec:connectionWithMSOL}.  In
particular we have observed in
Section~\ref{sec:msolEmbedding} that every formula in the
monadic second-order theory of the infinite binary tree
\cite[Page 317]{BoergerETAL97ClassicalDecisionProblem} has a
corresponding formula in the first-order theory of
structural subtyping of recursive types.  In that sense, the
decision problem for structural subtyping recursive types is
at least as hard as the decision problem for the monadic
second-order logic interpreted over the infinite binary
tree.  This observation is relevant for two reasons.  

First, it is unlikely that a minor modification of the
quantifier elimination technique we used to show the
decidability of structural subtyping non-recursive types can
be used to show the decidability of recursive types.
Because of the embedding in Section~\ref{sec:msolEmbedding}
such a quantifier-elimination proof would have to subsume
the determinization of tree automata over infinite trees.

Second, the embedding suggests even greater difficulties in
implementing a decision procedure for the first-order theory
of structural subtyping (provided that it exists).  While we
know at least one interesting example of \emph{weak} monadic
second-order logic decision procedure, namely
\cite{KlarlundETAL00MONA} we are not aware of any
implementation of the full monadic second-order logic
decision procedure for the infinite tree.

The relationship between the non-structural as well as
structural subtyping and monadic second-order logic of the
infinite binary tree and tree like structures
\cite{Walukiewicz02MonadicSecondOrderLogicTreeLikeStructures}
requires further study.  In that respect the work on feature
trees
\cite{MuellerNiehren00OrderingConstraintsFeatureTreesMSOL,
  MuellerETAL01OrderingConstraintsFeatureTrees} appears
particularly relevant.

%%% Local Variables: 
%%% mode: latex
%%% TeX-master: "main"
%%% End: 

% LocalWords:  axiomatizability Kleene Ehrenfeucht Fraisse determinization

\paragraph{Acknowledgements}  The first author would like to
thank Albert Meyer for pointing out to the work
\cite{FerranteRackoff79ComputationalComplexityLogicalTheories},
and Jens Palsberg and Jakob Rehof for useful discussions
about the subject of this paper.
% LocalWords:  Jens Palsberg Jakob Rehof

\bibliographystyle{plain}
\bibliography{pnew}

\end{document}